\documentclass[]{IEEEtran}
\usepackage{verbatim}
\usepackage[utf8x]{inputenc}
\usepackage[pdftex]{graphicx}
\usepackage{tabularx}
\usepackage{booktabs}
\usepackage[table]{xcolor}
\usepackage{multirow}
\usepackage{siunitx}
\usepackage{tabu}
\usepackage{booktabs}
\usepackage{ctable}
\usepackage[cmex10]{amsmath}
\usepackage{amssymb}
\usepackage{cite}
\usepackage{textcomp}
\usepackage{kbordermatrix}
\usepackage{array}
\newcolumntype{C}[1]{>{\centering\let\newline\\\arraybackslash\hspace{0pt}}m{#1}}

\graphicspath{{./Figures/}}

\ifCLASSOPTIONcompsoc
\usepackage[tight,normalsize,sf,SF]{subfigure}
\else
\usepackage[tight,footnotesize]{subfigure}
\fi

\begin{document}
\title{Reconfigurable Wireless Networks} 
\author{Amr El-Mougy,~ Mohamed Ibnkahla, Ghaith Hattab, and Waleed Ejaz\thanks{The authors are with the department of Electrical and Computer Engineering at Queen's University, Kingston, ON, Canada (email: \{mohamed.ibnkahla; 6ema; g.hattab; we5\}@queensu.ca).}}

\maketitle

\begin{abstract}
Driven by the advent of sophisticated and ubiquitous applications, and the ever-growing need for information, wireless networks are without a doubt steadily evolving into profoundly more complex and dynamic systems. The user demands are progressively rampant, while application requirements continue to expand in both range and diversity. Future wireless networks, therefore, must be equipped with the ability to handle numerous, albeit challenging requirements. Network reconfiguration, considered as a prominent network paradigm, is envisioned to play a key role in leveraging future network performance and considerably advancing current user experiences. This paper presents a comprehensive overview of reconfigurable wireless networks and an in-depth analysis of reconfiguration at all layers of the protocol stack. Such networks characteristically possess the ability to reconfigure and adapt their hardware and software components and architectures, thus enabling flexible delivery of broad services, as well as sustaining robust operation under highly dynamic conditions. The paper offers a unifying framework for research in reconfigurable wireless networks. This should provide the reader with a holistic view of concepts, methods, and strategies in reconfigurable wireless networks. Focus is given to reconfigurable systems in relatively new and emerging research areas such as cognitive radio networks, cross-layer reconfiguration and software-defined networks. In addition, modern networks have to be intelligent and capable of self-organization.  Thus, this paper discusses the concept of network intelligence as a means to enable reconfiguration in highly complex and dynamic networks. Key processes in network intelligence, such as reasoning, learning, and context-awareness are presented to illustrate how these methods can take reconfiguration to a new level.  Finally, the paper is supported with several examples and case studies showing the tremendous impact of reconfiguration on wireless networks.
\end{abstract}


\section{Introduction}
The last few years have witnessed an incredible revolution in wireless networks. The number of wireless devices worldwide continues to increase at an enormous rate and the applications of wireless networks are getting increasingly versatile, sophisticated, and ubiquitous. Thus, future wireless networks are expected to operate under several challenging conditions. First, they are expected to provide efficient always-on reliable high data rate access. Second, they have to be energy-efficient for the sake of user convenience and for environmental concerns, since the widespread of wireless devices leads to significant increase in their total energy consumption and greenhouse gas (GHG) emission. Indeed, information and communication systems consume over 10\% of the world’s total energy \cite{Mills1}. Third, wireless networks have to consider challenging and changing user demands (video traffic over mobile devices now accounts for over 50\% of global data traffic \cite{Cisco1}). Fourth, users expect their devices to be \emph{smart}, to learn about their interests, and to provide them with information that is timely and convenient (over 2 billion smartphone subscriptions have become available in 2014 \cite{Ericsson1}). Last but not least, the spread of cloud resources and big data centres, together with the advent of the Internet of Everything (IoE), promise to provide wireless users with unprecedented services and opportunities.

These challenges have forced designers of wireless systems to completely change their traditional design methodologies. Primarily, the the above challenges mean that modern wireless networks have to be able to consider a wide range of changing objects and conditions. Thus, modern wireless systems have to be able to adapt to the evolving nature of wireless networks and services. In other words, they have to be reconfigurable across multiple layers of the protocol stack. For example, modern devices have to adapt their transmission schemes at the physical (PHY) layer in order to operate in time-varying conditions or high mobility, they have to reconfigure their routing protocols to operate in heterogeneous or evolving networks, and they have to adapt to their application requirements in order to satisfy a wide range of user interests. Therefore, reconfiguration will be a key enabler for next-generation wireless systems and services.

\begin{figure}[!b]
\centering
\includegraphics[width=3.5in]{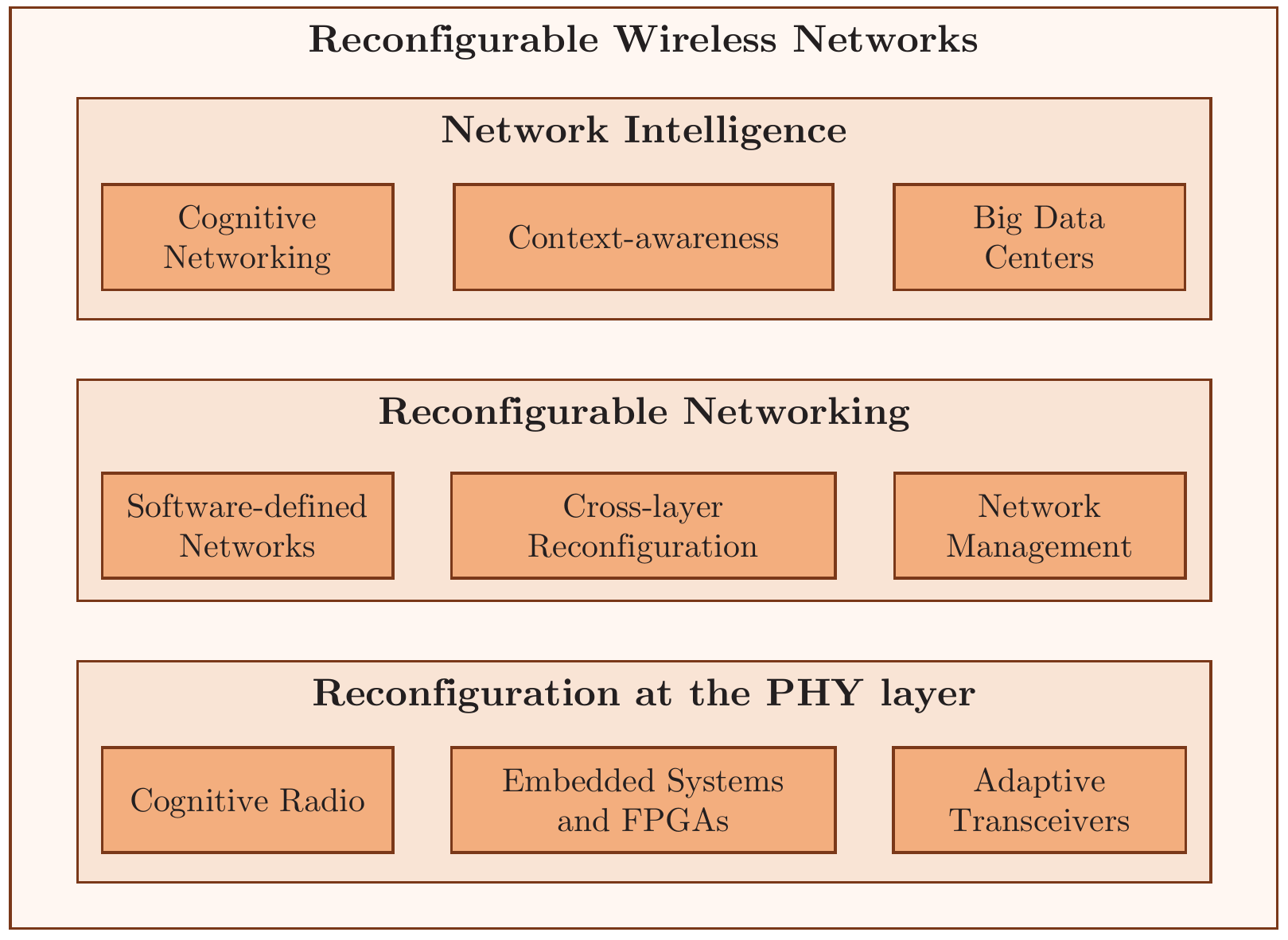}
\caption{Classification of research in reconfigurable networks and examples of research topics.}
\label{fig:Amr1}
\end{figure}

We can classify research in reconfigurable networks into three levels: Reconfiguration at the PHY layer, reconfigurable networking, and network intelligence. This hierarchy is shown in Fig. \ref{fig:Amr1} with a few examples of research topics at each level.

Reconfiguration at the PHY layer covers a broad range of topics such as adaptive modulation, antenna beamforming, and software-defined radio (SDR). Work in these functions has been covered comprehensively in the literature \cite{Ibnkahla3,Arslan1,Ibnkahla4,Clerckx1,Sun_Shaohui1,Sternad1,Mietzner1,Miridakis1,Ulversoy1}. Section II focuses on reconfiguration at the PHY layer, where network-wide considerations are taken into account during device reconfiguration. With the recent emergence of cognitive radio networks (CRNs), and the significant role CRNs are expected to play in next-generation wireless networks, a special discussion on CRNs is given in Section II.

Reconfigurable networking is concerned with building adaptive networking firmware that can address changing network topologies or application requirements. Thus, Section III studies methodologies of implementing reconfiguration at the networking level. Three particular aspects of reconfigurable networking are considered: Cross-layer design, network management, and software-defined networks (SDN). Several examples are thoroughly analyzed in order to demonstrate how reconfiguration can solve different network challenges. Performance analysis is also presented to show the gain achieved from reconfiguration. For illustration purposes, this paper  considers the layered model presented in Table \ref{tab:Layers}. The model is composed of the application, transport, network, medium access control (MAC), and PHY layers. Cross-layer reconfiguration  is the leading methodology in this area because it has demonstrated significant improvements in the network performance when inter-layer effects are taken into consideration \cite{Ibnkahla3,Edirisinghe1}.  For example, PHY layer parameters such as transmit power and bit error rate (BER) can have profound effects on several processes in other layers of the protocol stack such as scheduling at MAC layer and topology control at the network layer. Therefore, the single-layer design approach can tangibly limit the adaptations that can be implemented. In addition, network management is an important tool for reconfiguring network parameters and protocols from a network-wide perspective, which provides the potential to consider global network objectives. On the other hand, SDN is a promising method that can leverage capabilities in automation and virtualization of networks.

\begin{table}[!b]
\renewcommand{\arraystretch}{1.3}
\small
\caption{Reconfiguration can be performed at all layers}
\label{tab:Layers}
\centering
\begin{tabular}{cp{2.25in}}
\toprule
 \bfseries{Layers}                         & \bfseries{Examples of Possible Reconfigurations}                          \\     \toprule
\multirow{3}{*}{Application}                  &Context information     \\
                                               &User requirements        \\
                                               &User interface          \\ \hline
\multirow{2}{*}{Transport}                     &Number of retransmissions\\
                                                &Congestion control       \\ \hline
\multirow{3}{*}{Network}                       &Routing\\
                                                &Admission control\\
                                                &QoS management\\ \hline
\multirow{3}{*}{MAC}              &Transmission and sleep scheduling\\
                                              &Contention/sensing window\\
                                               &Transmission rate   \\ \hline
\multirow{1}{*}{PHY}                           &Transceiver and antenna reconfiguration \\
\bottomrule
\end{tabular}%
\end{table}

The third level of reconfiguration in wireless networks adopts intelligence and cognitive strategies throughout the wireless system and its surrounding environment \cite{Haykin2}. Intelligence and cognition include reasoning and learning, which aim at improving networking decisions while fulfilling network end-to-end goals. Of particular importance in this level is context-awareness (CA) \cite{Makris1}, which enables networking decisions to be made according to certain relevant information. Here, the goal is to reconfigure the network to suit system-wide architecture/deployment, user interests, or the environment, often through predictions and proactive actions. This level is discussed in Section IV.

It is clear from this introduction that research in reconfigurable wireless networks has quite a wide scope. It also demands expertise from multiple disciplines, mainly engineering and computer science. However, publications in this field have been quite scattered, often focusing on a limited set of issues. In this paper, we provide a unifying framework for research in reconfigurable wireless networks. The objective is to link together ideas from different fields and provide the reader with a holistic view of concepts, methods, and strategies in reconfigurable wireless networks.

The rest of the paper is organized as follows. Section II covers the PHY layer reconfiguration from a network perspective, focusing on CRNs. A case study is provided to examine the performance gains achieved when network-wide reconfiguration is implemented. Section III covers reconfigurable networking, with emphasis on: Cross-layer reconfiguration, network management, and SDNs. A case study is presented to show the performance gains archived when cross-layer reconfiguration is used in machine-to-machine (M2M) networks. Section IV is devoted to cognitive networking and context-awareness. A case study is presented which investigates network management using reasoning and learning in wireless sensor networks. Finally, the main conclusions and open research areas are presented in Section V.

\section{PHY Layer Reconfiguration: A Network Perspective}

PHY layer reconfiguration includes SDR, transceiver reconfiguration,  adaptive antennas and beamforming, adaptive modulation and coding (AMC), etc. \cite{Arslan1,Ibnkahla4,Al-Asady1}. Most publications in this field target device-level reconfiguration that often does not take into consideration the network aspect (even when wireless devices are part of a wider network). This section shows the importance of network-based device reconfiguration. Without loss of generality, and in order to illustrate the different concepts to the reader, this section explores the case of CRNs.

\begin{figure*}[!t]
\centering
\includegraphics[width=5in]{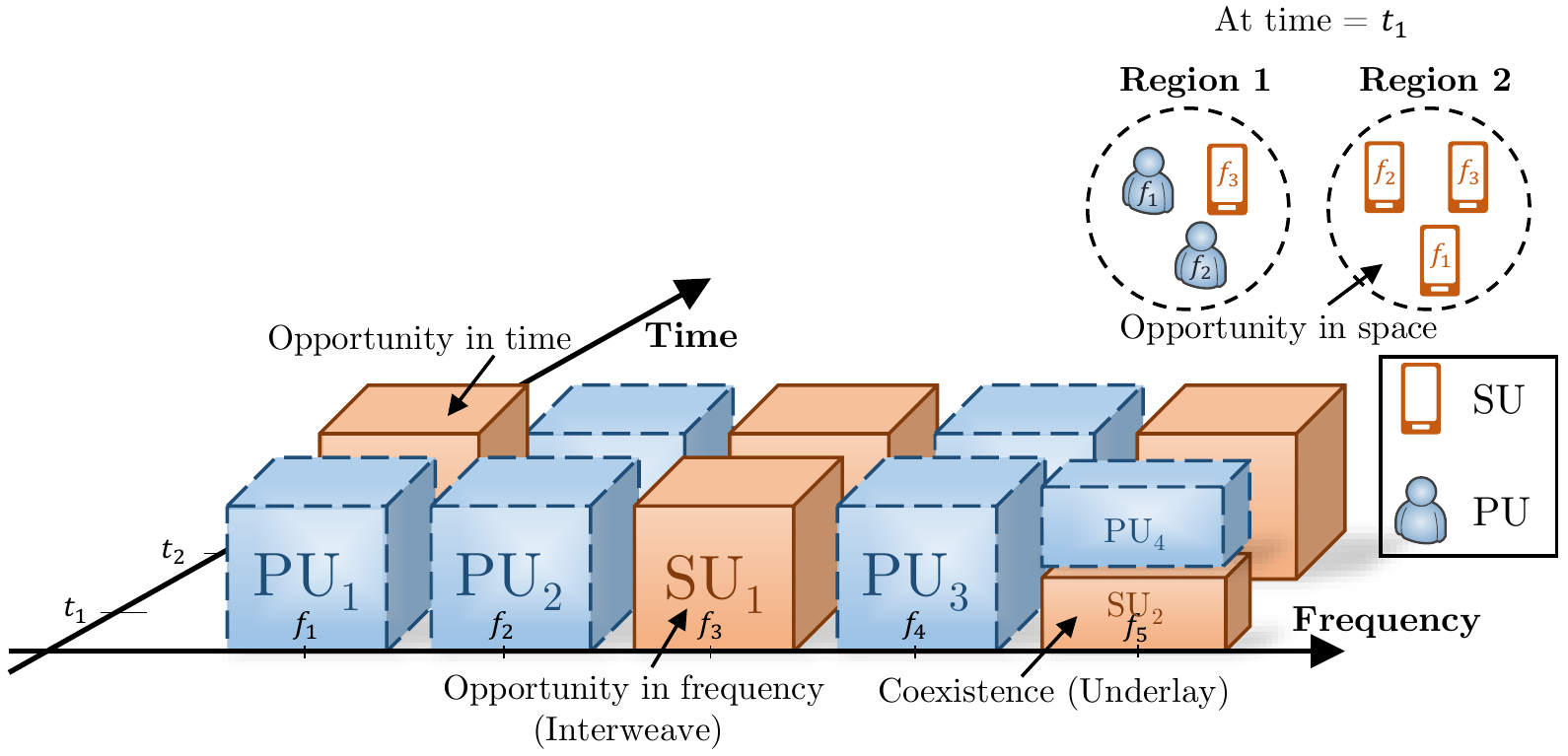}
\caption{There are unused spectral opportunities in time, frequency, and space. SUs can orthogonally exist with PUs (interweave paradigm), or coexist with them and share the resources (underlay paradigm).}
\label{fig:Spectrum}
\end{figure*}

CRNs are envisioned to tackle the problem of low utilization of the spectrum by exploiting spectral opportunities in time, frequency, and space. These opportunities can be used by secondary users (SUs), as illustrated in Fig. \ref{fig:Spectrum}, in situations where the incumbents (also known as  primary users (PUs)) of these resources are idle, or they can be shared by both PUs and SUs if the latter conform to stringent rules to protect the former from harmful interference.

Conventionally, a wireless network is primarily concerned with providing some services at certain performance levels to its users (e.g., throughput, fairness, etc.). However, the CRN must provide guarantees to both SUs and PUs. In particular, the CRN must not degrade the quality-of-service (QoS) of PUs. It must also provide QoS guarantees to SUs. The coexistence of CRNs with heterogenous PU networks (PUNs) is challenging because:
\begin{itemize}
\item Different PUNs may have different interference tolerance levels.
\item Some PUs are static (e.g., TV channels) while others are mobile (e.g., cellular users). Protecting the latter is more challenging.
\item Some PUs transmit at low powers (e.g., wireless microphones), and thus detecting the presence of these incumbents is challenging.
\end{itemize}
In addition, providing a certain QoS for SUs remains a challenging task mainly because:
\begin{itemize}
\item The spectrum access is opportunistic. Hence, the resources are limited, and they must be shared among multiple SUs efficiently and in a fair manner.
\item The PU activity may change. Thus, periodic monitoring of the spectral opportunity is vital, meaning that data transmission interruptions or spectrum handover to other available opportunities are inevitable.
\end{itemize}
To deal with these challenges, network level \emph{reconfigurability} and \emph{learning} are key enablers for effective CRNs.

In the subsequent subsections, we discuss the key functionalities of a CRN, and particularly, the cognitive management module, which acts as the brain of the CRN. Then, we examine different levels of reconfiguration and evaluate the throughput performance of a CRN in the absence and presence of network-based reconfiguration and learning. We show that significant gains can be achieved with proper reconfiguration at the SU device and the network level.

\subsection{Cognitive Management}
\begin{figure}[!t]
\centering
\includegraphics[width=3.5in]{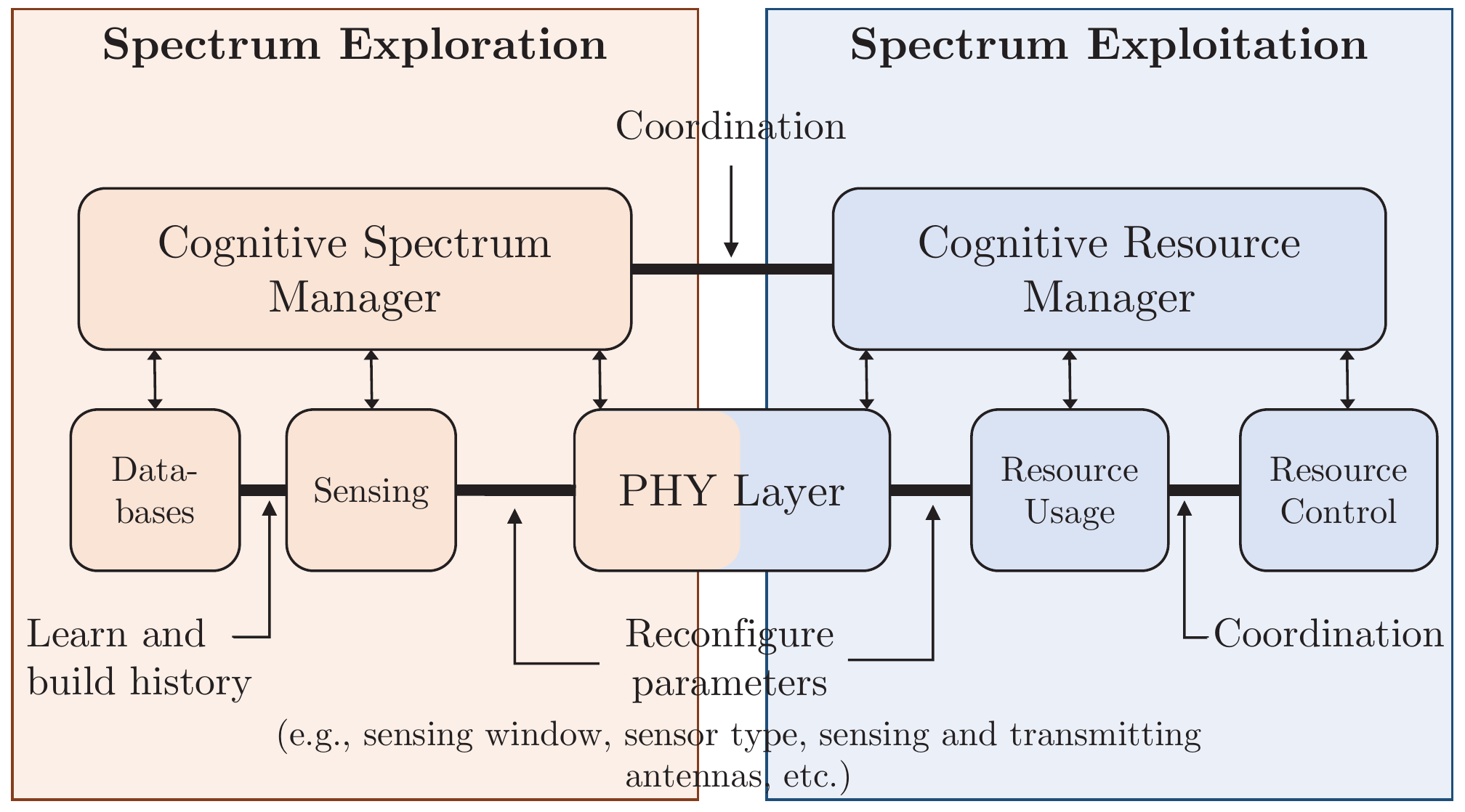}
\caption{Spectrum exploration and exploitation must work in tandem to utilize the spectral opportunities.}
\label{fig:CognitiveManagers}
\end{figure}

The CRN must be equipped with a cognitive management module, which is composed of a \emph{cognitive spectrum} manager and a \emph{cognitive resource} manager as shown in Fig. \ref{fig:CognitiveManagers} \cite{Arshad1,Akyildiz1}. The former helps \emph{explore} the spectral opportunities, and the latter helps \emph{exploit} spectral resources and allocates them to different SUs.

\subsubsection{The cognitive spectrum manager (CSM)}
It is responsible for exploring spectral opportunities over the multi-dimensional domain (time, frequency, and space). This can be realized through accurate databases (e.g., databases of the location of TV channels), spectrum sensing (e.g., energy detection), or a cooperation of both. In this section, multiband sensing is assumed because it identifies a wider range of spectral opportunities compared to single-band techniques \cite{Hattab1}. However, instead of exploring the entire spectrum by a single SU, multiple SUs can cooperate to efficiently identify the available channels, where those with the highest potential are explored first \cite{Hattab1}.

\subsubsection{The cognitive resource manager (CRM)}
It has several key responsibilities. First, it collaborates with the CSM to:
\begin{itemize}
\item Divide the multiband spectrum sensing task among different SUs, based on different possible factors (e.g., the SU's capability, battery level, location, etc.).
\item Sense specific parts of the spectrum instead of sensing the entire spectrum. For instance, to guarantee a certain level of QoS for SUs, the channels with the highest signal-to-noise ratio (SNR) as well as the channels with the lowest PU activity are sensed first.
\end{itemize}
Second, once the CSM identifies the spectral opportunities for current and future use, the CRM must divide these resources among the different SUs. This is realized through \emph{resource usage} (RU) and \emph{resource control} (RC) modules \cite{Arshad1}. The RU module is responsible for reconfiguring the PHY layer to exploit the resources (e.g., spectrum shaping, which can be done via orthogonal frequency division multiplexing (OFDM)). The RC module is responsible for resource allocation (and reallocation in the event of returning PUs) and mobility management (which includes spectrum handover).

\subsection{Cognitive Radio Layers Reconfiguration}

Intuitively, the CSM and the CRM must work in tandem and provide cross-layer optimization to guarantee QoS support for SUs and to protect PUs. Network reconfiguration is not solely used at the PHY layer, but across all layers, from the application layer down to the PHY layer as illustrated in Table \ref{tab:Layers}. For instance, the application layer determines the QoS requirements for the SUs (e.g., different applications have different bandwidth requirements, data rates, delay-tolerance, etc.). Therefore, the wideband receiver front-end is reconfigured to support the QoS requirement of a given application at the SU's device level. At the network level, the number of sensing SUs can be reconfigured to meet these requirements (e.g., more SUs can be asked to sense a channel that requires very high sensitivity for detecting a PU). At the transport layer, delays and SU data packet losses may not only be induced due to the channel,  mobility, or congestion, but may also be due to spectrum handover (e.g., when a PU reclaims a channel), or when an intermediate SU (acting as a hop) cannot properly relay information \cite{Chowdhury1,Hu3}. To overcome these unique challenges, the CRM must identify a list of candidate and backup spectral opportunities.

At the network layer, cognitive managers must learn and recognize the traffic patterns of the PUNs (also known as \emph{network tomography} \cite{Liang2}). This is necessary because it helps the network in routing reconfiguration and in understanding the interference and spectrum usage patterns related to PUs. Moreover, routing in CRNs must be spectrum-aware in order to consider the dynamic nature of the available spectral opportunities. For example, routing protocols must not only consider the shortest paths (to minimize energy) and the strongest links (to maximize throughput), but also paths that generate the minimal interference to PUs \cite{Liang2,Ibnkahla2}.

At the MAC layer, access protocols must be intelligently aware of the spectrum. Since the resources are dynamically changing, access control must be flexible and must consider practical constraints such as the availability of a common control channel for SUs to share the sensing results. In addition, the MAC layer directly controls the reconfiguration of the SU's transceiver. For instance, reconfiguring the sensing schedule by distributing the task of multiband sensing over multiple SUs is essential in order to reduce the high complexity of multiband sensing. Moreover, the MAC frame  includes a sensing slot and a transmission slot. Thus, an inherent-tradeoff between sensing accuracy and network throughput must be optimized by properly reconfiguring the sensing time to maintain a balance between QoS delivery for SUs and protection for PUs \cite{Liang1}.

At the PHY layer, several reconfiguration strategies can be adopted for both sensing and transmission:
\begin{itemize}
\item   The SU can reconfigure the spectrum sensor based on channel conditions,  prior channel state-information (CSI),  prior knowledge about the PU signal, or any combination of these. For example, energy sensors can be used in the absence of any prior knowledge about the PU signal and when the channel condition is good. Coherent and feature-based sensors can be used if the SU has prior information about the PU signal \cite{Hattab1, Ibnkahla2, Axell1}.
\item In the case of multiband sensing, spectrum sensing can be reconfigured to sense a subset of channels using, for instance, filter banks \cite{Farhang-Boroujeny1}.
\item The parameters of the spectrum sensor can be dynamically reconfigured (e.g., the number of collected samples, the decision threshold, etc.).
\item In the case of multiple-input multiple-output (MIMO) SUs, antennas can be reconfigured into transmitting antennas, sensing antennas, or both. In the last case, concurrent sensing and transmission is feasible, under certain conditions \cite{Heo1}.
\item For transmission, multiple SUs may access non-contiguous channels, and thus spectrum sculpting is essential to circumvent interfering with  PUs (see \cite{Ma3} for more details about spectrum sculpting techniques).
\end{itemize}
Besides all the aforementioned hardware reconfigurations, conventional reconfiguration can also be used such as power control and AMC schemes.

\subsection{Case Study}
In this section we study a CRN where network-based reconfiguration is envisioned and compared with a basic CRN (in the absence of network reconfiguration).

We consider a network that consists of $K$ SUs $\in\mathcal{K}=\{SU_1,SU_2,\ldots,SU_K\}$, and $M$ channels at center frequencies $\in\mathcal{F}=\{f_1,f_2,\ldots,f_M\}$. The PU activity (i.e., the probability that the PU is active) for each channel, $0\leq p_m\leq1,~m=1,2,\ldots,M$, is assumed to be known (it can be obtained using PU traffic estimation techniques \cite{Liu_Chun-Hau1}). For instance, $p_m=0.3$ means that the $m$-th channel is occupied $30\%$ of the time.

Each SU senses multiple channels to determine whether they are occupied or not. That is, $SU_k$ solves the following multiple binary hypothesis testing problem \cite{Hattab1}
\begin{equation}
\label{eq:MultibandDetectionProblem}
\begin{aligned}
\mathcal{H}_0^{m,k}:  &~  \mathbf{y}_m^k= \mathbf{w}_m^k                 \\
\mathcal{H}_1^{m,k}:  &~  \mathbf{y}_m^k= \mathbf{x}_m+\mathbf{w}_m^k,
\end{aligned}
\end{equation}
where $m=1,2,\ldots,I_k\leq M$ is the channel index, such that $SU_k$ senses $I_k$ channels, $\mathbf{y}_m^k=[y^k_{m,1},y^k_{m,2},\ldots,y^k_{m,N_m^k}]^T$ is the sensed signal at the $m$-th channel by $SU_k$, $\mathbf{x}_m$ is the transmitted PU signal, and $\mathbf{w}_m^k$ is a zero-mean additive white Gaussian noise (AWGN) with unit variance.  $SU_k$ decides $\mathcal{H}_0^{m,k}$ if the $m$-th channel is unoccupied or decides $\mathcal{H}_1^{m,k}$ otherwise.

In general, $SU_k$ forms a test statistic based on the likelihood ratio test. Thus, the decision rule of $SU_k$ for the $m$-th channel is
\begin{equation}
\label{eq:MBDecisionRule}
\Lambda(\mathbf{y}_m^k)    \overset{\mathcal{H}_1^{m,k}}{\underset{\mathcal{H}_0^{m,k}}{\gtrless}} \xi_m^k,
\end{equation}
where $\Lambda(\mathbf{y}_m^k)$ is the test statistic, and $\xi_m^k$ is the threshold that divides the decision region into $\mathcal{H}_1^{m,k}$ and $\mathcal{H}_0^{m,k}$. To form a test statistic for the $m$-th channel, $SU_k$ collects $N_m^k$ samples. This parameter is commonly known as the \emph{sensing window}. It plays a key role in the sensing-throughput tradeoff \cite{Liang1}.

We consider three spectrum sensors: An energy-based sensor and two feature-based sensors (FBS). The first FBS exploits pilot patterns in the frequency spectrum, and the second  exploits the cyclic prefix in an OFDM-based PU. The reader may refer to \cite{Hattab1,Axell1,Ibnkahla2} for a comprehensive discussion of state-of-the-art sensing techniques.

If the SU has no prior knowledge about the PU signal, or if the channel link is good, then the SU may use the energy detector (ED), which is expressed as
\begin{equation}
\label{eq:EnergyDetection}
\frac{1}{N_m^k}\big\|\mathbf{y}_m^k\big\|^2   \overset{\mathcal{H}_1^{m,k}}{\underset{\mathcal{H}_0^{m,k}}{\gtrless}} \xi_{m,ED}^k,
\end{equation}
where $\|.\|^2$ is the Frobenius norm. This spectrum sensor has the following receiver operating characteristic (ROC) performance under the AWGN channel condition \cite{Liang1}
\begin{equation}
\label{eq:EnergyROC}
P_{D_{k,m}}^{ED}   = \mathcal{Q}\bigg(\frac{\mathcal{Q}^{-1}(P_{FA})-\sqrt{N_m^k}\textsl{SNR}_k}{\sqrt{2\textsl{SNR}_k+1}}\bigg),
\end{equation}
where $P_{D_{k,m}}^{ED}$ is the ED detection probability of $SU_k$  at the $m$-th channel, $P_{FA}$ is the false alarm probability requirement, $\textsl{SNR}_k$ is the SNR at $SU_k$, $\mathcal{Q}(.)$ is  the complementary distribution function of the standard Gaussian, and $\mathcal{Q}^{-1}(.)$ is its inverse.

In many practical systems, the PU transmits a pilot pattern along with the data-carrying signal. If the SU has a prior knowledge about the pilot pattern,  $\mathbf{x}_{m}^P$, then it may use the pilot detector (PD), which is expressed as
\begin{equation}
\label{eq:PilotDetection}
\frac{1}{N_k}(\mathbf{x}_{m}^P)^H\mathbf{y}_m^k    \overset{\mathcal{H}_1^{m,k}}{\underset{\mathcal{H}_0^{m,k}}{\gtrless}} \xi_{m,PD}^k,
\end{equation}
where $(.)^H$ is the Hermitian operator. The ROC performance of this sensor (in the AWGN channel case) is expressed as \cite{Hattab2,Hattab3}
\begin{equation}
\label{eq:PilotROC}
P_{D_{k,m}}^{PD}    =   \mathcal{Q}\bigg(\mathcal{Q}^{-1}(P_{FA})-\sqrt{2\theta N_m^k\textsl{SNR}_k}\bigg),
\end{equation}
where $\theta$ is the power allocation factor for the pilot signal.

Finally, using the detector proposed in \cite{Axell4} for detecting OFDM-based PUs, the ROC performance can be approximated  by
\begin{equation}
\label{eq:OFDMROC}
P_{D_{k,m}}^{OFDM}   \approx \mathcal{Q}\bigg(\frac{\mathcal{Q}^{-1}(P_{FA})-\sqrt{2N_m^k}\textsl{SNR}_k}{\sqrt{4\textsl{SNR}_k+1}}\bigg).
\end{equation}

Fig. \ref{fig:Detection} illustrates the detection performance of the three spectrum sensors in the single SU single-band case. The sensing duration is $N=1000$ samples for all sensors to achieve a fixed false alarm probability of 10\%. It is observed that the pilot-based detector achieves the best detection performance at very low SNR, whereas the energy detector exhibits a poor performance because it does not incorporate any prior knowledge about the PU signal.

We present now a number of reconfiguration strategies for the CRN:

\begin{figure}[!t]
\centering
\includegraphics[width=3.5in]{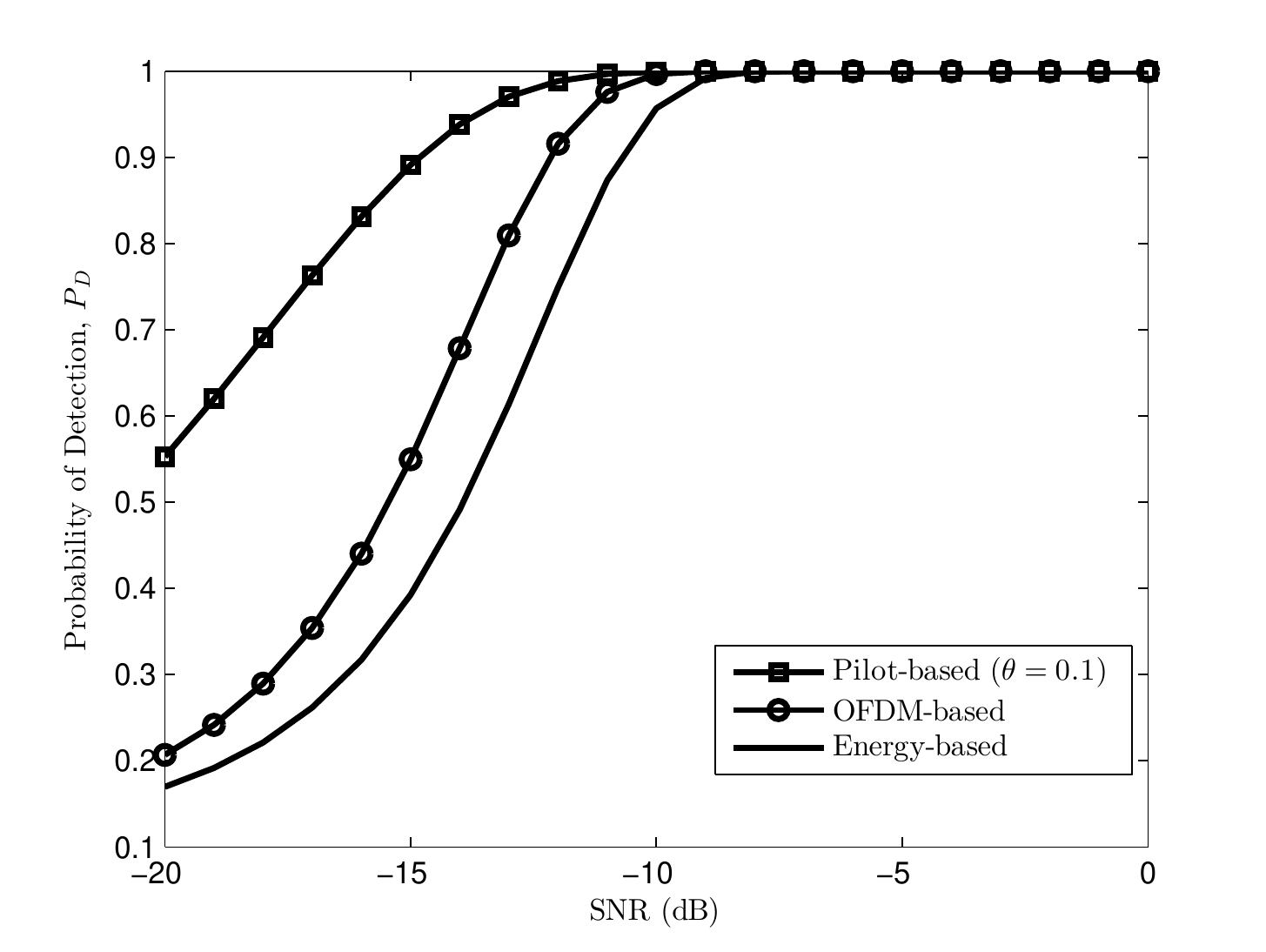}
\caption{Detection performance of the three spectrum sensors.}
\label{fig:Detection}
\end{figure}

\subsubsection{Basic CRN}
This is the simplest scenario where the network arbitrarily allocates the sensing tasks to the SUs. Here, each individual SU merely configures its transmission frequency according to the spectrum availability. Particularly, in this scenario:
\begin{itemize}
\item Each SU may have to sense all channels in the absence of any coordination. Thus, $I_k=M~\forall k=1,2,\ldots,K$ regardless of the PU activity profile. Note that this implies that the SU may sense a channel with $p_m\approx1$ (i.e., almost always busy), which is clearly a waste of resources. Moreover, sensing the entire spectrum by an individual SU is expensive and complex.
\item All SUs sense for a fixed duration (i.e., $N_m^k=N~\forall k=1,2,\ldots,K,~\forall m=1,2,\ldots M$). That is, the SU does not reconfigure its sensors based on the channel quality. This may lead to throughput losses especially if a channel is in a  good condition and the sensing window is not reduced for that channel.
\item Each SU implements a \emph{non-reconfigurable multiband detector} (NMD), as shown in Fig. \ref{fig:NMD}. It consists, in this case, of $M$ EDs. Here, the same threshold is used at $SU_k$, i.e., $\xi_{m,ED}^k=\xi^k_{ED}$, regardless of PU activities and interference protection levels.
\end{itemize}
Once the sensing results are obtained for all channels, the SUs  send their local decisions to a fusion center (FC) to make a global decision on the occupancy of the $M$ channels.

\subsubsection{CRN with device-based reconfiguration}
\begin{figure}[!t]
\centerline{\subfigure[]{\includegraphics[width=2.5in]{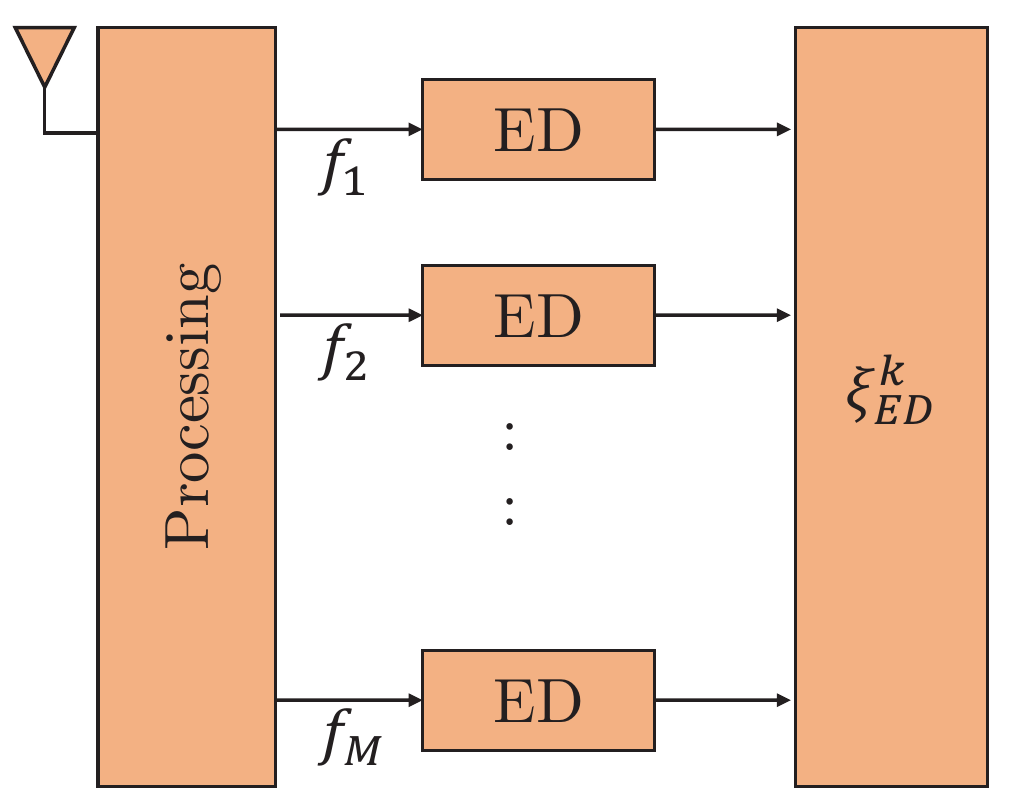}
\label{fig:NMD}}}
\centerline{\subfigure[]{\includegraphics[width=2.5in]{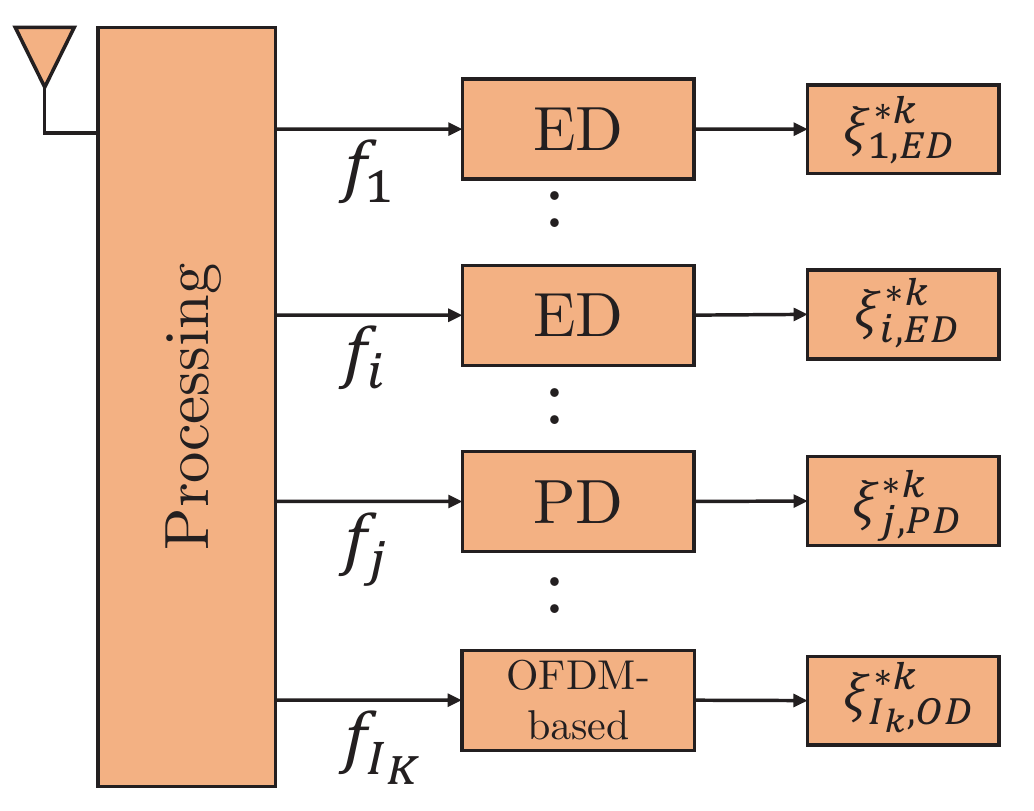}
\label{fig:RMD}}}
\caption{(a) The NMD consists of multiple single-type detectors (in this case, all detectors are EDs); (b) The RMD may consist of multiple detectors of multiple types (in this case, there are EDs, PDs, and OFDM-based detectors).}
\label{fig:Detectors}
\end{figure}

Here, each SU optimizes the thresholds and the sensing duration of each branch of its multiband detector to adapt to the interference levels, channel quality, etc. For instance, the \emph{multiband joint detector} (MJD), which consists of energy sensors, jointly optimizes $\xi_{m,ED}^{k}$ and $N_m^k$ (see \cite{Quan1,Paysarvi-Hoseini1, Paysarvi-Hoseini3} for more details) to maximize the CRN throughput. A higher level of reconfiguration is obtained when the multiband detector consists of sensors of various types, as illustrated in Fig. \ref{fig:RMD}. We refer to it as the \emph{reconfigurable multiband detector} (RMD).

\subsubsection{CRN with centralized network-based reconfiguration}
Here, the FC reconfigures the network's parameters and coordinates the spectral sensing and exploitation tasks of SUs. In particular, the FC may group the SUs into multiple clusters. Each cluster is  responsible of sensing a subset of the wideband spectrum. Furthermore, the FC can assign the sensing tasks among the SUs based on various factors (e.g., the FC can ask an SU to only sense the channels with the highest SNR at its location).

\subsubsection{CRN with distributed network-based reconfiguration}
The centralized CRN has several limitations. First, the amount of information exchange and network overhead between the SUs and the FC is tremendous due to continuous spectrum sensing and decision updates performed by the SUs in real-time. Second, in practice the communication links between the SUs and the FC are not error-free. Indeed, the network's performance can be limited if the reporting channel is in a deep fade \cite{Chaudhari2}. Third, many  network architectures lack a common central unit (e.g., ad hoc networks). For these reasons, distributed network-based reconfiguration  is appealing for CRNs.
Due to the absence of the FC, \emph{learning} becomes prominent in order to improve the reliability of spectrum sensing \cite{Li_Zhiqiang1,Ibnkahla2,Hu1,Hu2,Ejaz2,Sayed1}. Here, a consensus-based learning scheme is considered. It consists of two stages: In the first stage, each SU senses its corresponding list of channels using either the NMD or the RMD (depending which device reconfiguration is in effect). In the second stage, each SU communicates with its neighboring SUs for information exchange. Each $SU_k$ checks with its neighboring SUs, and if some of them are sensing some channels that $SU_k$ is sensing, then it will use the information obtained from these neighbours to update its own test statistics as follows
\begin{equation}
\label{eq:Learning}
\Lambda(\mathbf{y}_m^k(i+1)) = \Lambda(\mathbf{y}_m^k(i))+\mu\sum_{l\in\mathcal{N}_k}\Lambda(\mathbf{y}_m^l(i))-\Lambda(\mathbf{y}_m^k(i))
\end{equation}
where $i$ is the iteration index, $\mathcal{N}_k$ is a set of $SU_k$'s neighbours, and $\mu$ is the learning step size.

For example, let $\Lambda(\mathbf{y}_m^k)=E(k,m)$ be the energy of the samples collected by $SU_k$ for the $m$-th channel. Then, $SU_k$ will exchange information about this energy measurement with its neighbors until all corresponding SUs reach a consensus about the energy level of that channel (i.e., until $E(k,m)$ converges to some constant $\forall k\in\mathcal{N}_k$). See \cite{Sayed1} for an in-depth analysis of the convergence properties of consensus-based algorithms.

\subsubsection{Performance analysis}
The average aggregate throughput of the CRN is investigated for different levels of reconfiguration and compared to the basic CRN (see \cite{Hattab1} for more details about the throughput in multiband CRNs). The case of $M=5$ channels and $K=10$ users is considered. Each user senses $I_k=I=2~\forall k$ channels (the same assumption is made for the basic CRN for a fair comparison) and transmits its own data over one available channel with transmit power of 10dB. The SNR of each channel at each SU is based on a Rayleigh fading model, with an average SNR of $\bar\gamma$. Moreover, the PU activity over the $m$-th channel is assumed to follow a uniform random process, i.e., $p_m\sim\mathcal{U}(0,1)$, and it is independent of the other channels.

\begin{figure}[!b]
\centerline{\subfigure[]{\includegraphics[width=3in]{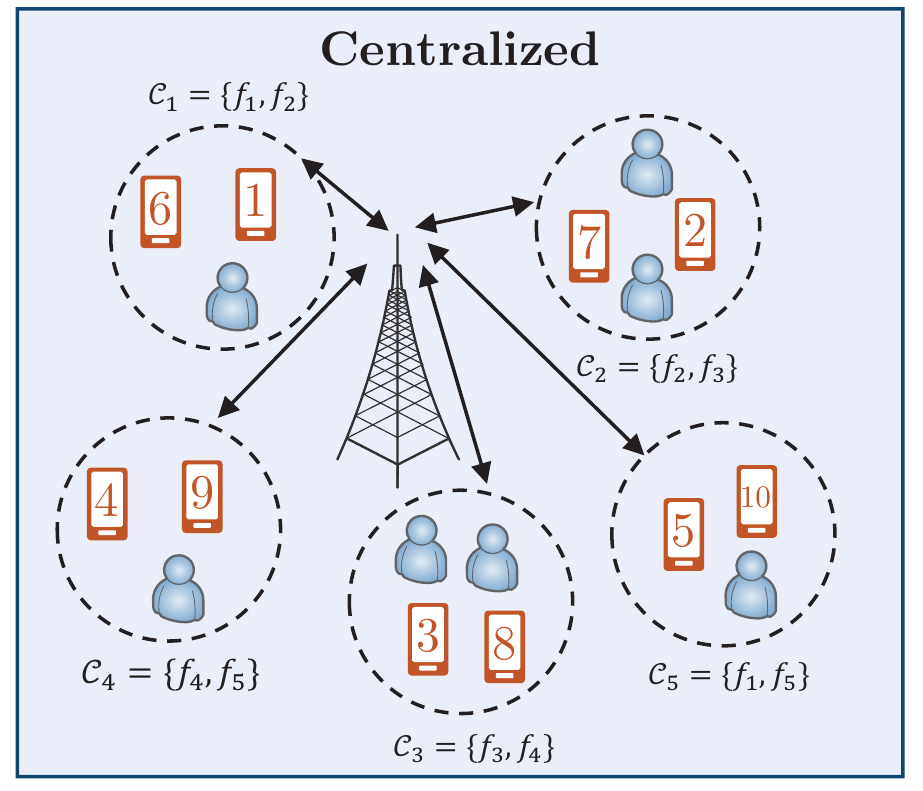}
\label{fig:CentralizedCRN}}}
\centerline{\subfigure[]{\includegraphics[width=3in]{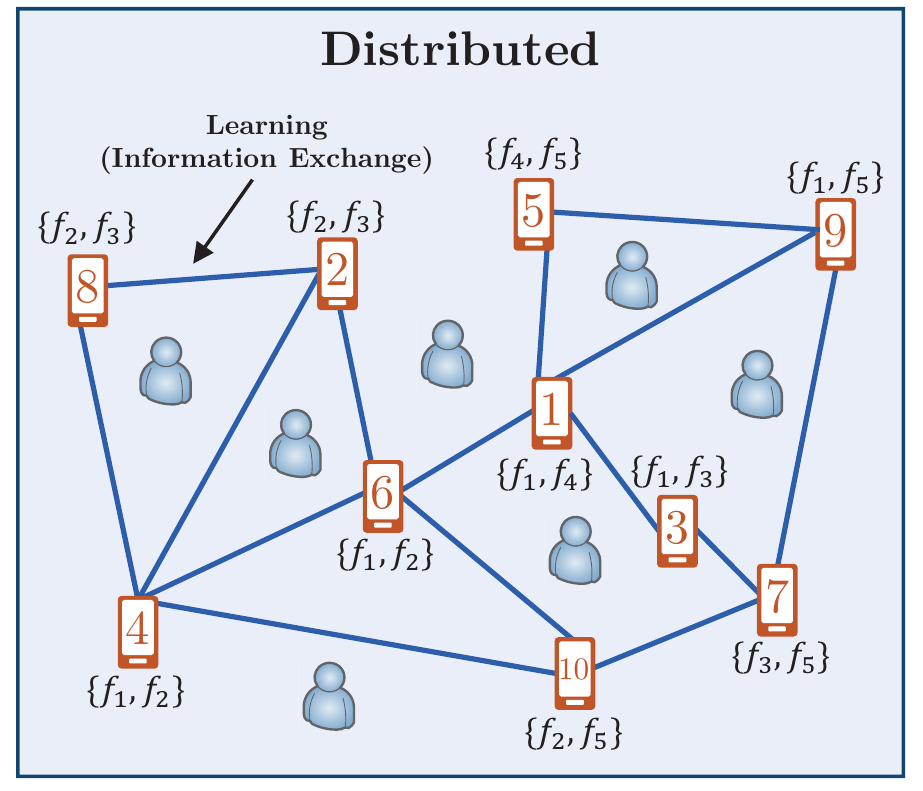}
\label{fig:DistributedCRN}}}
\caption{Network-based reconfiguration: (a) In the centralized approach, the FC allocates the channels to the SUs. For instance, in the basic CRN, the FC allocates $f_1$ and $f_2$ to $SU_1$ and $SU_6$, as shown. In the network-based reconfiguration, the FC will allocate the best channels at each SU's location; (b) In the distributed approach,  SUs will sense the best channels. For instance, it is shown here that $f_1$ and $f_5$ are the best channels at $SU_9$.  Then, $SU_9$ will exchange information with SUs that sense these channels. That is, $SU_9$ will exchange information about $f_1$ with $SU_1$ (in a single-hop) and $SU_3$ (in multiple hops), and so on.}
\label{fig:CRNs}
\end{figure}

In the centralized case (Fig. \ref{fig:CentralizedCRN}), SUs send their decisions to the FC, which combines the results using OR-logic rule \cite{Letaief1}. The resulting decisions are then sent by the FC to the SUs for spectrum access.

For the basic CRN scenario, each SU senses two pre-assigned channels using the ED-based NMD (i.e., it is assumed that the FC allocates two channels for each SU such that each channel is sensed by $(I\times K)/M$ SUs as shown in Fig. \ref{fig:CentralizedCRN}).

For the CRN with device-based reconfiguration, the channel allocation is similar to that of the basic CRN case. However, the SUs jointly optimize the sensing duration and the threshold of each channel (to maximize the throughput over that channel subject to a false alarm constraint of 10\%). Here, two cases are studied: In the first one, the SUs use the MJD, and in the second case they use the RMD. Without loss of generality, let us assume that an SU with the RMD uses a PD for $f_1$ and $f_2$, an ED for $f_3$ and $f_4$, and an OFDM-based detector for $f_5$.

For the CRN with network-based reconfiguration (centralized and distributed), the sensing tasks are no longer randomly allocated. In this study, each SU senses the channels that have the highest SNRs at their locations.

For the distributed network-based reconfiguration case  (Fig. \ref{fig:DistributedCRN}), it is assumed that each SU is aware of its neighbours. In addition, each SU collects a fixed number of samples, $N$, for the sensing process. The duration of the learning process is set to 10 iterations, with a step size of $\mu=0.25$. It is reasonable to use a fixed duration for the sensing and learning processes because synchronization among SUs is essential in distributed networks.

\begin{figure}[!b]
\centerline{\subfigure[]{\includegraphics[width=3.5in]{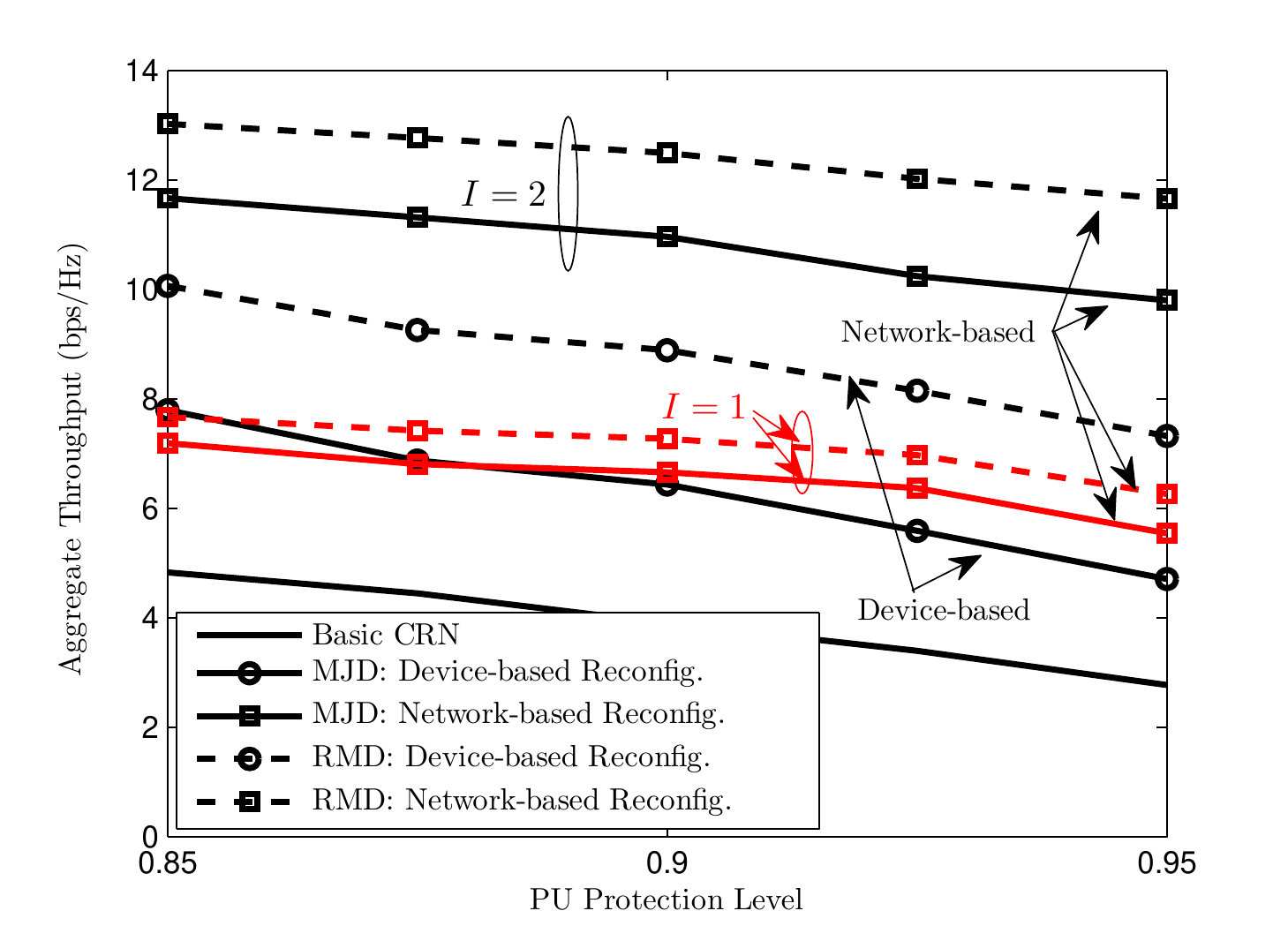}
\label{fig:CentralizedThroughput}}}
\centerline{\subfigure[]{\includegraphics[width=3.5in]{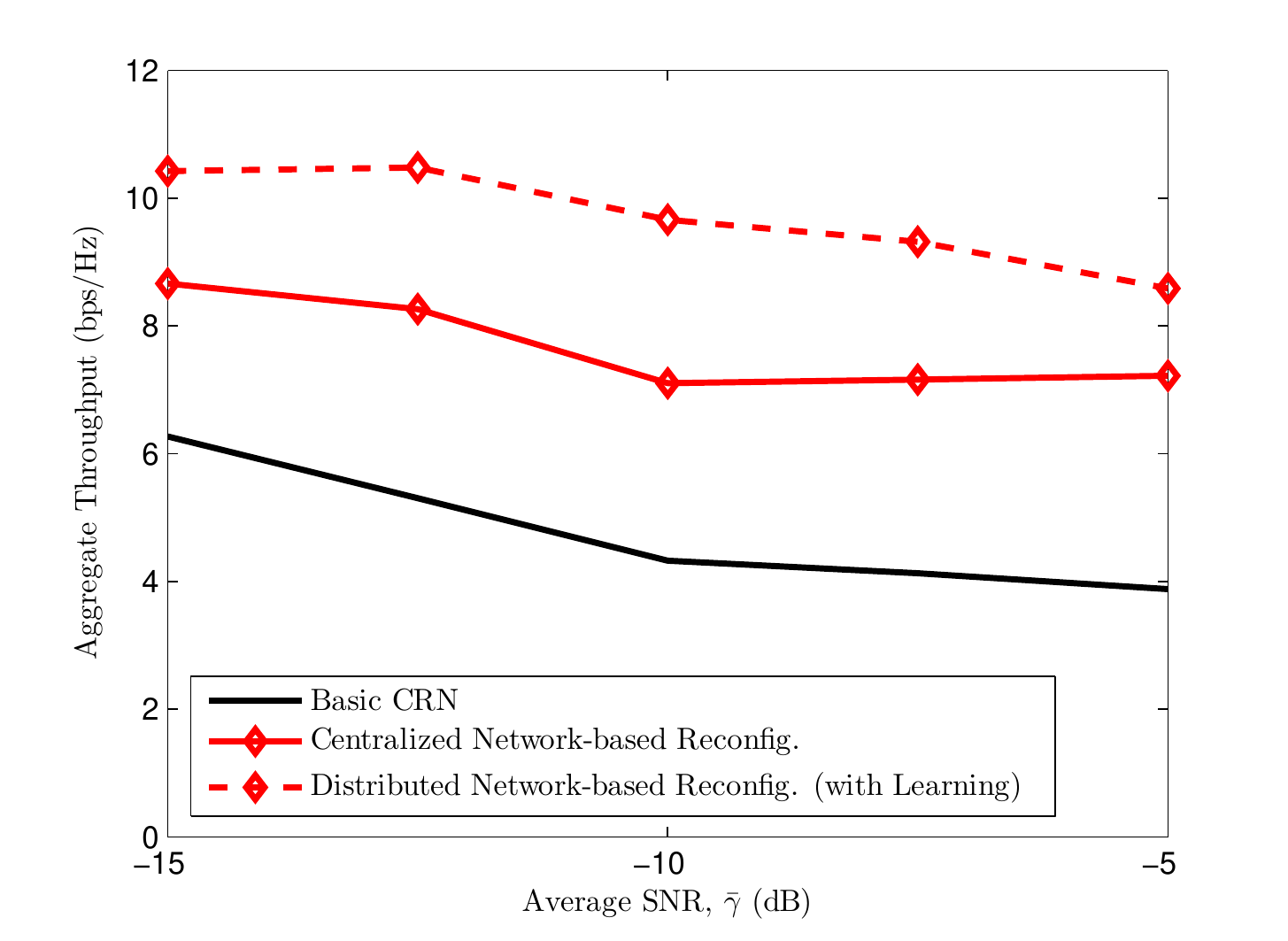}
\label{fig:DistributedThroughput}}}
\caption{(a) The aggregate throughput of the centralized CRN with different levels of reconfiguration versus the PU protection level; (b) Comparison of the aggregate throughput between the centralized and distributed network-based reconfiguration versus the average SNR.}
\label{fig:ThroughputAnalysis}
\end{figure}

The aggregate throughput of the centralized CRN versus the PU protection level is illustrated in Fig. \ref{fig:CentralizedThroughput} for $\bar\gamma=-15$dB. Here, a 0.90 protection level is equivalent to a global probability of detection of 90\%.  The sample budget is set at $\tau=10000$, and the sensing window is set such that:
\begin{itemize}
\item Each SU uses $N=2500$ samples in the basic CRN scenario.
\item Each SU uses $0\leq N_m^{k*}\leq 2500$ samples in the device-based and centralized network-based reconfiguration scenarios, where $N_m^{k*}$ is  optimized together with the thresholds to maximize the throughput of $SU_k$ at the $m$-th channel.
\end{itemize}
It is clear from Fig. \ref{fig:CentralizedThroughput} that the higher the reconfiguration level in the network, the higher the aggregate throughput. For instance,  SUs that use the MJD achieve a lower throughput than SUs that use the RMD. This is because the former uses a single-type sensor (i.e., only the parameters of each sensor are reconfigured), whereas the latter uses sensors of multiple types (i.e., the sensor type and its specific parameters are reconfigured). More throughput gains can be accomplished when the FC reconfigures the sensing tasks for the SUs. Furthermore, for the same level of network reconfiguration, it is observed that there is a significant gain when  SUs sense two channels (i.e., $I=2$) compared to the single-channel case (i.e., $I=1$). For instance, when the PU protection level is 90\%, the throughput can be improved by 67\% when multiband sensing is used in the network-based reconfiguration compared to the single-channel case. This illustrates the superiority of multiband spectrum sensing over single-band techniques.

Fig. \ref{fig:DistributedThroughput} shows the aggregate throughput of the network-based reconfiguration as a function of $\bar\gamma$. The false alarm requirement for each channel is 10\%. In all scenarios, a fixed sensing window is implemented, where $N=5000$ samples. In the network-based reconfiguration, it is assumed that the SUs use the RMD, where only the thresholds are optimized. It is clear that the distributed reconfiguration approach outperforms the centralized reconfiguration. This is expected since the latter implements the OR-logic rule.

\begin{figure}[!b]
\centering
\includegraphics[width=3.5in]{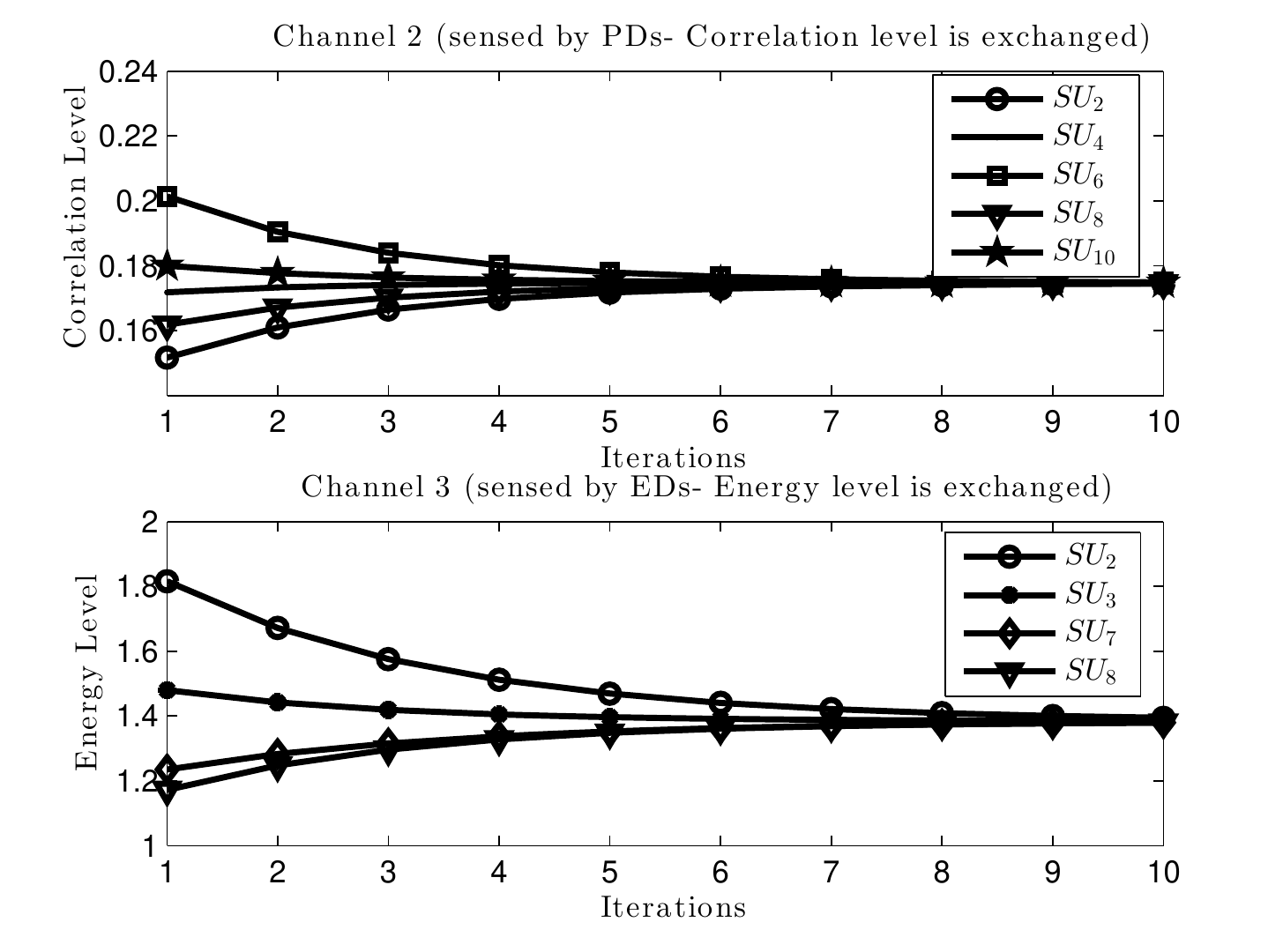}
\caption{Learning curves for different iterations ($\mu=0.1$).}
\label{fig:Learning}
\end{figure}

Fig. \ref{fig:Learning} illustrates how SUs reach consensus for different channels even though the initial measurements vary between  SUs due to the random condition of the wireless environment (a snapshot of the network is shown in Fig. \ref{fig:DistributedCRN}). Because $f_2$ is being sensed by SUs using PDs, they will exchange information about the correlation level (shown in the top plot), whereas SUs sensing $f_3$ will exchange information about the energy level because this channel is being sensed using EDs (shown in the bottom plot). In both cases, it can be observed that SUs quickly reach a consensus.

\subsubsection{Complexity}
Distributed network-based reconfiguration is more complex than the other approaches, yet no centralized units are required. Because leaning is done in a distributed way, the burden is shared among the nodes, and hence the additional complexity is worth it given the tremendous saving in terms of throughput and energy consumption.

In what we have discussed so far, network-based reconfiguration has been limited to the PHY layer. The following section covers reconfigurable networks where all layers of the protocol stack are jointly reconfigurable.

\section{Reconfigurable Networking}
Networking functionalities need to support a variety of issues such as mobility, QoS, security, energy-efficiency, etc. Several research directions have tried to achieve these goals. In this section, we focus on three areas of reconfigurable networking. These are: Cross-layer-based reconfiguration, network management, and SDNs.

\subsection{Cross-layer-based Reconfiguration}
Cross-layer design reconfigurations break the boundaries between layers by exchanging parameters or jointly designing layers in order to overcome the limitations and the rigidity of the traditional protocol stack (or single layer design) \cite{Ibnkahla3}. This opens significant possibilities for reconfiguration in networks, primarily because it allows for the consideration of information and objectives from multiple layers. For example, packet loss may be attributed to many factors such as link breakages, which can be detected at the PHY and MAC layers, or congestion, which can be detected at the network and transport layers. Even though the resulting event in all these cases is the same (packet loss), the reasons behind each case is quite different, and so the way to deal with the event must also be different.

Thus, in order to design a reconfigurable system which is capable of  producing the appropriate decisions in response to some event, information from multiple layers may be needed.  Cross-layer protocols can be classified into two main categories according to how information is exchanged between layers and how the network is organized \cite{Fu1}.

In the first category, cross-layer protocols exchange information in one of two ways: Either by utilizing a database where parameters from all layers are stored and can be accessed by any layer, or by allowing direct exchange of information between any two layers. Having a database of parameters simplifies information exchange and reduces the chance of conflict when two layers are modifying the same parameter. However, the design of such a multi-layer database may require modifications at multiple layers. On the other hand, allowing direct information exchange allows for more localized cross-layer implementations, but may produce conflicts when the number of layers and parameters involved increases.

In the second category, cross-layered protocols are designed according to the type of network architecture. Here, cross-layered approaches can be implemented using centralized or distributed network architectures. In the centralized approach, network-wide information is used, which may result in more optimized reconfigurations. However, this may come at the cost of extra overhead and latency associated with the required information gathering. On the other hand, distributed approaches are faster and thus more adaptive. However, the decisions produced can only be optimum in localized parts of the network.

We can classify reconfigurations within each layer into processes that are controlled or information that is passed between layers. At the PHY layer, two important processes to be controlled are transmit power and data rate. In addition, MIMO techniques have introduced other important processes such as beamforming and precoding \cite{Clerckx1}, whereas CRNs require spectrum control at the PHY layer. On the other hand, information that can be passed from the PHY layer to other layers include BER, CSI, SNR, interference temperature, mobility information, remaining energy, etc.

At the MAC layer, processes to be controlled include scheduling, source rate control, hybrid automatic repeat request (HARQ), maximum number of retransmissions, and contention window size, etc. Information that can be passed from the MAC layer to other layers include packet loss rate, queue length, packet types, frame length, etc.

At the network layer, the main process to be controlled is routing, which includes path discovery and maintenance. Other processes include admission control, which is particularly challenging in distributed networks due to the lack of complete information about interfering nodes. Congestion control and QoS support also coincide at the network layer. Parameters that can be passed from the network layer include path length and cost, session blocking and dropping rates, and path quality (in terms of any parameters under consideration). There is also some consideration of the transport layer, where processes include transmission window, flow control, and information passed includes error rate and congestion detection. This list of processes and parameters is not exclusive, but intended to provide an idea about what can be controlled at each layer.

To illustrate, several cross-layer implementations that enable reconfiguration in networks can be found in long-term evolution (LTE) and LTE-Advanced (LTE-A) technologies \cite{Cox1, Taha1}. One example is coordinated multipoint transmission and reception (CoMP) \cite{Lee_Daewon1}, expected to appear in Rel. 11 of LTE. This technology tries to improve performance at the cell edges by taking advantage of the fact that users are receiving signals from more than one eNodeB. There are mainly two types of CoMP: Coordinated scheduling/beamforming (CS/CB) and joint processing (JP) \cite{Pateromichelakis1}. In CS/CB, the user is still associated with a single eNodeB, but the neighboring eNodeBs exchange information in order to coordinate their activities and reduce inter-cell interference. In CS, the exchanged information is used to coordinate scheduling to different users at the cell edges, while CB uses the information exchange to adjust the antenna beams to users. On the other hand, in JP, the neighboring eNodeBs exchange information in order to coordinate simultaneous transmissions to the same user.

\begin{figure}[!b]
\centerline{\subfigure[]{\includegraphics[width=3.5in]{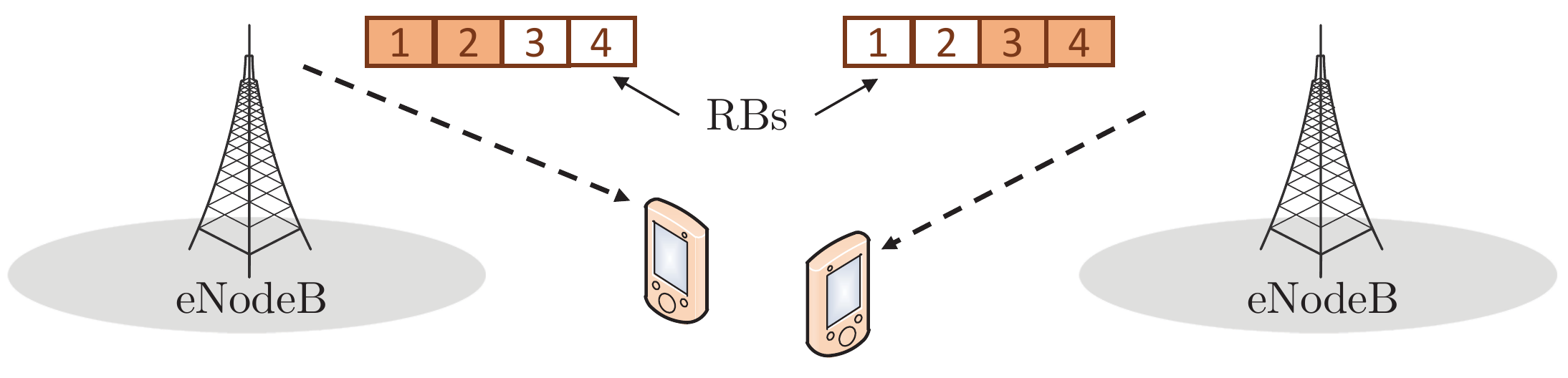}
\label{fig:Amr2a}}}
\centerline{\subfigure[]{\includegraphics[width=3.5in]{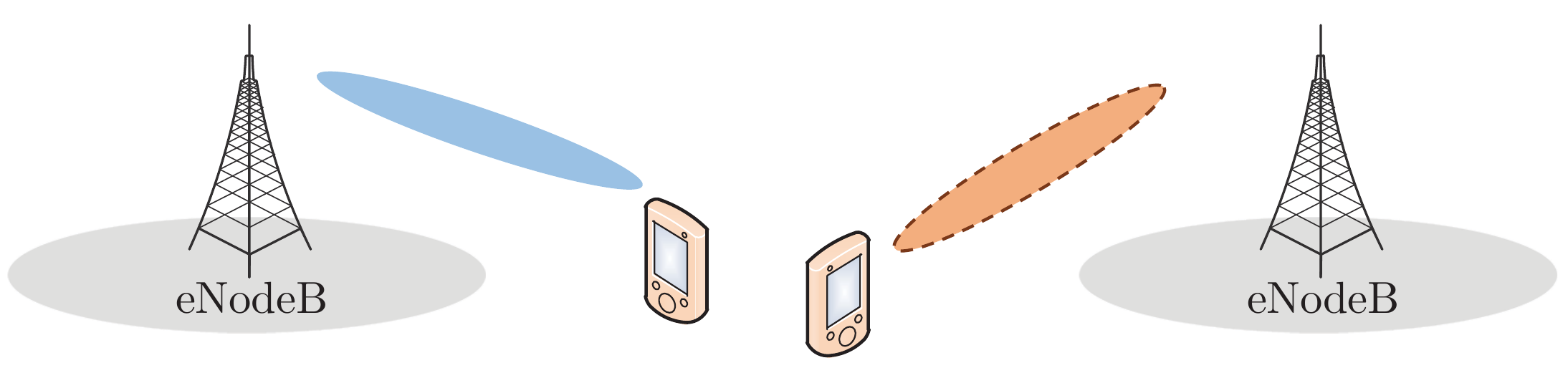}
\label{fig:Amr2b}}}
\centerline{\subfigure[]{\includegraphics[width=3.5in]{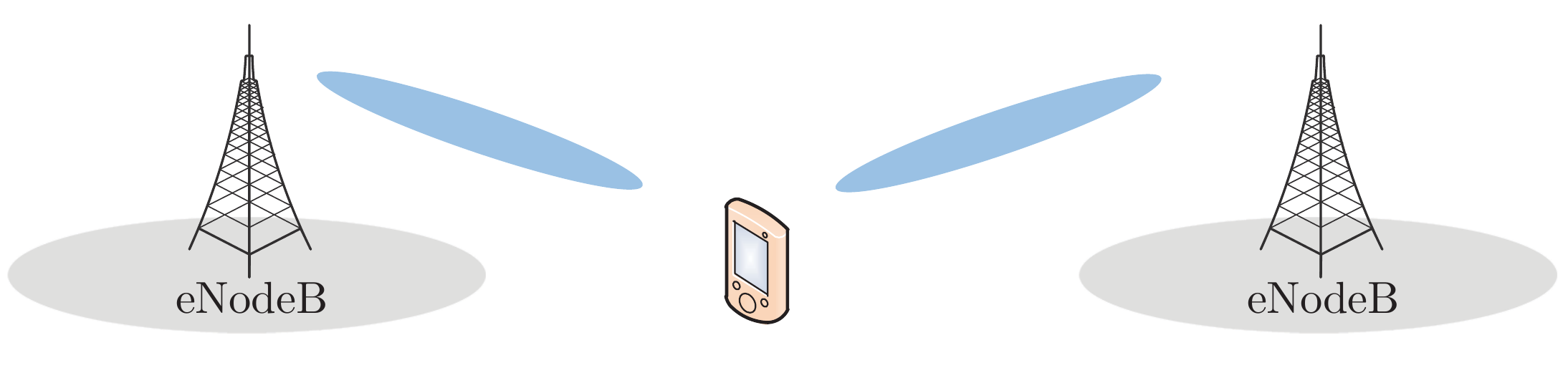}
\label{fig:Amr2c}}}
\caption{CoMP techniques: (a) In CS, the scheduling is coordinated; (b) In CB, antenna beams are coordinated; (c) In JP, the user simultaneously receives information from the neighboring eNodeBs.}
\label{fig:Amr2}
\end{figure}

The reconfigurations done in different CoMP types are shown in Fig. \ref{fig:Amr2}.  These techniques represent a clear example of cross-layer reconfiguration of network parameters and systems in order to reduce interference and improve performance \cite{Sun_Shaohui1}. This is done based on feedback from the user equipment (UE) that indicates CSI to the eNodeBs. This CSI can be used to assign resource blocks (RBs - the smallest individual scheduling unit that can be assigned to users and consists of a chunk of adjacent frequencies) in CS, reconfigure precoding matrices for beamforming in CB, or perform JP. In any case, multiple layers are involved in the process and the system behaves differently according to the number of UEs and their CSI.

LTE has other examples where cross-layer reconfiguration improves user experience such as downlink and uplink packet scheduling \cite{Pedersen1}, which relies on feedback for the UE about CSI to perform time-domain and frequency-domain scheduling (deciding the set of users that will transmit in a particular time slot and the RBs that will be assigned to each one of those users). This is also an example of joint design between the PHY and MAC layers. Note that this is possible only for UEs with low mobility. In high mobility, CSI is changing quite rapidly and the eNodeBs will not be able to change the assignment of RBs for all users that quickly. Thus, the protocol is reconfigured according to the statuses of the UEs: If mobility is low, then frequency-domain scheduling is applied in order to assign the optimum RBs to each UE according to their CSI; otherwise, RBs are assigned over a broader part of the bandwidth to utilize frequency diversity in case of high mobility.

The above examples represent centralized implementations of cross-layer reconfiguration without the existence of a database of parameters. Centralized implementations are straightforward in cellular networks due to the presence of base stations (BSs) that perform most of the processing on behalf of all UEs in the network. Thus, fine-grained optimizations, such as assigning RBs to users based on their CSI, can be performed. For example, \cite{Prasad1} proposes an uplink scheduler for LTE-A networks that exploits multi-user-MIMO (MU-MIMO) techniques. For cross-layer reconfiguration between the PHY and MAC layers, MU-MIMO reconfigures matrix precoding techniques to allow multiple users to use the same RBs without interfering with each other. Here, the eNodeB receives feedback from every UE about CSI for each individual RB. Based on this feedback, an optimization problem is formed where the objective function is to maximize the rate assigned to every user for every RB. This is subject to multiple constraints such as decodability (the rates have to be decodable by the eNodeB), one precoder matrix per user, finite buffers per user, limits on inter-cell interference, and a maximum of two chunks of contiguous RBs for every user in any time slot. The output of the optimization function is the set of RBs and the precoding matrix that will be assigned to each  user. Therefore, the assigned RBs and the antenna precoding matrices are reconfigured according to the CSI of each user.

\begin{figure}[!b]
\centering
\includegraphics[width=3.5in]{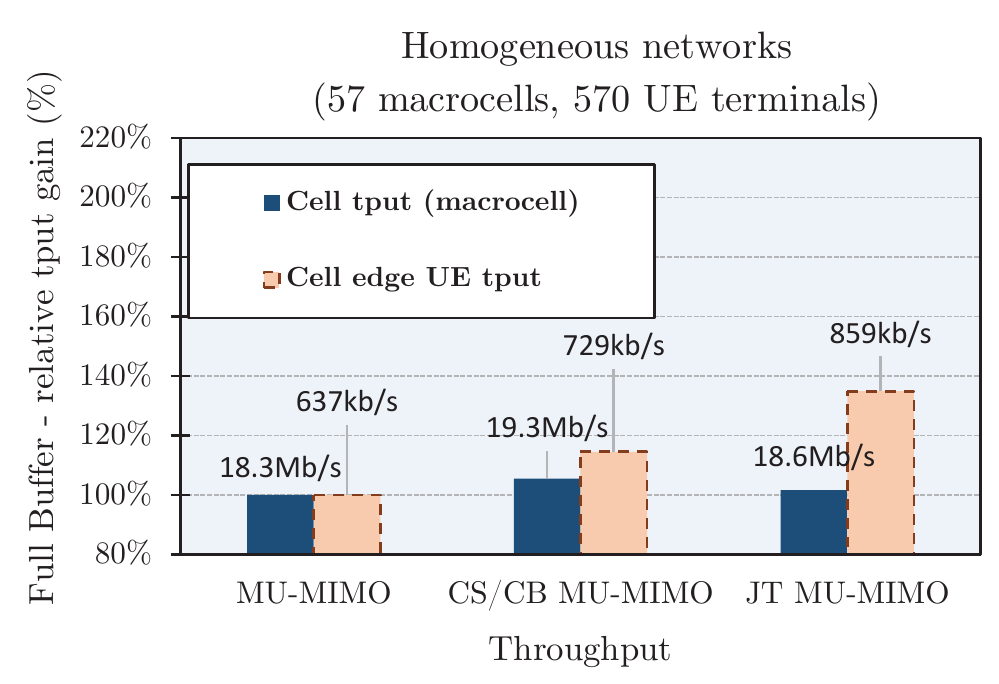}
\caption{Performance analysis of MU-MIMO and CoMP techniques \cite{Lee_Daewon1}.}
\label{fig:Amr3}
\end{figure}

The performance of MU-MIMO and CoMP techniques is analyzed in \cite{Lee_Daewon1}. Two particular cases are considered: Homogeneous and heterogeneous networks. In homogeneous networks, all BSs belong to the same power class. On the other hand, heterogeneous networks use low-power nodes known as pico cells. The percentage gain in throughput is evaluated in case of MU-MIMO, CS/CB, or joint transmission (JT), which is the most common JP scheme. All nodes are assumed to have a full buffer and thus can continuously transmit data. The case of the homogenous network is illustrated in Fig. \ref{fig:Amr3}. As the figure shows, all techniques achieve a minimum of 100\% throughput gain. JT achieves the highest gain, particularly at the cell edges. The gain is higher in heterogeneous networks, since the introduction of pico cells means that there is a larger number of cells and thus a larger number of cell edges \cite{Lee_Daewon1}.

Additional examples include the work in \cite{Araniti1}, which proposes a scheduler for multicast services in LTE systems. This work exploits multi-user diversity and uses AMC to maximize throughput. Another example is the work in \cite{Chang_Ben_Jye1}, which is a cross-layer channel selection scheme for WiMAX networks. It addresses the performance degradation associated with high mobility and proposes a channel selection scheme that utilizes AMC. A reward-based scheme is used for power control in order to reduce interference. Prioritization for users is considered as well.

Centralized cross-layer design has also been used in networks without infrastructure such as ad hoc networks. However, one or more nodes have to assume a central or manager role. Thus, in order for this central node to perform the required processing, information typically has to be collected from the nodes under its service. This is not as straightforward as in cellular networks, where dedicated channels are used to relay the required information. In addition, nodes in cellular networks are synchronized, and all communications are single-hop (from UE to BS or BS to UE). Some of the challenges in ad hoc networks include: First, the lack of synchronization could mean collisions when relaying information, especially when the number of nodes served by the central manager increases. Second, multi-hopping complicates several issues such as routing, admission control, multiple access, and third the lack of dedicated channels mean that control information has to compete for bandwidth with data. This means that the protocols may not be highly reactive and latency may be incurred in reconfiguring system parameters.

\begin{figure}[!b]
\centerline{\subfigure[]{\includegraphics[width=3.5in]{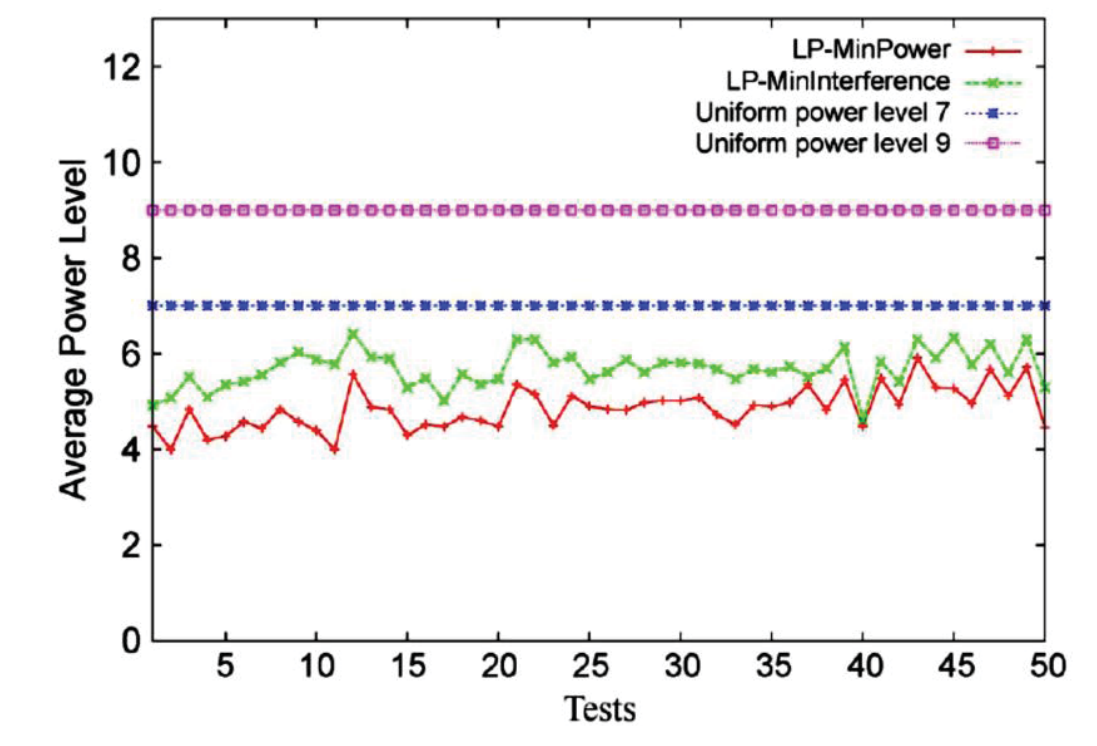}
\label{fig:Amr4a}}}
\centerline{\subfigure[]{\includegraphics[width=3.5in]{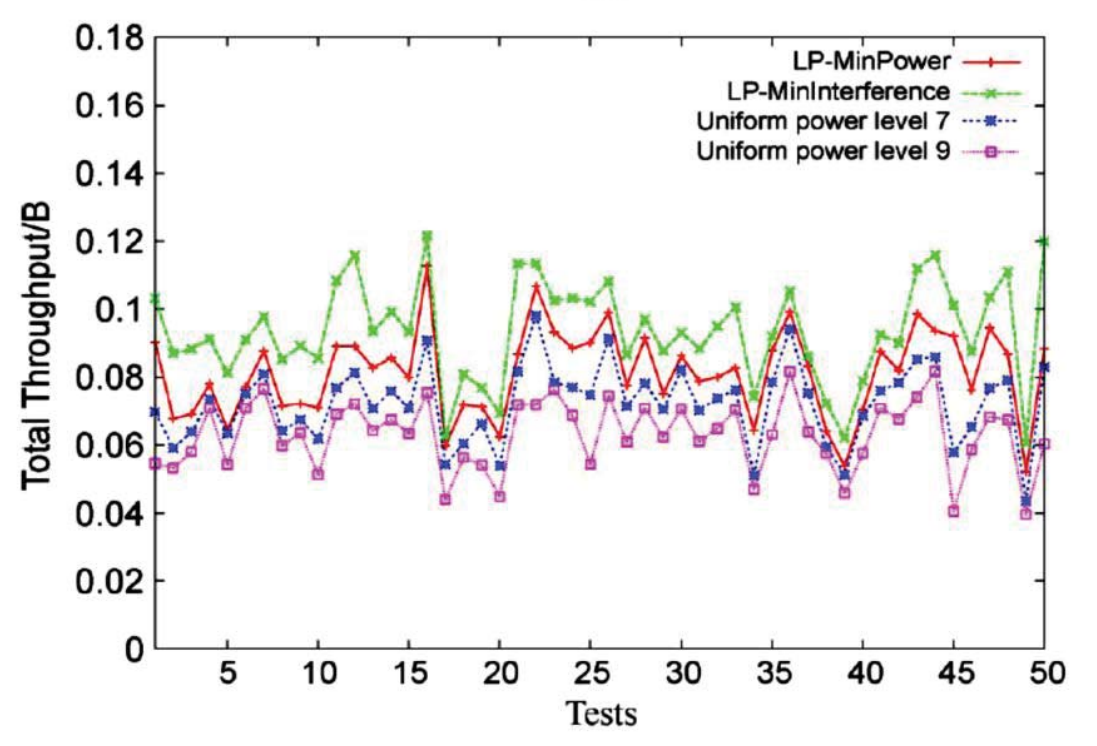}
\label{fig:Amr4b}}}
\caption{Performance of the power control algorithms in metrics of (a) average total power and (b) total throughput \cite{Cheng_Maggie1}.}
\label{fig:Amr4}
\end{figure}

For these reasons, centralized cross-layer implementations in ad hoc networks often target long-term planning or processes that do not require quick or frequent reconfigurations. An example is the cross-layer protocol proposed for ad hoc networks in \cite{Cheng_Maggie1}, which targets the optimization of throughput in sensor networks, where the existence of one or more BSs simplifies the implementation of centralized protocols. The protocol in \cite{Cheng_Maggie1} is divided into two parts: The first one performs topology control while the second one performs joint routing and rate control to maximize throughput. Two algorithms are proposed for topology control based on linear programming problems that utilize power control. In the first algorithm, the linear programming problem is solved to minimize total transmission powers at all nodes while guaranteeing connectivity. The second algorithm configures transmit powers to minimize the total interference in the network. After the topology has been configured, the second part of the algorithm runs and another linear programming problem is solved to maximize data rates over all links. The algorithms are compared to the case where there is no power control. The results in Fig. \ref{fig:Amr4} show that these algorithms have better energy efficiency and throughput.

As we can see, the protocol can only be implemented in a centralized node, where network-wide information is available. In addition, each time the algorithms are used, they are reconfigured according to the status of the nodes in the network. However, the work in \cite{Cheng_Maggie1} does not consider the latency involved in gathering this information and does not indicate how quickly the system may react to network changes.

From the above discussion, it is clear that centralized cross-layer implementations may not be suitable for highly dynamic networks, where protocols are required to be highly adaptive. For example, in vehicular ad hoc networks (VANETs), high vehicle mobility complicates the design of networking protocols. For example, routing protocols have to recover quickly due to frequent link breakages (links may not be alive for more than 30 seconds), which makes heuristic approaches such as greedy routing \cite{Ammari1} more favorable than optimized solutions, even if greedy routing produces suboptimal results. One protocol that employs greedy routing is the grid-based predictive geographical routing (GPGR) protocol \cite{Cha_Si-Ho1}. This protocol uses global positioning system (GPS) devices on board vehicles to route packets between road segments. In each segment, the protocol tries to forward data to the farthest node within its communication range. GPGR uses information about position, velocity, and direction of nodes to choose the next relay node along the path in a dynamic way that reduces link breakages and packet loss. A similar approach is adopted by reliability-improving position-based routing (RIPR) \cite{Ryu1}. Optimization is used to study cooperative routing, but no analysis of the adaptability of the protocol is given. In all these routing protocols, the devices have to reconfigure their routing paths faster than the speed at which the network topology is changing.

Reconfiguration is particularly important in heterogeneous networks where topologies, architectures, devices, and goals are different. In the following section, we detail the case of cross-layer reconfiguration in M2M networks, which are expected to be an important part of next-generation networks.

\subsection{Case Study: Cross-layer Reconfiguration in M2M Networks Based on Game Theory}
M2M networks have many applications which can lead to new business models and opportunities. These applications include smart grid, intelligent transportation systems, healthcare, smart houses, and environmental monitoring. The deployment of indoor/outdoor M2M devices depends on the specific application. Recent studies show that about 80\% of M2M communication take place indoor and 20\% outdoor \cite{Chandrasekhar1, Wang_Cheng-Xiang1}. M2M communications face many challenges such as the massive deployment of devices, transmissions in dynamic environments and architectures, high data traffic, and strict QoS requirements (such as coverage, latency, bandwidth, lifetime, and scalability) \cite{Boswarthick1}. Moreover, machines have to operate autonomously with minimal human intervention. Therefore, system reconfiguration and self-organization are required in order to meet user requirements  in a cost effective way \cite{Lo_Anthony1, 3GPP2, IEEE6}. This section presents a cross-layer framework for M2M device reconfiguration based on a game-theoretic approach.

\subsubsection{Network architecture}
\begin{figure}[!b]
\centering
\includegraphics[width=3.5in]{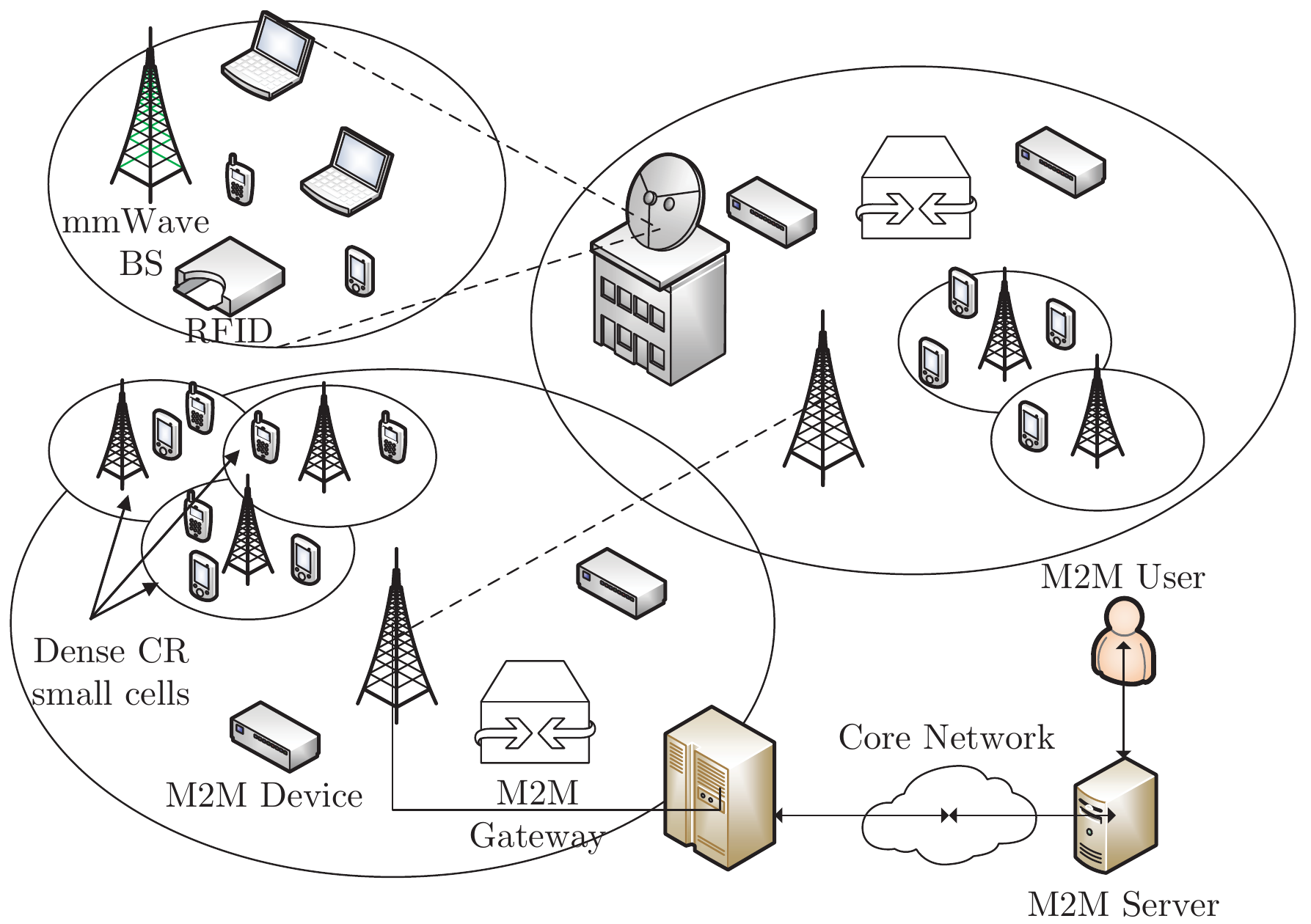}
\caption{Network architecture for support of M2M communication.}
\label{fig:Waleed1}
\end{figure}

This section adopts the wireless M2M architecture presented in Fig. \ref{fig:Waleed1}, where all devices are assumed to use cognitive radio (CR) technology. For indoor communication, many technologies can be used for short range communications such as WiFi, ultra wide band, mm-Wave communications, radio frequency identification (RFID), femtocell, Bluetooth, ZigBee, visible light communication, etc. \cite{Chandrasekhar1, Wang_Cheng-Xiang1,Bleicher1}. For outdoor communication, indoor BSs for each access network can directly or indirectly communicate with antennas installed outside, on the top of buildings, or in  streets in the case of residential areas. These antennas will then communicate with the outdoor macro cell BSs (eNodeBs) \cite{Rusek1}.

\subsubsection{System model}
Let $N_{P}$ be the number of cognitive M2M devices and $N_A$ the number of access networks available for uplink transmission in M2M communications. Each access network $b$ has $\alpha_b$ channels in each cell, where the total number of cells in each access network is $\gamma_b$. The system capacity for the combination of all access networks can be written as
\begin{equation}
\mathbb{C}^S    =   \sum_{b=1}^{N_A}\sum_{a=1}^{\gamma_b}\sum_{c=1}^{\alpha_b}   B_c\log_2(1+\zeta^b_{a,c}),
\end{equation}
where $B_c$ is the bandwidth of channel $c$ and $\zeta^b_{a,c}$ is the signal-to-interference-plus-noise ratio (SINR) at the M2M device on channel $c$ belonging to cell $a$.

The total energy efficiency is defined as
\begin{equation}
\mathbb{E}^S    =   \sum_{b=1}^{N_A}\sum_{a=1}^{\gamma_b}\sum_{c=1}^{\alpha_b}    \frac{B_c\log_2(1+\zeta^b_{a,c})}{P_A (d_{a,c})^\nu+P_{RF}},
\end{equation}
where $P_{RF}$ is the energy consumed by the radio frequency circuit, $P_A$ is the amplifier energy required at the transmitter, $d_{a,c}$ is the distance between the M2M device using channel $c$ and BS $a\in\gamma_b$, and $\nu$ is the path loss exponent.

M2M applications have diverse QoS requirements including data rate, latency, reliability, etc. Moreover, due to the massive deployment of M2M devices, the environment and economic impact must be considered in the system design. This is because an increase in service price will increase the overall cost of the M2M communication system, and an increase in energy consumption is considered as a major threat to the environment as it is directly linked to GHG emissions.

Let $SC$ be the set of service classes in the M2M communication system. Each class $sc\in SC$ has the following QoS parameters: Data rate ($R$), latency ($L$), reliability ($RE$), economic impact ($EC$), and environment impact ($EN$). Without loss of generality and for illustration purposes, we consider  a limited set of service classes with  QoS parameter attributes are as indicated in Table \ref{tab:Waleed1}. The parameters are defined over the levels 0 to 3 according to the significance, with 0 as unimportant, 1  as low, 2 as important and 3 as most important.

\begin{table}[!b]
\renewcommand{\arraystretch}{1.3}
\small
\caption{QoS classes for M2M applications}
\label{tab:Waleed1}
\centering
\begin{tabular}{m{1.5in}ccccc}
\toprule
\bfseries{Service}                         & $R$    &$L$  &$RE$  &$EC$     &$EN$                                \\\toprule
Real-time/Reliable (RtRe)              & 2    &3  &3   &0      &1                                     \\\hline
Non-real-time/reliable (NRtRe)         & 1    &0  &3   &3      &3                                     \\\hline
Real-time/non-reliable (RtNRe)         & 2    &3  &1   &2      &3                                      \\\hline
Non-real-time/non-reliable (NRtNRe)    & 0    &0  &1   &3      &3                                       \\
\bottomrule
\end{tabular}%
\end{table}

\subsubsection{Reconfiguration based on game theory}
Here a cross-layer reconfigurable (CLR) scheme is adopted for M2M device reconfiguration and resource allocation. The scheme is based on potential games where the M2M devices act as players.

A potential game is a type of strategic non-cooperative game in which all players autonomously learn their strategies \cite{Monderer1}. There are three fundamental components of non-cooperative games which include players, strategies, and utilities. A game can be represented as $G=\{N_P, \{S_n\}_{n\in N_P}, \{U_n(.)\}_{n\in N_P}\}$, where $S_n$ is the set of strategies of the $n$-th player, $S=\prod_{n\in N_p} S_n$, and $U_n$ is the utility of the $n$-th player, $U_n:S\rightarrow \mathbb{R}$ which maps strategy profiles $S$ into a real value $\mathbb{R}$. The utility function $U_n$ shows the satisfaction of the $n$-th player while considering its own strategy $S_n$ and the strategies of other players denoted by $S_{-n}=\{S_1, S_2, \ldots,S_{n-1},S_{n+1},\ldots,S_{N_P}\}$.

In the CLR scheme, each player (or M2M device) is capable of reconfiguring its operating parameters according to the best available access network (Table \ref{tab:Waleed2}). The PHY layer is responsible for the CR module, operating frequency, and transmit power. The MAC layer is responsible for access network selection and resource allocation according to the requirement of service class mentioned in Table \ref{tab:Waleed1}. It is assumed that the network layer uses spectrum-aware routing protocols that can be classified into four classes: Power-based, delay-based, throughput-based, and reliability-based where the routing protocol targets the energy efficiency, end-to-end delay, achievable throughput, and link stability, respectively \cite{Cesana1}\cite{Youssef1}.

\begin{table}[!b]
\renewcommand{\arraystretch}{1.3}
\small
\caption{Reconfigurable M2M device}
\label{tab:Waleed2}
\centering
\begin{tabular}{cp{2in}}
\toprule
\bfseries{Layers}                         & \bfseries{Reconfigurable Parameters/Protocols}                          \\     \toprule
\multirow{1}{*}{Application}                  &QoS requirements\\ \hline
\multirow{1}{*}{Network}                     &Spectrum aware routing protocol   \\ \hline
\multirow{2}{*}{Data Link \& Mac}              &Network selection\\
                                              &Channel allocation \\ \hline
\multirow{2}{*}{PHY}                           &CR module \\
                                                &Operating frequency \\
                                                &Transmit power\\
\bottomrule
\end{tabular}%
\end{table}

Reconfiguration is performed according to the diverse QoS requirements for classes RtRe, NRtRe, RtNRe, and NRtNRe (Table \ref{tab:Waleed1}) based on weighted instantaneous date rate, latency, reliability, economic impact, environment impact, and cross-co-tier interference. The utility function of the resource allocation and reconfiguration problem for each M2M device can be formulated as:

\begin{equation}
\begin{aligned}
\label{eq:Waleed3}
U_n(s_n,s_{-n}) &=   \omega_n^R R_n^{b,a}+\omega_n^L L_n^{b,a} + \omega_n^{RE} RE_n^{b,a}\\
                &+   \omega_n^{RP} h_{n,c}^{b,a} P_n^{b,a} - \omega_n^{EC} EC_n^{b,a} B_n^{b,a}\\
                &-   \omega_n^{EN} EN_n^{b,a} P_n^{b,a}-\sum_{m=1,m\neq n}^{N_P} h_{m,c}^{b,a}\sigma_{c_nc_m}\\
                &-   \sum_{m=1,m\neq n}^{N_P} h_{n,c}^{b,a}\sigma_{c_mc_n},
\end{aligned}
\end{equation}
and the potential function used for the potential game is
\begin{equation}
\label{eq:Waleed4}
  \Phi(S)   =   \frac{1}{2} \sum_{n=1}^{N_P}  U_n,
\end{equation}
where $R_n^{b,a}$ is the data rate, $L_n^{b,a}$ is the latency, $RE_n^{b,a}$ is the reliability, $EC_n^{b,a}$ is economic price for bandwidth $B_n^{b,a}$ corresponding to particular application class, and $EN_n^{b,a}$ is the environment cost of power $P_n^{b,a}$ corresponding to a particular application class of the $n$-th player in cell $a$ of access network $b$, respectively. $\omega_n^R$, $\omega_n^L$, $\omega_n^{RE}$, $\omega_n^{EC}$, and $\omega_n^{EN}$ are the coefficients which intensify the QoS requirements according to the service classes mentioned in Table \ref{tab:Waleed1}. $\omega_n^{RP}$, the reward/penalty coefficient is set to 1 when the access network is cost effective and have less impact on environment (e.g., small cells and WLAN) and is set to $-1$ otherwise. $h_{m,c}^{b,a}$ is the channel gain between the $m$-th M2M device and BS $a$ which serves the $n$-th M2M device. $\sigma_{c_n,c_m}$ is the interference function which is one if channel $c$ is same for the $n$-th and the $m$-th M2M device and zero otherwise.

In this game theoretical framework,  each player executes the three-step cycle shown in Fig. \ref{fig:Waleed3}:

\begin{figure}[!b]
\centering
\includegraphics[width=3.5in]{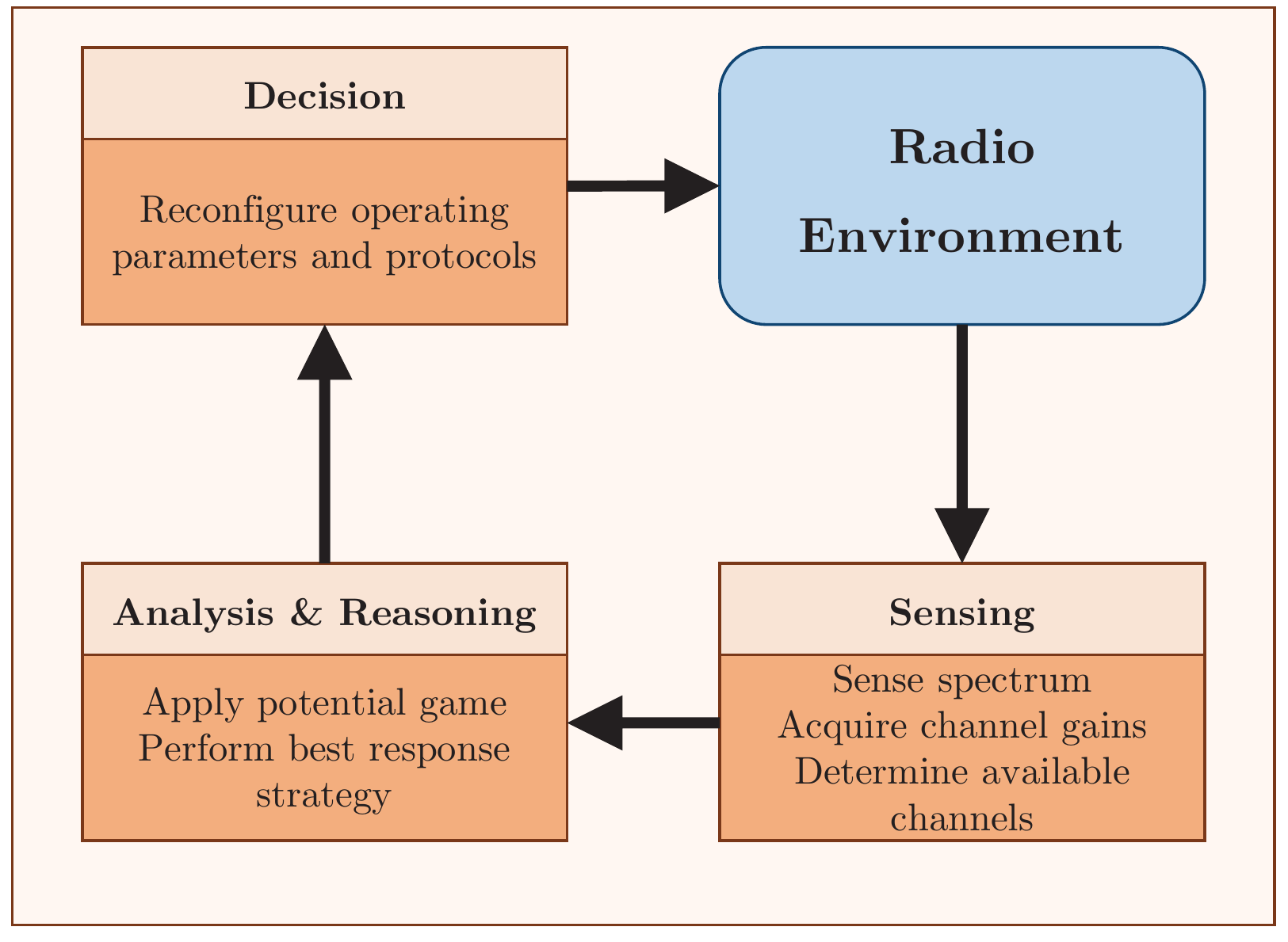}
\caption{Main steps in the CLR scheme.}
\label{fig:Waleed3}
\end{figure}

\begin{itemize}
\item In the \emph{sensing step}, each player interacts with the radio environment for spectrum sensing to determine available channels and acquire their link gains.

\item During the \emph{analysis and reasoning} step, each player maximizes its utility by choosing appropriate strategies while considering the current strategies of other players. Nash Equilibrium is considered as a steady state point after which the players cannot change their strategies to maximize their utilities. It has been demonstrated that for the potential function in (\ref{eq:Waleed4}), the players reach Nash Equilibrium after a certain number of iterations. For an in-depth study of the Nash Equilibrium and the convergence of this game, the reader is referred to  \cite{Monderer1, Sandholm1}.

\item In the \emph{decision} step, each device decides to update its strategy according to the best response update strategy. If the device updates its strategy then it will share this strategy with its neighboring devices.
\end{itemize}

\subsubsection{Performance analysis}
The performance of the system is evaluated through extensive computer simulations. Four service classes  are considered with 25\% of total M2M devices in each service class. For illustration purposes, it is assumed that cellular, WLAN, and mmWave channels are available for data transmission where the parameters for  cellular bands are according to 3GPP specifications \cite{3GPP3}, WLAN are according to WiFi \cite{Rangan1}, and mmWaves are according to specifications in \cite{Rangan1, Bai1}. It is assumed also that four routing protocols are available for M2M devices and these are power-based \cite{Pyo1}, delay-based \cite{Yang_Zongkai1}, throughput-based \cite{Sampath1}, and reliability-based \cite{Abbagnale1}. The reconfigurable approach is compared to a user-optimized scheme called joint resource allocation and network selection (JRANS), where the resources are allocated  based on the interference factor and the channel gain only  \cite{Mustika1}. The performance metrics considered in the simulations are throughput, energy efficiency, economic cost, and environment impact.

\begin{figure}[!b]
\centering
\includegraphics[width=3.5in]{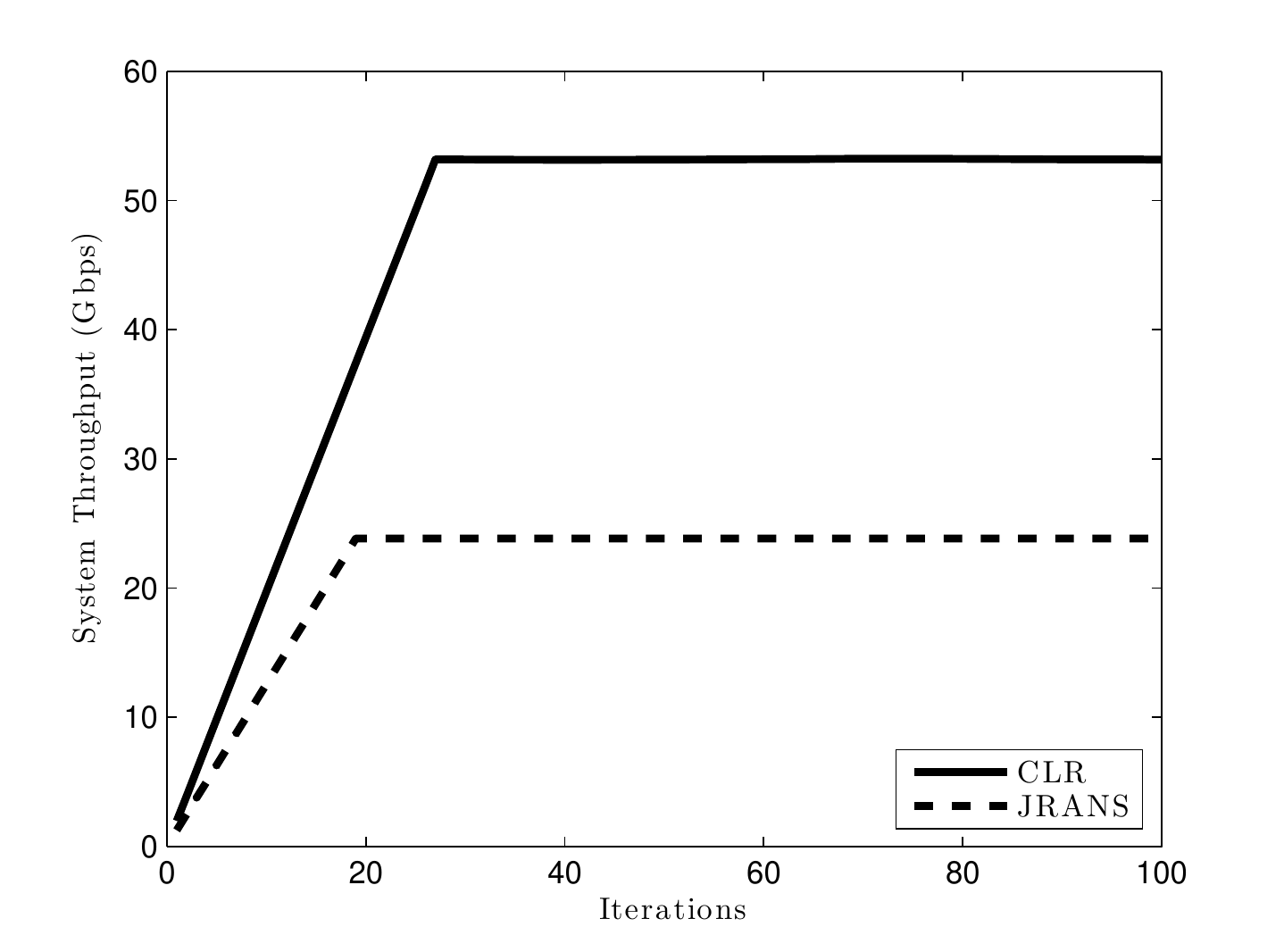}
\caption{Convergence of the CLR  and the JRANS schemes in terms of system throughput (the number of M2M devices is 300).}
\label{fig:Waleed4}
\end{figure}

Fig. \ref{fig:Waleed4} illustrates the convergence  in terms of throughput of the CLR scheme and comparison is carried out with JRANS scheme. In the CLR scheme, each M2M device selects suitable strategy to improve its utility while considering strategies of other players. After some iterations, a steady state (Nash Equilibrium) is reached from where no one can increase its utility. The JRANS scheme converges faster  because it only considers interference and channel gains for channel allocation and network selection. However, the CLR outperforms the JRANS in terms of total throughput.

\begin{figure}[!t]
\centering
\includegraphics[width=3.5in]{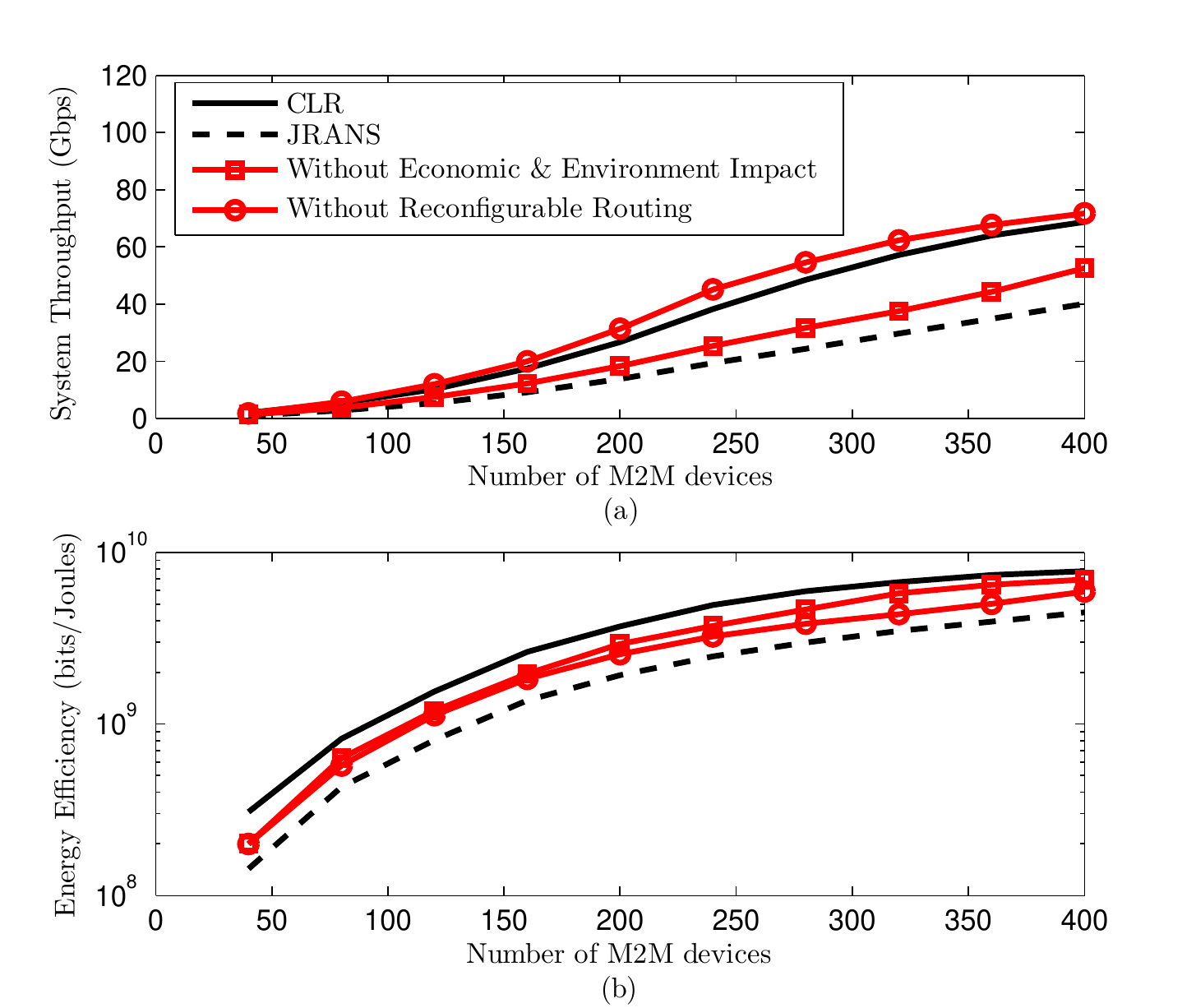}
\caption{(a) Throughput and (b) energy efficiency with variations of the number of M2M devices.}
\label{fig:Waleed46}
\end{figure}

Fig. \ref{fig:Waleed46}(a) compares the system throughput versus the number of M2M devices for the CLR scheme, JRANS scheme, and two other schemes that consider different variants of  (\ref{eq:Waleed3}) (without economic and environment impact, and without reconfigurable routing). It should be noted that the higher throughput in the case of non-reconfigurable routing is due to the fact that in that scheme, throughput-based routing protocols are used for all M2M devices. However, the CLR based scheme reconfigures the routing protocols according to the corresponding service classes. The CLR scheme outperforms the JRANS scheme and the other scheme that excludes economic and environment impact from the  utility function. The economic and environment impact pushes M2M devices to unlicensed and low power resources such as mmWaves, which offer high bandwidth in indoor areas while reducing energy consumption and satisfying other QoS requirements. This explains the degradation of system throughput when economic and environment impacts are excluded from utility function.

Fig. \ref{fig:Waleed46}(b) shows energy efficiency comparison for the aforementioned schemes. The CLR scheme outperforms all other competitors. The performance degrades in terms of energy efficiency when  the economic and environment impacts are removed from the utility function because in that case the system will choose channels without considering their energy footprint. Moreover, energy efficiency without considering reconfigurable routing further degrades the performance because in that case the system uses the routing protocol with the highest throughput regardless of its energy cost.

\begin{figure}[!t]
\centering
\includegraphics[width=3.5in]{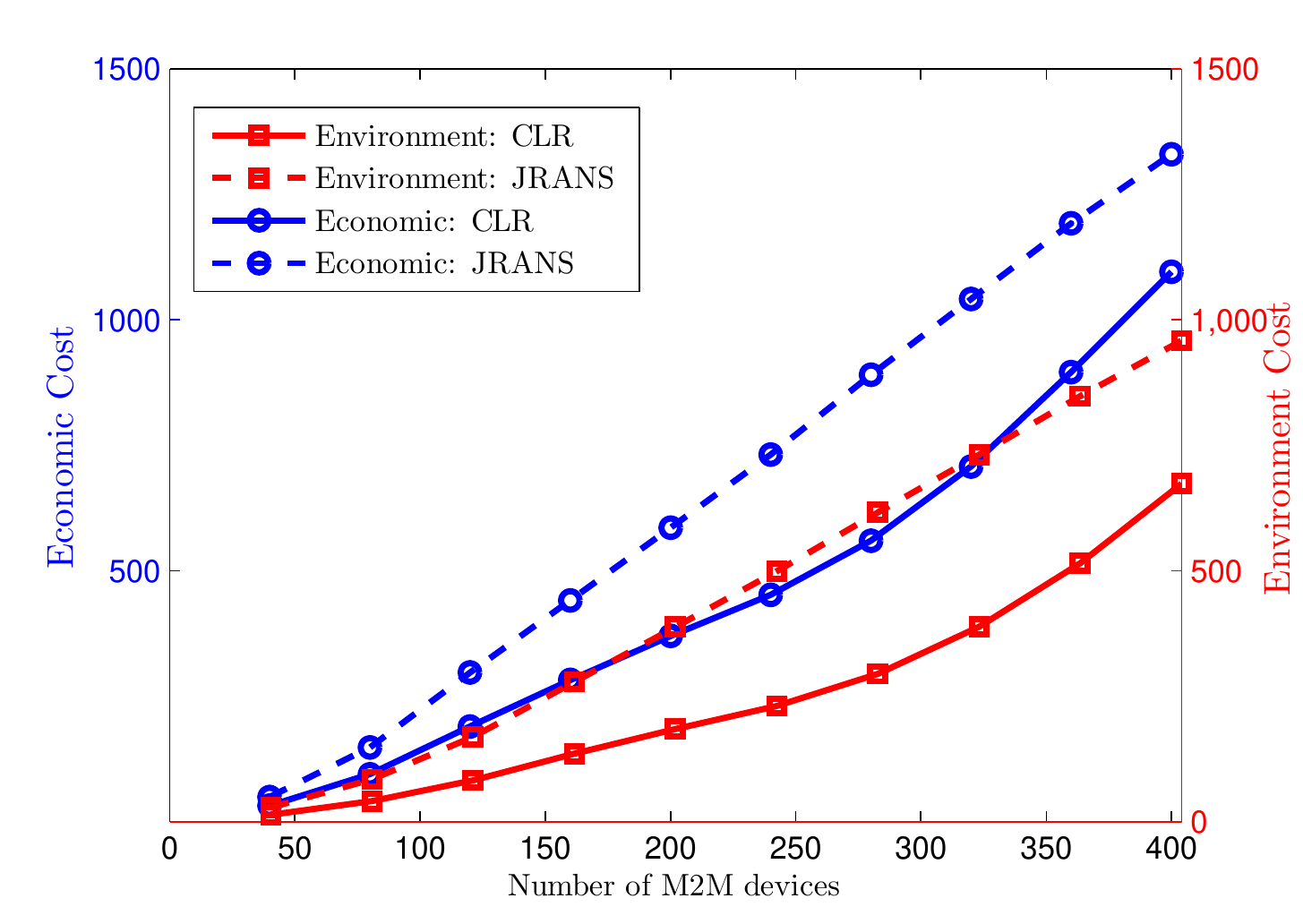}
\caption{Economic cost and environment impact versus the number of M2M devices.}
\label{fig:Waleed9}
\end{figure}

Fig. \ref{fig:Waleed9} compares economic cost and environment impact of the CLR and JRANS schemes versus the number of M2M devices, respectively. It can be seen that the CLR scheme has about 25\% less economic cost and 27\% less environment cost in comparison with the JRANS scheme when number of M2M devices is 400.

\subsubsection{Complexity}
It should be noted that the computational complexity of CLR is higher than JRANS. However, CLR uses a decentralized scheme where the complexity cost is distributed among all M2M devices. For the communication overhead (due to information exchange required by the game-theoretical approach), it can be significantly reduced if M2M devices use existing common control channels. This complexity cost is fully justified given the significant performance improvement that CLR offers in terms of throughput, energy efficiency as well as economic and environment costs compared to JRANS.

\subsection{Network Management}
Network management covers a broad range of functions that generally have to do with network reconfiguration. These functions can generally be classified into three categories: Parameter configuration, connectivity management, and network monitoring. Parameter configuration consists of choosing the appropriate network parameters to achieve a given set of objectives. For example, topology control algorithms typically focus on choosing the appropriate transmit power that achieves network-wide connectivity \cite{Li_Mo1}. This can be done in a centralized way \cite{Li_Mo1} or in a distributed way \cite{Aziz1}. It can also be homogeneous (all nodes transmit using the same transmit power) or heterogeneous.
On the other hand, connectivity management encompasses a wide variety of tasks such as mobility management \cite{Fernandes1} \cite{Liou1}, radio resource management (RRM) \cite{Pedersen1} \cite{Lee_Ying1}, and QoS support \cite{Pedersen1}. Finally, network monitoring refers to monitoring the links between users and the serving stations or between users. It is a function used to collect information that serves other networking functions such as CSI or mobility information.

Clearly, network management is cross-layer in nature. However, the main difference between network management protocols and other cross-layer protocols is that network management focuses on functions that have network-wide effect, while other cross-layer protocols may optimize individual links or a small set of functions. In this sense, network management acts as the brain behind the network. In this section, we will consider two examples that illustrate how network management is used across different layers. Particularly, we consider the cases of RRM in LTE, and network management in WLANs.

In LTE, RRM is performed using several protocols \cite{Cox1, Pedersen1}. At the topmost layer of the LTE protocol stack, the network layer, the radio resource control (RRC) protocol performs three main functions: QoS management, admission control, and semi-persistent scheduling. QoS management builds a profile for the requirements of the upcoming communication session, admission control decides if these requirements can be satisfied given the current load, and semi-persistent scheduling is used for periodic allocation of resources (used for deterministic communication sessions such as video streaming). Then, at the MAC layer, the radio link control (RLC) and MAC protocols perform three other functions: HARQ, dynamic scheduling, and link adaptation. HARQ is an efficient error control mechanism that adjusts the number of error control bits according to link quality. On the other hand, dynamic scheduling and link adaptation are performed jointly according to the algorithm presented in Section III.A. Finally, at the PHY layer we find mainly the channel quality indicator (CQI) manager, which is basically a tool to monitor link quality. Thus, these protocols perform network management functions in the categories of parameter configuration (HARQ, link adaptation, dynamic scheduling, and semi-persistent scheduling), connectivity management (QoS management and admission control), and monitoring (CQI management). The reconfigurations performed by these protocol stack are numerous. For example, whether the system will apply semi-persistent scheduling or not depends on the type of application. In addition, admission control considers existing connections as well as new requests to determine the ones that will be dropped. Furthermore, dynamic scheduling may switch between scheduling techniques according to the type of applications requested by the users. Generally speaking, the main objective of the reconfigurations in the LTE protocol stack is QoS support. The protocols are designed to consider a large number of users with varying application requirements.

RRM functions in LTE are performed at the eNodeBs, which control the operations of the UEs in their service areas. In addition, in LTE-A, eNodeBs communicate with each other and coordinate their transmissions, as explained before. In any case, RRM functions are coordinated and can be optimized even across cells. However, this is not the case in WLANs with multiple interfering access points (APs). Here, management functions are performed independently at each AP, with minimum collaboration with the other APs. Nevertheless, there are provisions for network management within the IEEE 802.11 family of standards. For example, 802.11r \cite{IEEE1} allows for fast negotiations between APs by allowing multiple negotiation sets to occur in parallel, 802.11k \cite{IEEE2} defines how APs exchange management information, while 802.11s \cite{IEEE3} allows APs to operate in a mesh topology. However, each AP still manages the users within their cells independently of any other APs. These management functions include channel monitoring, priority management, and power configuration. However, there are two particular management functions that require collaboration between APs: Channel selection and AP association. The first is the process by which the AP chooses the channel over which it will operate, while the second is the process by which the user chooses a serving AP. Traditionally, channel selection is done by considering the channels utilized by the neighboring APs, in order to reduce inter-AP interference. Once the AP chooses the channel, it starts transmitting its beacon messages to announce its presence. Then, users typically associate with the AP with the highest signal strength. It is worth noting that other protocols for AP association have been proposed that consider metrics such as packet error rate \cite{Fukuda1} or throughput \cite{Kauffmann1}, which are more efficient than signal strength alone.

\begin{figure}[!b]
\centerline{\subfigure[]{\includegraphics[width=3.5in]{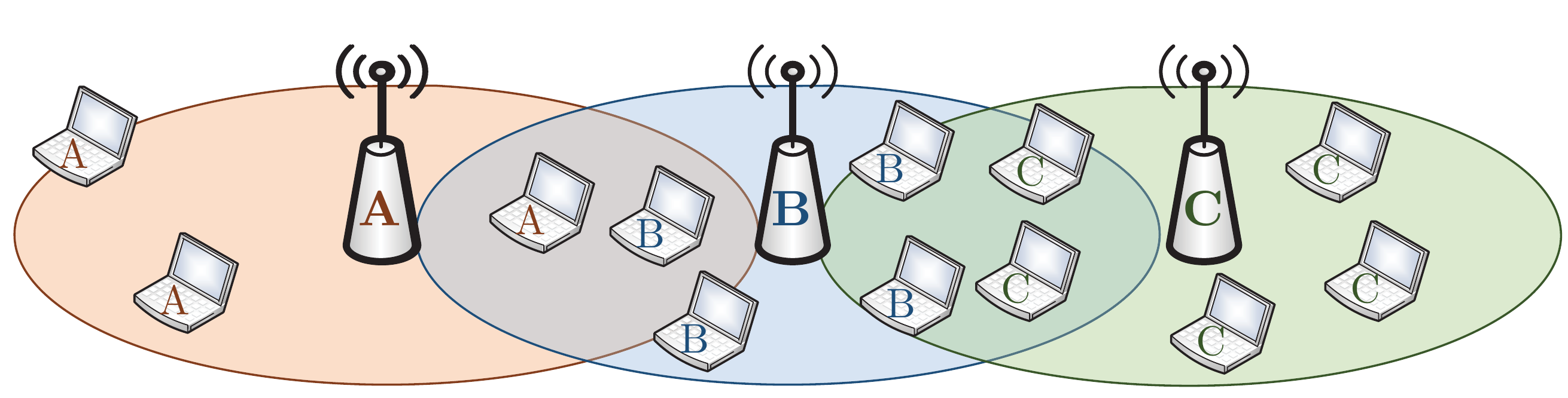}
\label{fig:Amr6a}}}
\centerline{\subfigure[]{\includegraphics[width=3.5in]{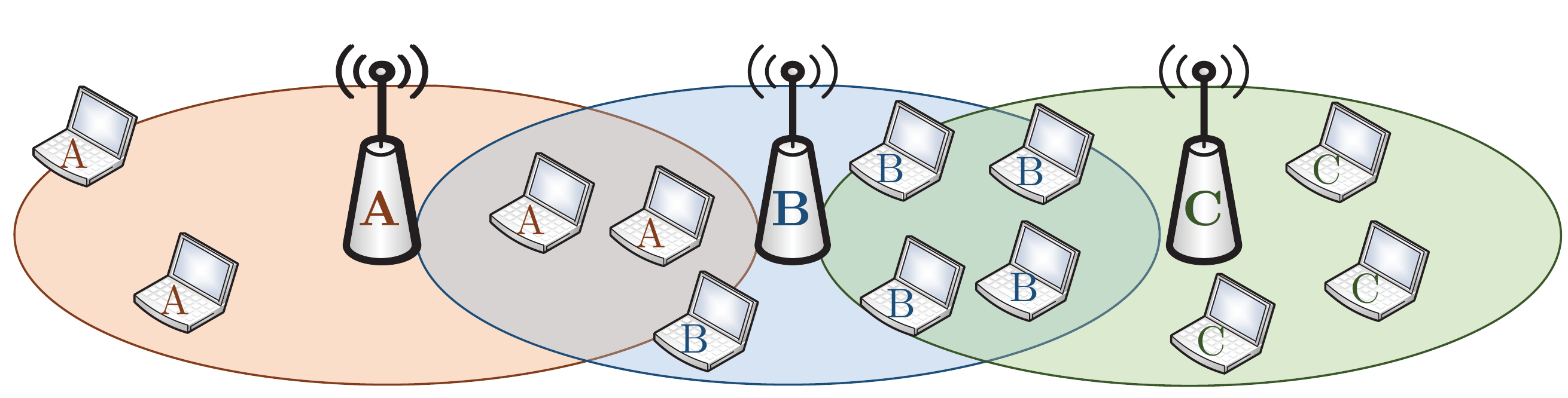}
\label{fig:Amr6b}}}
\caption{Topology reconfiguration in three WLANs: (a) After initial associations; (b) After reporting their CSIs. Some device associations have been reconfigured to achieve load balancing.}
\label{fig:Amr6}
\end{figure}

According to \cite{Zhu1}, channel selection and AP association must be done in a user-centric way in order to improve throughput. This means that the channel conditions seen by the user must be considered, and load balancing must be performed. Thus, the authors in \cite{Zhu1} propose a user-centric protocol for network management in WLANs. This protocol starts in the traditional way by having the AP select a channel and send beacons. Then the user sends its CSI seen with every AP from which it is receiving beacons. The APs then calculate the potential throughput according to this CSI and send the information back to the user. Using this information, the user can associate with the AP that provides the maximum throughput. In addition, each AP periodically collects user performance statistics from its users and may trigger a channel adjustment operation if the aggregate throughput drops below a certain threshold. Since AP association is based on throughput, load balancing is implicitly included since loaded cells will provide users a potentially low throughput estimation, discouraging it from associating with it. These reconfigurations are illustrated in Fig. \ref{fig:Amr6}.

\begin{figure}[!b]
\centering
\includegraphics[width=3.5in]{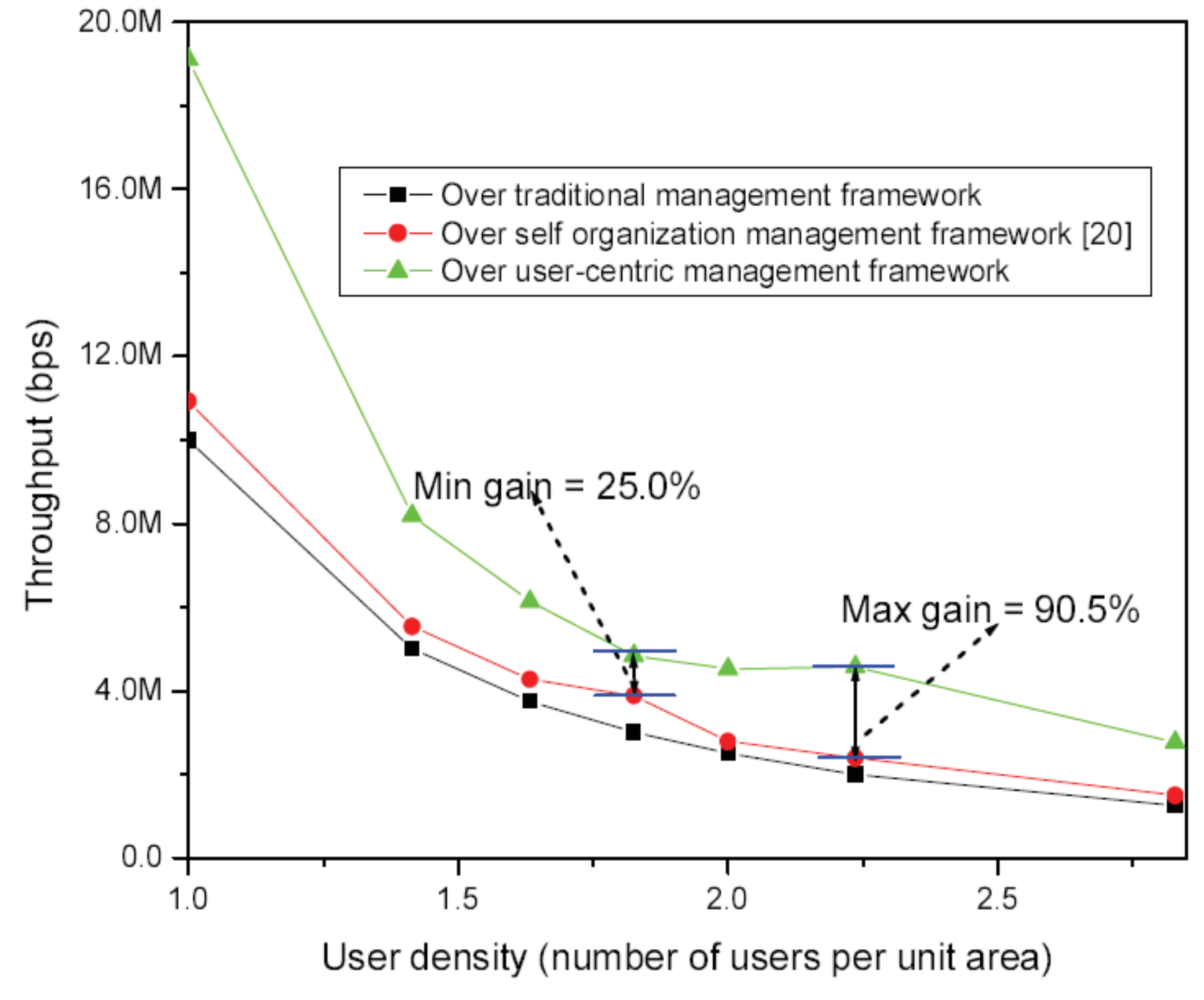}
\caption{Performance results for the AP association scheme.}
\label{fig:Amr7}
\end{figure}

The performance of the user-centric algorithm is evaluated in \cite{Zhu1}  from the perspective of the throughput observed by the user and the throughput observed by the AP. Several experiments are conducted. In one experiment, one user is tagged as the observing user and its throughput is measured for different number of users per unit area. This experiment evaluates the performance of the AP association scheme proposed in \cite{Zhu1}. Fig. \ref{fig:Amr7} shows that the proposed scheme achieves performance gains from 25\% to about 90\% over the self-organization management framework, and the gain is higher in lower user densities.

Network management protocols have also been proposed for ad hoc networks. However, due to the distributed nature of these networks, it is difficult to propose protocols that can have a network-wide effect. In addition, the multi-hop nature of these networks make collecting information costly. Thus, network management protocols for ad hoc networks typically focus on a narrow set of objectives such as topology control \cite{Aziz1}, admission control and scheduling \cite{HanzoII1} \cite{Gabale1}, or congestion control \cite{Kafi1}.

\subsection{Software-Defined Networking (SDN)}
Conventional networks rely on protocols that implement a particular set of functions that target some specified objectives, as we have seen in previous sections. However, modern networks are becoming increasingly complex in terms of challenging requirements such as changing user demands, reliability, scalability, and security. These demands may even be time-varying. Thus, protocols (especially those that are implemented in hardware) may not adapt so easily. This has led to the revolutionary idea of SDNs. The main concept in SDN is to decouple the control plane from the data plane \cite{Lara1} \cite{Nunes1}. Thus, all network policy rules and configurations, forwarding rules, packet handling rules and priorities, etc., are implemented using reprogrammable software called the \emph{controller} \cite{Hu_Fei1}. These rules are forwarded to special SDN-compliant switches and routers, where they are stored in tables called \emph{flow tables}.

\begin{figure}[!b]
\centering
\includegraphics[width=3.5in]{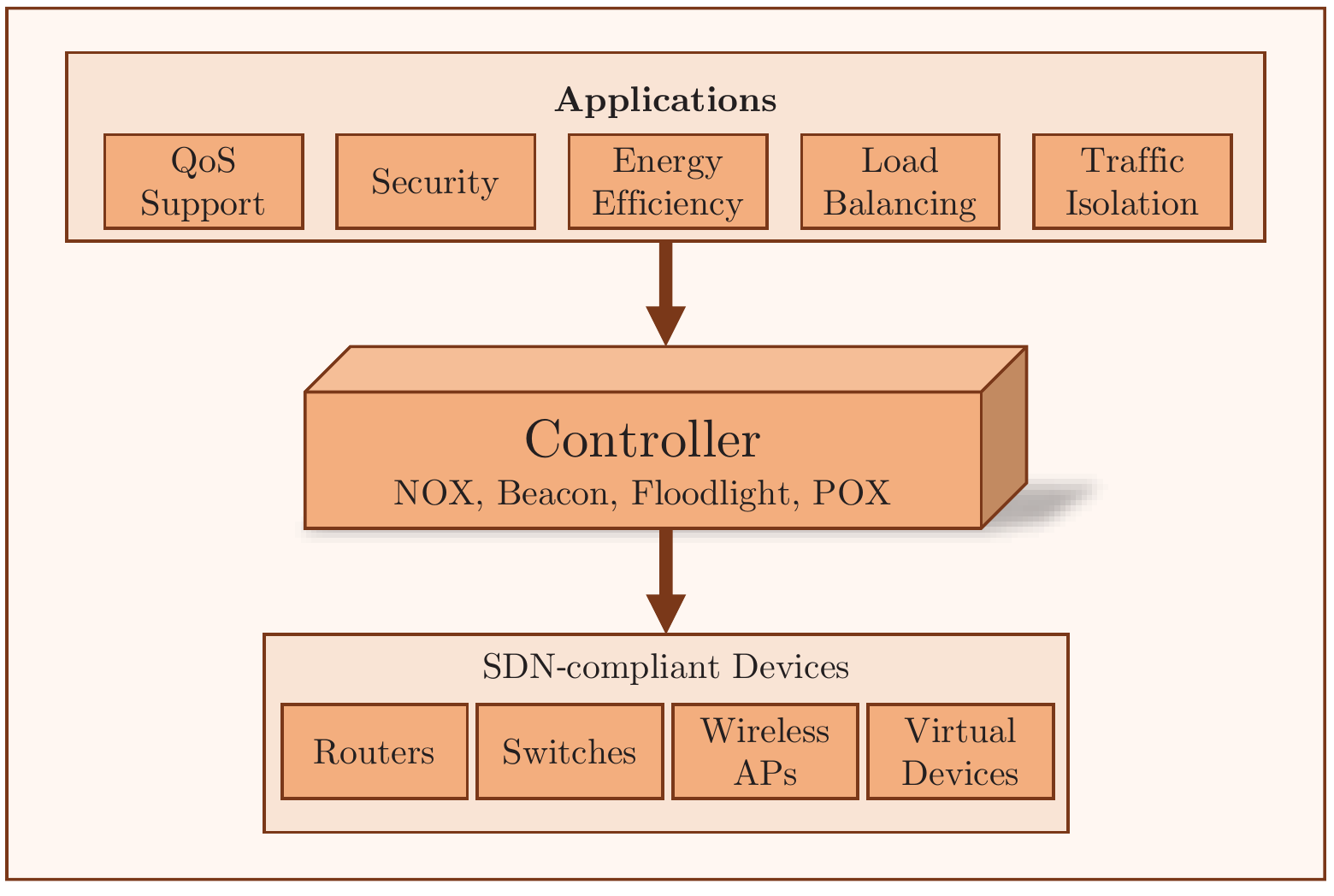}
\caption{SDN architecture.}
\label{fig:Amr8}
\end{figure}

SDN architecture is illustrated in Fig. \ref{fig:Amr8}, where there  are several controller implementations such as POX \cite{Noxrepo1}, NOX \cite{Noxrepo1}, Beacon \cite{Beacon1}, and Floodlight \cite{Floodlight1}. Each one provides a base on which any application can be implemented. On the other hand, communication between the controller and the SDN-complaint devices is based on an SDN protocol such as forwarding and control element specification (ForCES) \cite{Doria1} or interface to the routing system (IRS) \cite{IWGroup1}. Nevertheless, the most prominent SDN protocol is OpenFlow \cite{Lara1,ONFoundation1}. Using, OpenFlow, entries in the flow table contain match fields (forwarding rules), counters and instructions (prioritization, dropping rules, etc.). Upon receiving a packet, the SDN-compliant device checks its flow table for a match. If found, the match will be used to forward the packet. However, if a match is not found, the packet is forwarded to the controller for analysis. The controller then uses software to decide what to do with the packet, and may then forward a new matching rule to the device. Since the controller is reprogrammable, all networking policies can be re-adjusted as needed.

SDN can be implemented in many ways \cite{Sezer1}. For example, a centralized implementation is to have one controller managing multiple devices, while a distributed implementation is to have one controller per device and to have the controllers communicate with each other to coordinate their operations. Hybrid and hierarchical implementations (one controller managing multiple controllers, each managing one or more switches) are also possible.

Due to this flexible architecture, SDN can have many possible applications. For one thing, it allows for experimenting with protocols without having to implement on hardware. It also has huge implications on interoperability in multi-vendor networks \cite{Kim_Hyojoon1} \cite{Martinez1}. For this reason, major cloud providers (such as Google and Amazon) are highly in favor of SDNs. Using SDNs, these cloud providers can purchase equipment from any vendor, and they can operate together using instructions from the controller if they are SDN-compliant \cite{Crisan1}. SDNs also simplify network management significantly and allow for large scale automated management. They also allow for easy management of VLANs \cite{Crisan1}, since managing network IDs and the size of each network becomes simple using the software-based controller.

\begin{figure*}[!t]
\centerline{\subfigure[]{\includegraphics[width=4.5in]{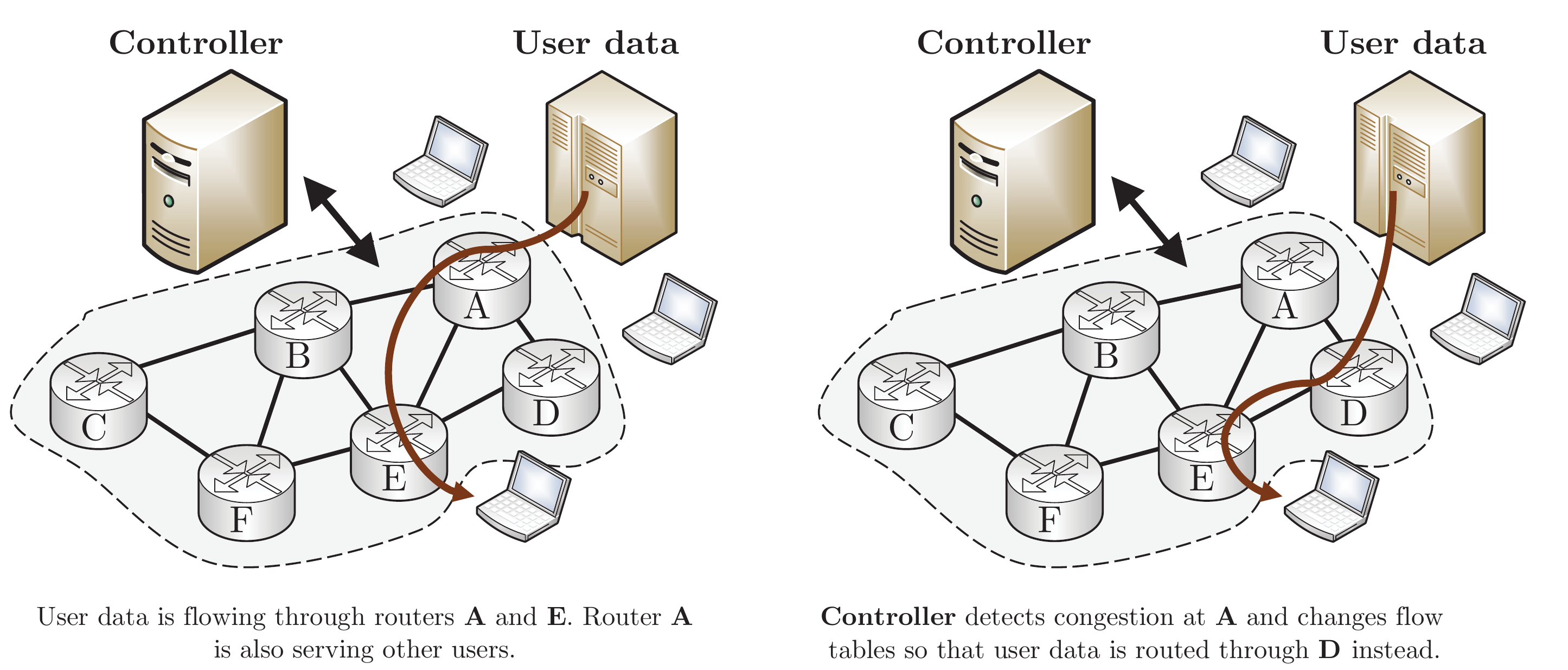}
\label{fig:Amr9a}}}
\centerline{\subfigure[]{\includegraphics[width=4.5in]{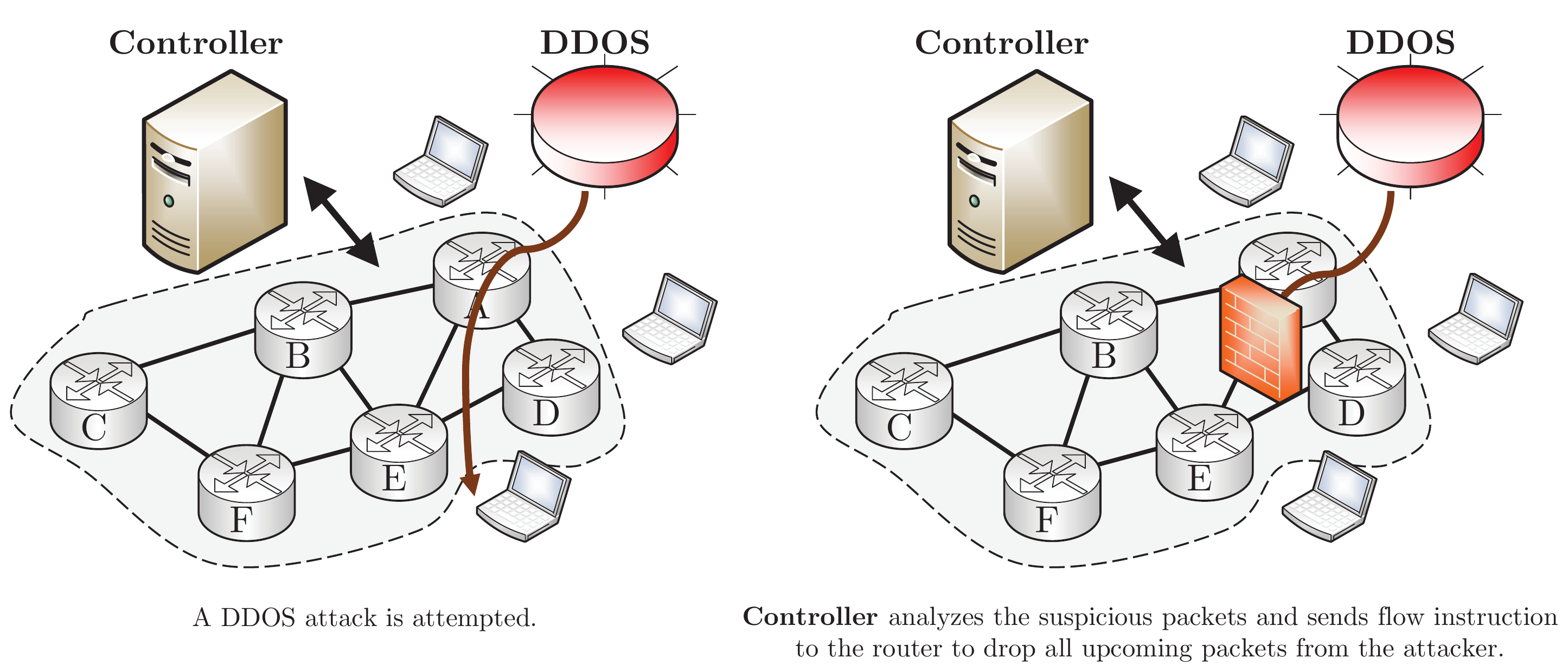}
\label{fig:Amr9b}}}
\centerline{\subfigure[]{\includegraphics[width=4.5in]{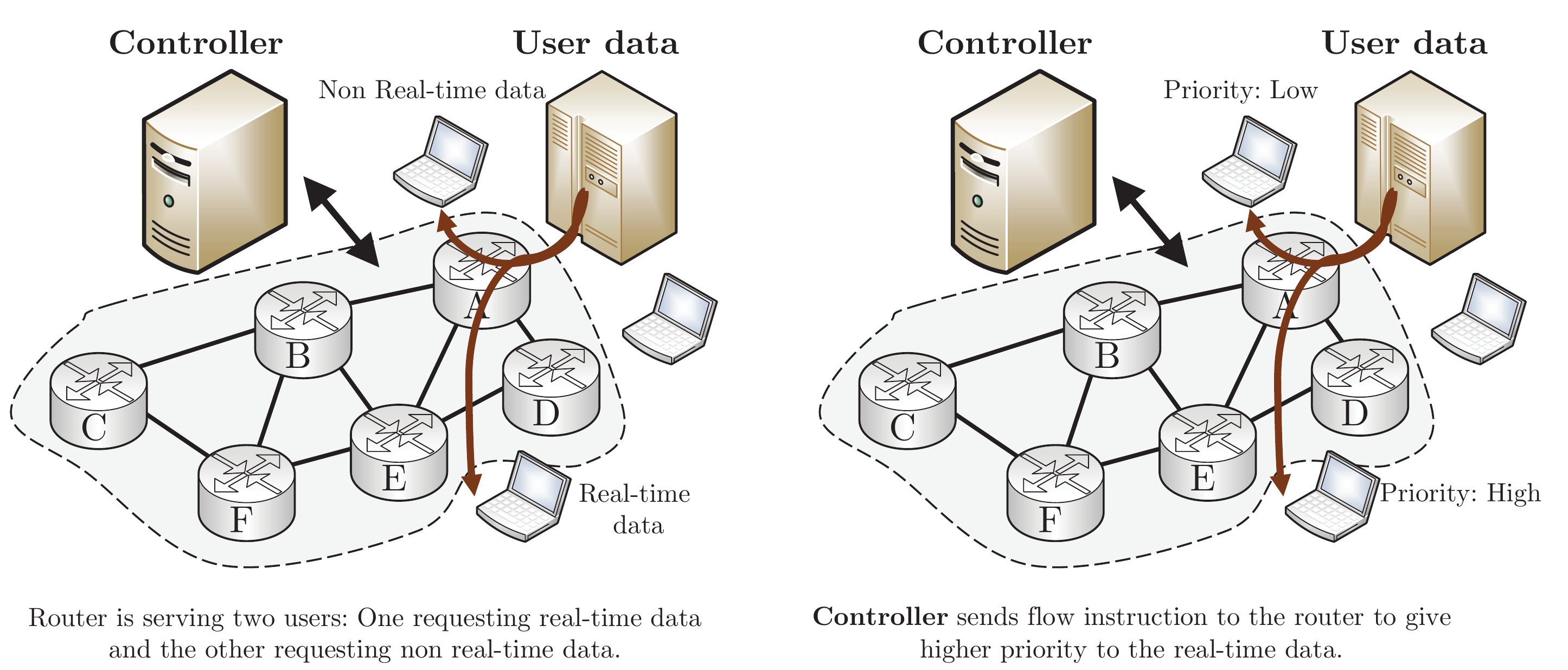}
\label{fig:Amr9c}}}
\caption{Examples of SDN applications: (a) Load balancing using SDN; (b) Security using SDN; (c) QoS support using SDN.}
\label{fig:Amr9}
\end{figure*}

SDNs also make load balancing and congestion control quite simple since the controller can get statistics from the devices and optimize network traffic distribution \cite{Gruen1} \cite{Hu_Yannan1}. Another important application is network security \cite{Gutz1} \cite{Scott-Hayward1}. This is because the presence of the controller provides powerful tools for traffic analysis. The software can also be modified and adapted to consider new forms of attacks. Thus, security attacks such as distributed denial of service (DDoS) can be handled more easily \cite{Lara1}. The controller also provides means for traffic isolation, simply by adjusting the forwarding and dropping rules. Finally, SDNs provide significant tools for flexible QoS support \cite{Egilmez1} \cite{Bari1}. Controllers can inspect packet types and payload, and adjust priorities and queues accordingly. If demand changes, the controller can re-adjust its configuration.  Note that all these reconfigurations can be performed automatically and in software, a capability that cannot be easily achieved outside the scope of SDN. An illustration of some SDN applications is shown in Fig. \ref{fig:Amr9}.

It is worth mentioning that there are multiple versions of the OpenFlow protocol \cite{Lara1}. The earliest version is OpenFlow 1.0.0 and is the most widely deployed version. Here, the controller has a single flow table and there is no support for IPv6, multi-protocol label switching (MPLS), or simultaneous communication with multiple controllers. MPLS matching and multiple flow tables are included in version 1.1.0. Then, version 1.2.0 adds support for IPv6 and inter-controller communication. Finally, the latest version is 1.3.0 and has more flexible capabilities for communications between controllers.

SDN can be used in wired and wireless networks and can provide a revolutionary way for network reconfiguration and adaptability. It is certainly a very promising technology that is expected to continue rising in the near future. However, there are challenges that need to be considered, and several of them stem from the dependency of the network on the controller. This presents reliability and scalability threats. If the controller fails, the network will not function. Moreover, if the controller is managing a large number of nodes, it may only be able to handle a limited amount of packets being forwarded to it for analysis. This also makes the controller an attractive target for attackers not only because it represents a single point of failure, but also because it holds critical information about the network. Thus, security measures have to be in place such as intrusion detection, policy adaptation, and network forensics. Finally, recovery time of SDNs needs investigation. In order for a node to recover from an error, it needs to send the packet to the controller and receive a decision back. This may cause some latency. This has to be considered in the controller design.

\section{Network Intelligence}

Cross-layer reconfiguration has some limitations \cite{Thomas1,Vijay_Gayathri1}. For example, there is a risk known as adaptation loops, where improving one aspect may lead to downgrading another. There are concerns regarding the coexistence of several cross-layer designs. Since cross-layer designs are generally application-specific, it may not be straightforward to integrate multiple solutions. Thus, wide-scale implementations are challenging.

In this section, we focus on cognitive networking and context-awareness as means of improving cross-layer adaptation and adaptability.

\subsection{Cognitive Networking}
\begin{figure}[!b]
\centering
\includegraphics[width=3.5in]{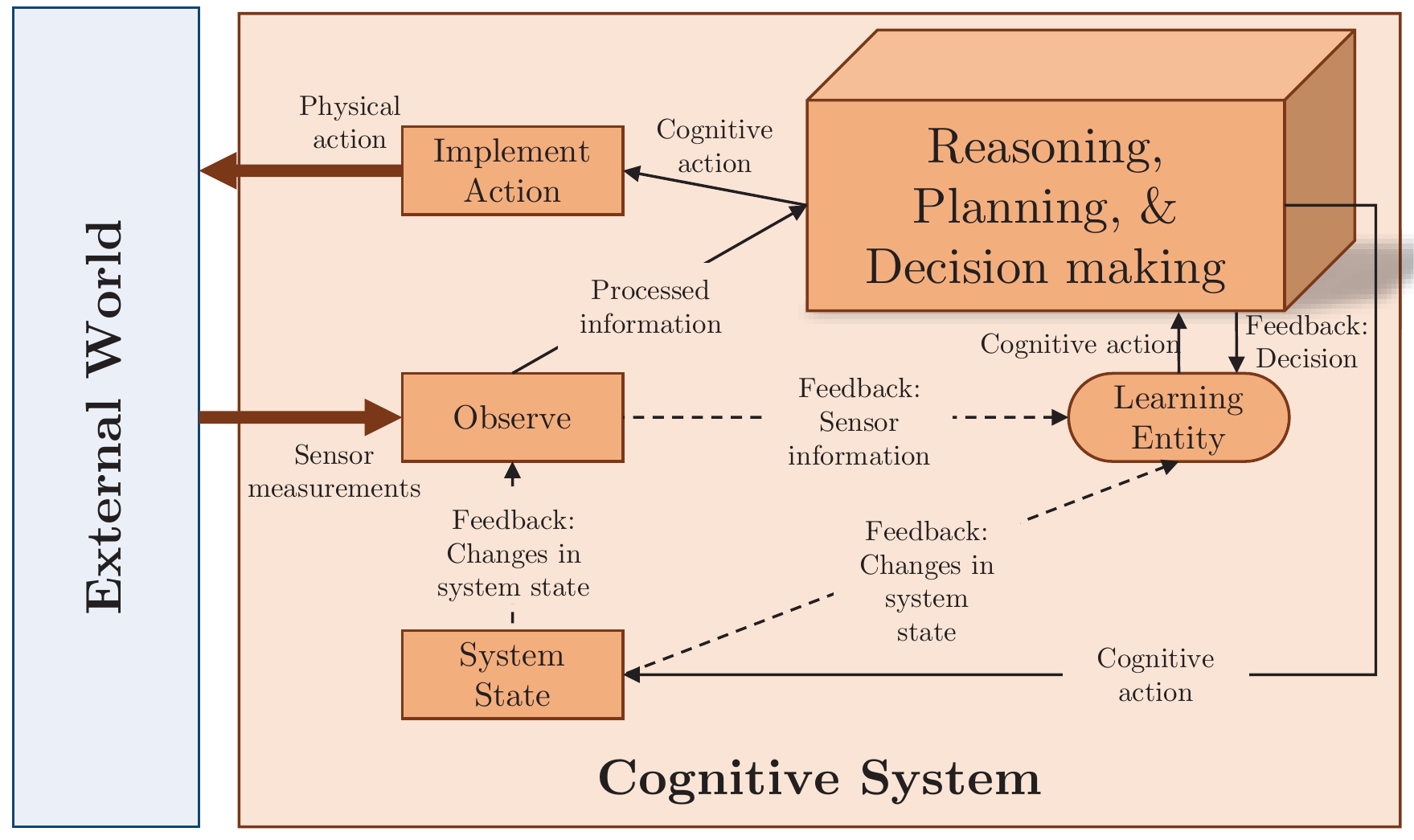}
\caption{Cognitive loop \cite{Mitola3}.}
\label{fig:Amr10}
\end{figure}

Cognitive networking is based on the cognitive loop \cite{Mitola3}, shown in Fig. \ref{fig:Amr10}. It contains an observation entity responsible for monitoring the external world through sensory measurements. It also collects information about the internal state of the system. All this information is processed and passed to a reasoning machine. Here, the information is analyzed and decisions are made. These decisions can be short-term or long-term (requires planning). Furthermore, the decisions may change the internal state of the system and may require implementation using physical actions. In addition, feedback information from different parts of the system is passed to a learning entity that builds a knowledge database. This database is used to enhance the decisions produced by the reasoning machine over time.

The cognitive loop implements degrees of intelligence in the network. Rather than targeting specific applications or algorithms, cognitive networking focuses on methods of analyzing information, whatever it may be, and producing decisions, while focusing on some specific goals or objectives. Since learning is a highly important cognitive tool in humans, it is also implemented in the cognitive networking loop. Just as humans improve their decision-making by learning from their past experiences, networks too can optimize their decisions by observing the results of their previous actions and of the changes in the external world.

The importance of cognitive networking has led standardizing bodies to develop frameworks for management and control based on the cognitive loop. Three standards in particular are worth mentioning \cite{Tsagkaris1}: Reconfigurable radio systems (RRS) \cite{ETSI1}, IEEE 1900.4 \cite{IEEE4}, self-organizing networks (SON) \cite{3GPP1}.

Cognitive networking has also found implementations in ad hoc networks, where the lack of centralized nodes poses challenges. Several tools of computational intelligence (CI) have been utilized for this task. For example, evolutionary algorithms such as particle swarm optimization and ant colonies have been used in \cite{Chen_Hongbin1, Krishna1} to perform topology maintenance and congestion control in wireless sensor networks (WSNs). Particle swarm optimization considered node positions and data transfer rate, while ant colony optimization considered the time taken to traverse a path and the number of successful iterations in the search process. Simulation results show that the proposed algorithms are able to reduce congestion and delay and increase throughput. Other CI tools that have been used include game theory \cite{Han1, Wang_Qiang1, Ren_Hongliang1} and Bayesian Networks \cite{Holmes1, Chin_George1, Lin_Chih-Kuang1}.

Learning transforms the operation of the system from policy-based to intelligence-based \cite{Yau_Kok-Lim1}. Policy-based systems follow static rules that are predefined and hardcoded, while intelligence-based systems have the ability to learn new states and evolve the operation of the system to achieve optimal or near optimal performance. These systems also have the ability to detect statistical patterns in the environment, such as regions with good or bad channel conditions or the nature of the traffic transmitted and adapt to them. Since most wireless systems exhibit dynamic properties, it is clear that learning can be a powerful tool for improving system performance.

There are mainly three categories of learning algorithms \cite{Sutton1, Alpaydin1}: Supervised learning, unsupervised learning, and reinforcement learning (RL). In supervised and unsupervised learning, examples or training sequences are used to accomplish the learning process. Learning agents use these sequences to identify patterns or recurring behavior and learn about the system. In supervised learning, each entry of the training sequence consists of an input signal and a desired output value, and the job of the learning protocol is to find an inferred function in all the sequences that can be used for calculating new values. On the other hand, the desired output is unknown in unsupervised learning. The learning algorithm simply observes the input and tries to detect patterns. In RL, the training sequence is not needed. The learning agent relies on feedback from the normal interactions of the system and identifies the actions that yield the most reward or least cost.

RL is the most prominent learning tool in wireless networks. This is because RL is a simple and model-free algorithm \cite{Yau_Kok-Lim1}. This means that RL does not need to model the entire operating environment in order to perform learning, only the desired parameters are monitored. This also means that RL can focus on metrics that enclose a wide range of networking issues, rather than having to learn about all the parameters in the network. For example, the RL algorithm can monitor end-to-end delay in order to discover the states that minimize this metric rather than monitoring a wide range of issues such as queue length, routing metrics, channel quality, etc. Delay, not only encompasses all these issues, but also has direct impact on the QoS observed by the user.

RL has been used in various ways in wireless networks and at different layers. For example, RL is used in \cite{Bhorkar1} to design an opportunistic routing protocol. The learning protocol rewards paths with good transmission success probabilities in order to minimize the cost of relaying a packet. In addition, this protocol balances the exploitation of paths that have been given previous rewards versus the exploration of new paths in order to improve the efficiency of routing decisions. This is done using probe packets that are sent periodically to explore new paths. In another example \cite{Naddafzadeh-Shirazi1}, RL is used to design a MAC protocol where nodes cooperate to maximize the throughput per unit of consumed energy. Nodes exchange the RL values in order to explore the efficiency of the decisions from a network-wide perspective. Thus, nodes adjust their transmission power and transmission probabilities according to the accumulated rewards. RL has also been used in \cite{Jiang_Tao1} to improve the efficiency of spectrum sharing in CRNs. A warm-up stage is defined where nodes test the quality of different channels. Afterwards, the nodes will only learn about the set of channels that are discovered to be promising during the warm-up stage. This reduces the sensing frequency and improves the quality of transmission by focusing on good channels. In \cite{Prabuchandran1}, RL is used in defining energy management policies for sensor nodes with finite buffers. The goal of this scheme is to maximize network lifetime in the presence of energy harvesting sources.

It is worth mentioning that RL has challenges. First, exploration versus exploitation has to be addressed. Exploration may lead to better solutions, but it also means slower algorithm convergence. In addition, traditional RL algorithms have to be implemented either in a centralized way or at every node in the network. Some modifications have been proposed in order to enable distributed implementation. Examples are the distributed value function (DVF) \cite{Schneider1} and the distributed reward and value function (DRV) \cite{Yau_Kok-Lim1}, which combine the values learned at different agents. Nevertheless, these functions require the exchange of such values, thus incurring higher overhead. Finally, there is a trade-off between learning time and granularity of the learning process. Having more states may mean that accuracy of the learning process is increased. However, it also means that it may take longer for the protocol to visit all these states and learn about their performance.

There have also been limited efforts on using supervised and unsupervised learning in WSNs. For example, supervised learning is used in \cite{Ray1} to design a neural network that exploits the spatial correlation of sensed events in nodes with close proximity in order to reduce packet loss. However, the algorithm has to be implemented in a centralized way. On the other hand, unsupervised learning is used in \cite{Clancy1} to classify SUs and PUs in CRNs. The work also considered malicious attacks from SUs that misclassify themselves as PUs. In another paper \cite{Han2}, Bayesian learning is used as a method for unsupervised learning in support to a game-theoretical spectrum sensing technique. An auctioning game is proposed where nodes entered \emph{bids} (to a centralized coordinator) on using the available spectrum holes. Bayesian learning is used to determine whether a node should enter a bid in each round or not, based on previous experience. However, as mentioned before, supervised and unsupervised learning algorithms often do not adapt quickly enough in dynamic environments. In addition, if goals or conditions change, the training sequences may have to be redesigned.

\subsection{Case Study: Network reconfiguration based on reasoning and learning for WSN}
In WSNs, there are numerous issues that have to be managed while performing planning and negotiations \cite{Ericsson1,Vijay_Gayathri1}. Connectivity, coverage, sensing frequency, QoS requirements, routing, packet loss rate, and congestion are among those issues. All tasks have to be performed while ensuring energy efficiency. Furthermore, WSNs may be highly dynamic: Sensor nodes may be mobile, duty cycling may continuously change network topology, the wireless environment may be changing, etc. Thus, managing network resources and achieving the goals become highly challenging. Since most of these issues may involve multiple layers, cross-layer approaches have been typically involved. However, despite the different approaches that have been tried, there is still a need for a tool that can address these challenges as well as the dynamic nature of the network. The characteristics of such a new tool may include the ability to adapt to changing goals, addressing conflicting constraints, supporting QoS requirements, and being able to be implemented in a distributed way.

\subsubsection{Basic concepts}
A mathematical tool known as weighted cognitive map (WCM) has been recently introduced \cite{Dickerson1,IEEE5}. It presents significant promises in achieving the aforementioned requirements. A WCM is a graphical model that can represent any dynamic network using the underlying causal relationships between the different parts of the system. Vertices of the map represent concepts (such as transmit power control or BER from the real system). A characterizing scalar, $A_i$, is associated with each vertex, $C_i$, which identifies its activation level. $A_i$ can take any real value from [-1, 1]. Negative values imply a decreasingly active concept (for example packet loss ratio (PLR) becoming lower), positive values imply an increasingly active concept, while a value of zero imply that a concept is inactive. On the other hand, edges of the WCM connect concepts that are causally related. Weights are associated with each edge to identify the strength of the causality. These weights can also take any real value from [-1, 1]. Changes in the activation levels of the concepts, which can occur due to changes in the environment (such as an increase in PLR), trigger the WCM to operate. The inference properties then enable the WCM to calculate new activation levels using simple mathematical operations. The new activation levels translate into actions that are executed by the network (for example, an increase in the activation level of PLR may lead to an increase in the activation level of transmit power, which causes the system to use the next higher transmit power).

\begin{figure}[!b]
\centering
\includegraphics[width=3.5in]{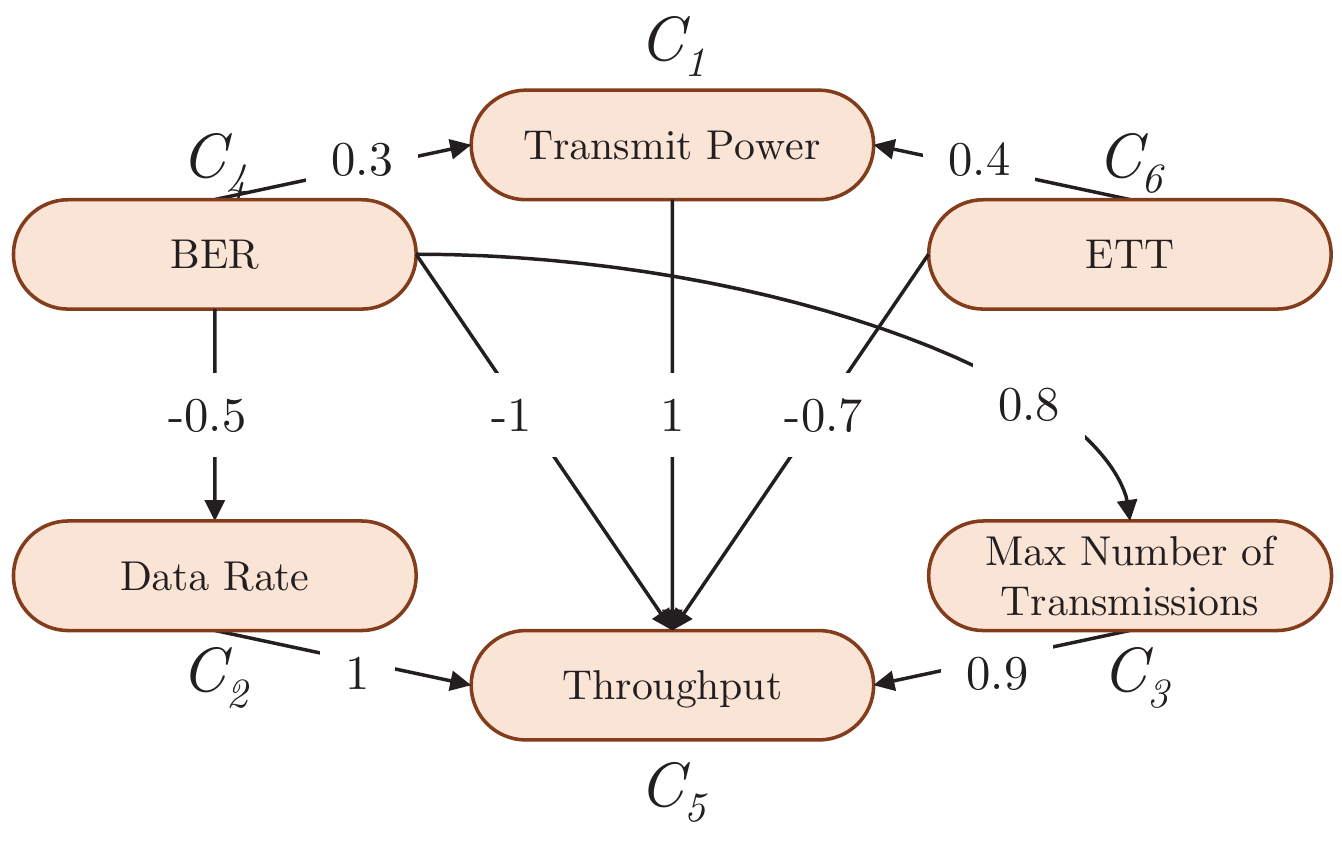}
\caption{WCM representing six processes within a wireless node.}
\label{fig:Amr13}
\end{figure}

As an example, the WCM in Fig. \ref{fig:Amr13} models the relationships between six processes (concepts) affecting a  node, namely, transmit power, data rate, maximum number of retransmissions allowed before a packet is dropped, BER, throughput, and the expected transmission time (ETT). The edge weights shown represent the strength of causality between concepts. We can classify concepts in any WCM into end-to-end goals, environment variables, and processes. All processes interact to achieve the end-to-end goals. On the other hand, environment variables cannot be manipulated directly, but only as a result of other actions. It can be seen from Fig. \ref{fig:Amr13} that the end-to-end goal is throughput (C5), since all other concepts can cause changes to it. It can also be seen that there are two environment variables considered, namely BER and ETT, since they cause changes in other processes but cannot be changed directly by them. They can only be affected by changes in the environment or as a result of actions taken by the WCM. The processes available to the WCM are transmit power, data rate adaptation, and changing the maximum number of retransmissions.

The WCM can be represented in a matrix form as
\begin{equation}
W=\bordermatrix{          \text{} &C_1    &C_2    &C_3    &C_4&C_5    &C_6\cr
                            C_1&    0       &0      &0      &0  &1      &0  \cr
                            C_2&    0       &0      &0      &0  &1      &0  \cr
                            C_3&    0       &0      &0      &0  &0.9     &0  \cr
                            C_4&    0.3     &-0.5   &0.8    &0  &-1     &0  \cr
                            C_5&    0       &0      &0      &0  &0      &0  \cr
                            C_6&    0.4     &0      &0      &0  &-0.7   &0}.
\end{equation}
For a WCM with $N_C$ concepts, its status at time $t$ can be given by
\begin{equation}
A(t)    =   [A_1,A_2,A_3,\ldots,A_{N_C}].
\end{equation}
The inference process of WCMs is the one by which the values of concepts change according to their underlying causal relationships. Thus, according to the inference properties of WCMs, the status at time $t+1$ can be given by
\begin{equation}
\label{eq:Amr3}
  A(t+1)    =   f\big(A(t)W\big),
\end{equation}
where $f(t)$ is a threshold function that determines the type of the WCM. The inference process is initialized when a particular concept is triggered, causing its activation value to change (e.g., a sudden increase in BER). Therefore, the input to the WCM at time $t$ is $A(t)$, including the new value of the concept that is just triggered. The triggered concept influences other concepts according to $W$, thus producing an output to the WCM, $A(t + 1)$, as given by (\ref{eq:Amr3}). Thus, WCMs can be implemented in a distributed way and has powerful reasoning capabilities. In addition, since WCMs are based mainly on the causal relationships between concepts, conflicting constraints can be easily considered. Furthermore, if goals or constraints change, modifying the edge weights is sufficient to allow the WCM to adapt to the new requirements.

\begin{figure}[!b]
\centering
\includegraphics[width=3.5in]{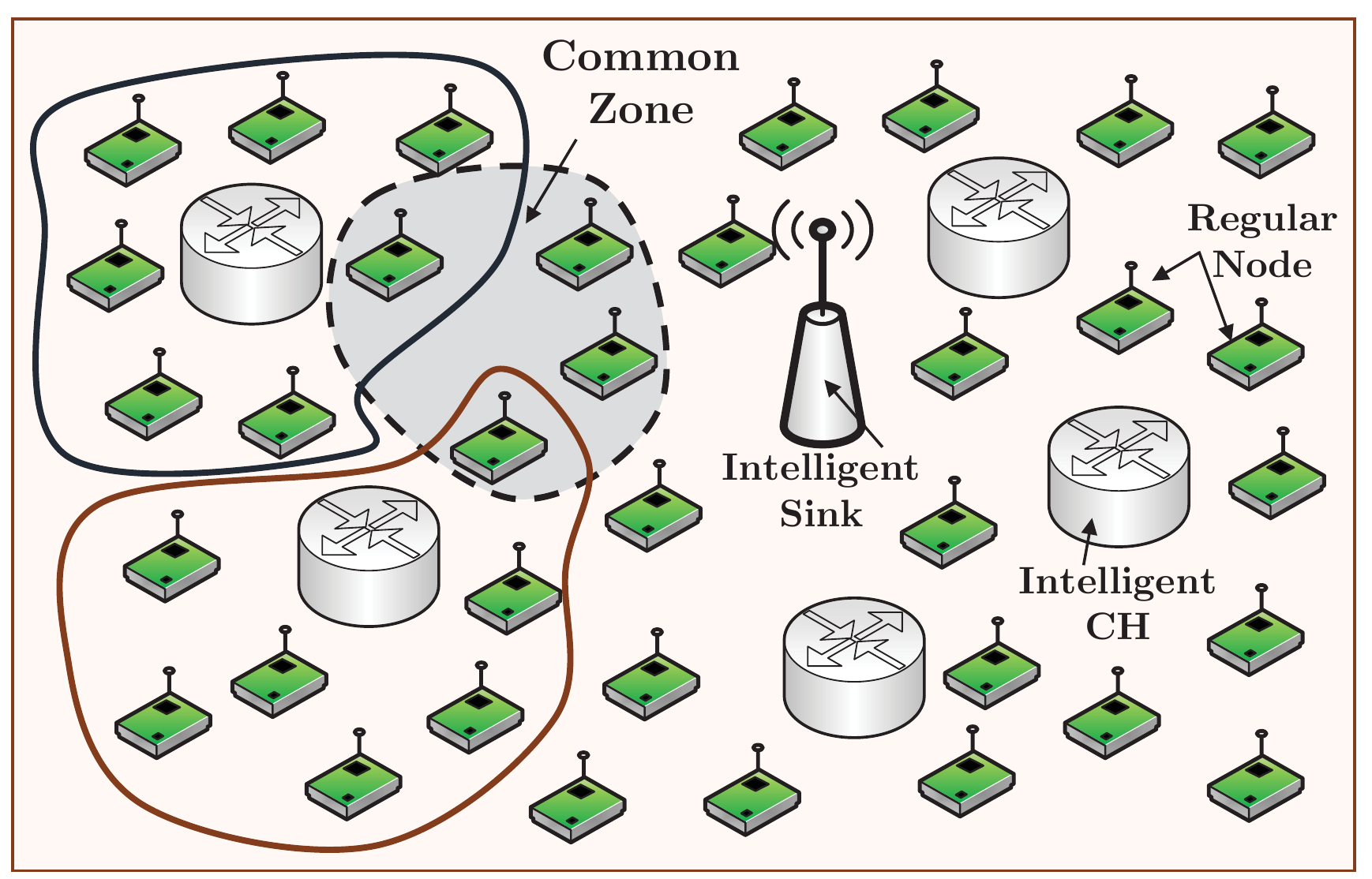}
\caption{Network topology with random deployment of nodes. The WCM is implemented in a distributed way on the CHs which act as intelligent nodes.}
\label{fig:Amr18}
\end{figure}

In this case study, we will illustrate how WCMs can be used to design a reasoning machine for WSNs \cite{El-Mougy1,Baladron1}. For the WSN architecture, a cluster hierarchy is employed for its popularity and simplicity. The WCM is implemented at the cluster heads (CHs) and a sink node,  which are assumed to have higher energy and processing capabilities than regular sensor nodes (Fig. \ref{fig:Amr18}). Sensor nodes are deployed in uniform random positions across the area. Thus, the WCM at each CH is responsible for determining the parameters to be used by all nodes in its cluster. It is assumed that nodes are synchronized and that time is slotted. Every $T$ consecutive time slots, the WCM determines the appropriate number of nodes that must be activated in order to guarantee connectivity and coverage, using the theorem that is proposed in \cite{Roy1}. These nodes remain active for the next $T$ time slots. On the other hand, the WCM monitors the cluster every $T/\eta$ time slots to check if actions need to be taken. The value $\eta$ depends on the dynamics of the network. In highly dynamic networks, $\eta$ can be equal to $T$ so that the WCM is activated in every time slot.

\begin{figure}[!b]
\centering
\includegraphics[width=3.5in]{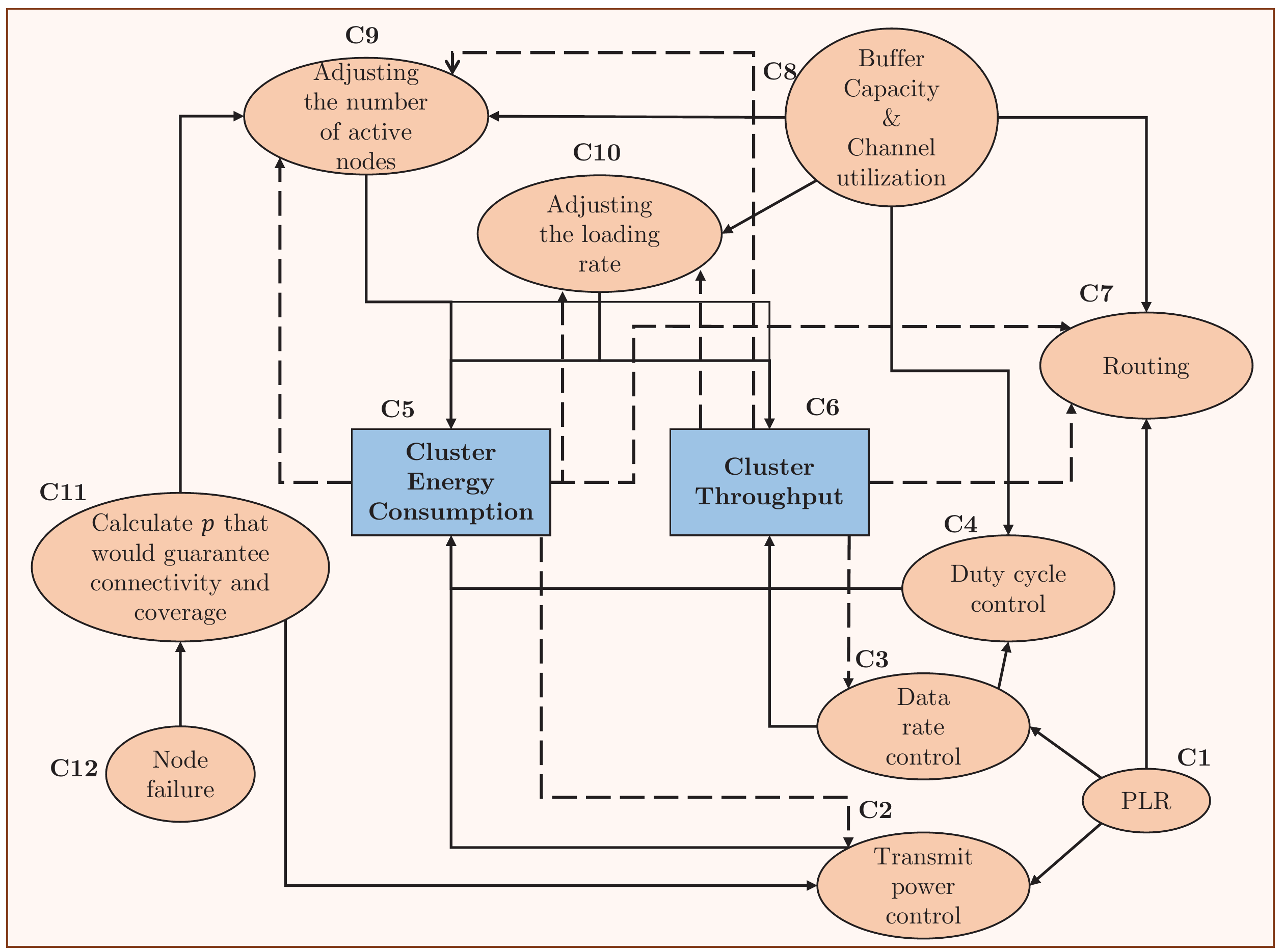}
\caption{Internal structure of the WCM engine as implemented in the cluster heads.}
\label{fig:Amr15}
\end{figure}

The first step in designing the WCM is to specify the end-to-end goal(s). Since network lifetime and throughput are the two main goals of the WSN, they must be directly incorporated as goals of the WCM. This way the WCM can constantly monitor these goals and make sure they are constantly achieved. Therefore, in order to consider network lifetime, one goal of the WCM will be energy consumption. Given that the initial battery power of nodes is $E$ mAhr, and that a target lifetime, $X$ time slots, is defined by the user, the task of the WCM is to make sure that the rate of energy consumption does not exceed $E/X$ mAhr per time slot per node. However, due to the expected variation in the energy consumption of nodes, forcing the WCM to react whenever there is an increase in energy consumption in a time slot will result in frequent changes, which is undesirable. Thus, if $N_C$ is the number of nodes within the cluster, the WCM at the CH will monitor the energy consumption, $EnC$, within its cluster and makes sure that the total energy consumption satisfies
\begin{equation}
  \sum_{i=1}^{N_C} EnC_i < \frac{T\cdot N_C\cdot E}{4X},
\end{equation}
mAhr per $T/4$ time slots, where monitoring takes place every $T/4$ time slots so that the WCM has enough time to take action before the next scheduling round takes place. Similarly, if $N_{WSN}$ is the total number of nodes in the network, the sink node is responsible for making sure that the total energy consumption of the network satisfies
\begin{equation}
  \sum_{i=1}^{N_C} EnC_i < \frac{T\cdot N_{WSN}\cdot E}{X},
\end{equation}
mAhr per $T$ time slots.  Note that monitoring at the sink node takes place every $T$ time slots. On the other hand, the end-to-end throughput can be monitored at the sink node by counting the number of packets successfully received per time slot. Thus, if $R_{WSN}$ is the throughput required by the user, the task of the WCM at the sink is to make sure that the number of packets successfully received at the sink, $\rho_S$, in $T$ time slots satisfies
\begin{equation}
  \rho_S \geq T\cdot R_{WSN}.
\end{equation}
Similarly, since $N_C/N_{WSN}$ is the ratio of the number of nodes within a cluster to the total number of nodes, the WCM at the CH can calculate its throughput by counting the number of successfully received packets per time slot and making sure that the number of packets successfully received at the CH, $\rho_{CH}$, in $T/4$ time slots satisfies
\begin{equation}
  \rho_{CH} \geq \frac{T\cdot N_C \cdot R_{WSN}}{4N_{WSN}}.
\end{equation}
Thus, the designed WCM considers $EnC$ and $R_{WSN}$ as the end-to-end goals. It considers transmit power, data rate, duty cycle control, routing, source loading rate control, and scheduling active nodes as tools for achieving these goals. The environment variables that trigger the WCM to operate are PLR, channel utilization, buffer capacity, and node failure. Fig. \ref{fig:Amr15} shows the overall WCM that is installed at the CHs and sink node.

The dashed edges in Fig. \ref{fig:Amr15} are the ones going in/out of the goals of the WCM. Thus, every $T/\eta$ time slots the WCM checks if any environment variable or goal violate their predefined threshold in order to take appropriate actions. For example, if cluster energy consumption is too high, the WCM has the option to decrease transmit power, decrease the number of active nodes, decrease the source loading rates, or invoke routing to find more energy efficient paths. This action is determined based on the current status of the network. For instance, if energy consumption is too high and the throughput is too low, then the WCM will not decrease the number of active nodes or loading rate because this will decrease throughput further. Instead routing may be invoked and transmit power reduced. This way, the WCM is able to consider multiple conflicting issues and deal with several challenges simultaneously.

\subsubsection{Performance analysis}
In order to study the performance of the WCM system, extensive computer simulations are performed. The WCM system is compared to TABU-RCC \cite{Anagnostopoulos1}, which is a prominent network planning protocol, and Symphony \cite{Al-Rabayah1}, which is an energy-efficient transmit power and data rate control protocol. Different mobility scenarios are considered in order to examine the performance of the WCM system in dynamic environments. In the simulation set-up, sensor nodes have an average speed of 1.5 m/s and the network performance is tested under different network sizes. The WSN is assumed to be running an application with target network lifetime of at least 1500 time slots and target throughput of at least 50 kbps.

\begin{figure}[!b]
\centerline{\subfigure[]{\includegraphics[width=3.5in]{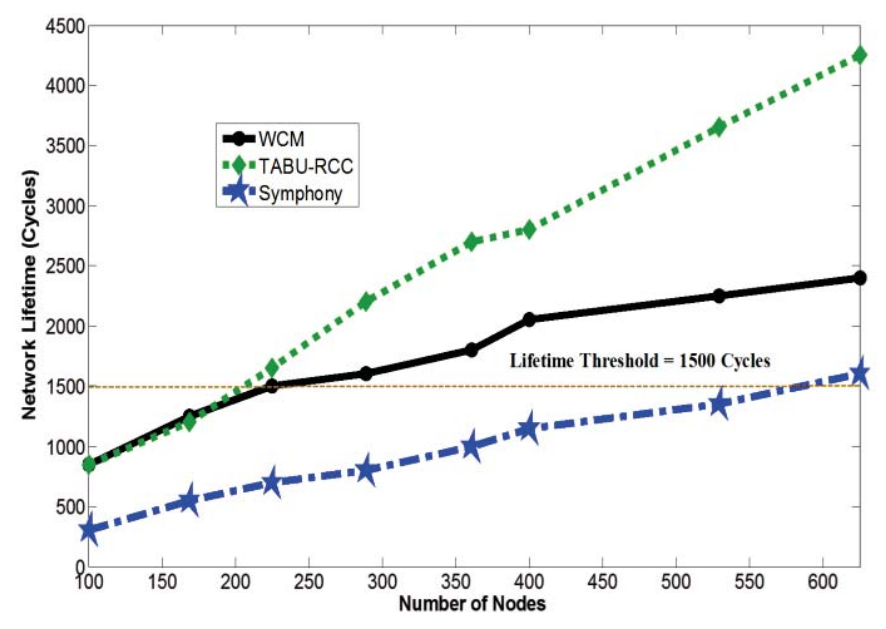}
\label{fig:Amr16a}}}
\centerline{\subfigure[]{\includegraphics[width=3.5in]{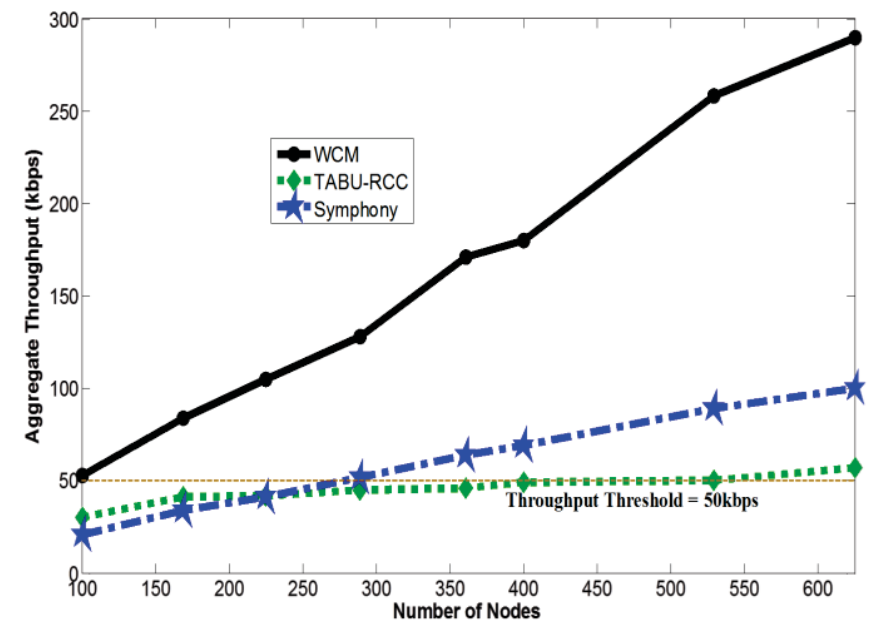}
\label{fig:Amr16b}}}
\caption{Simulation results of WCM: (a) Network lifetime; (b) Network throughput.}
\label{fig:Amr16}
\end{figure}

It can be seen from the simulation results in Fig. \ref{fig:Amr16} that the WCM system is the only one capable of achieving both (conflicting) goals at most network sizes. This is because of the ability of the WCM system to find compromises between conflicting goals and consider multiple issues simultaneously. TABU-RCC achieves good network lifetime results at the expense of throughput, while Symphony achieves only good throughput results at larger network sizes.

The performance of the WCM can be further improved through RL. As mentioned earlier, RL can be implemented in a distributed way and requires limited computations. This section presents $Q$-learning \cite{El-Mougy1} to illustrate how to design a distributed learning process. Here the learning process will observe the actions of the WCM system (i.e., the reconfigurations made) and assigns rewards to those reconfigurations that achieve the predefined goals.

\subsubsection{Learning}
Since the WCM engine is implemented at the CHs, the learning process is also implemented there. The $Q$-learning algorithm consists of: Policy, reward system, and value function. Thus, the state of any node will be represented using the combination \{transmit power ($T_x$), data rate ($R_{WSN}$), next hop index ($NI$)\}. These parameters are chosen because they can be controlled by the WCM and can have significant impact on the performance of several nodes in the network. This means that the set of actions (or reconfigurations) observed by the learning protocol are \{increase transmit power, decrease transmit power, increase rate, decrease rate, change next hop\}. The action \emph{change the next hop} occurs when routing is invoked.

To determine how the rewards are assigned, the application requirements are used to specify thresholds that can be monitored by the learning process. Therefore, if the reconfigurations made by the WCM lead to values of the aforementioned goals that are within the thresholds, a reward will be given. Otherwise the reward is zero. We specify an additional constraint that the thresholds of all goals have to be satisfied in order for the rewards to be assigned. If any goal is violated, the total reward is zero. This is to avoid accumulating rewards at actions that only achieve a subset of goals.

To implement distributed learning, DVF is used to combine knowledge from different CHs. Using DVF, the Q-values for each \{state, action\} pair is calculated as
\begin{equation}
\mathcal{Q}(S_t,a_t)    =   r_{t+1} + \sigma \sum_{j\in common\_zone} \omega_Q\times V_t^j (S_{t+1})
\end{equation}
where
\begin{equation}
V_t^j (S_{t+1}) =   \max_{a_{t+1}}  \mathcal{Q}(S_{t+1},a_{t+1})
\end{equation}
and $\omega_Q$ is a weight factor that can be used to give more importance to information from certain nodes. In this example, equal importance is given to all nodes. Thus, $\omega_Q = 1/|common\_zone|$, where $|common\_zone|$ is the number of nodes in the common zone. However, in order to limit the overhead incurred from exchanging values from the $Q$-table, only the values concerning the nodes on the boundaries of each cluster are exchanged. This is because these boundary nodes are the ones expected to interfere with the performance of other nodes in neighboring clusters as illustrated in Fig. \ref{fig:Amr18}.

\begin{figure}[!b]
\centerline{\subfigure[]{\includegraphics[width=3.5in]{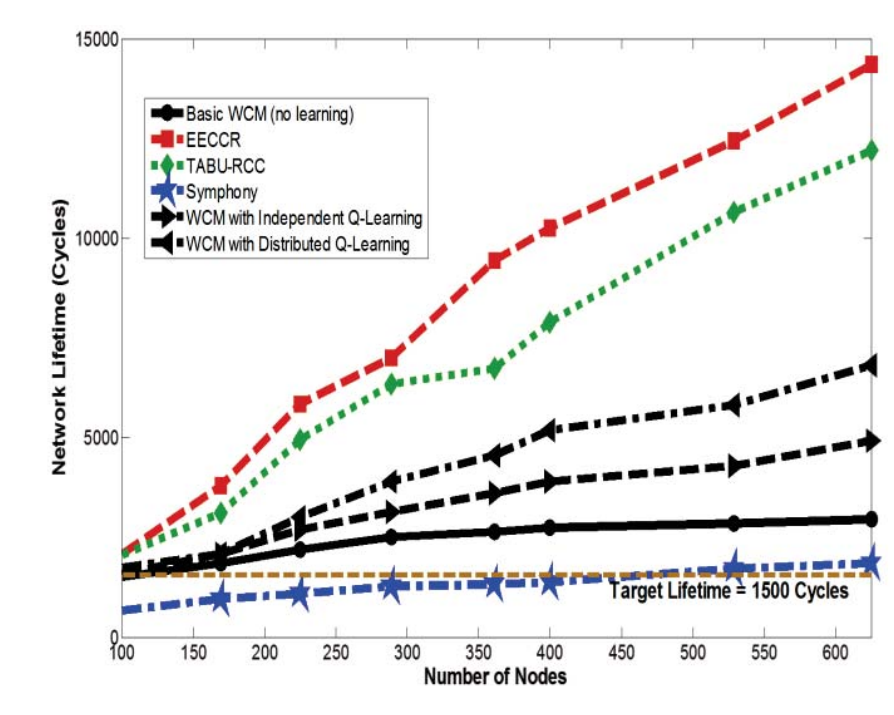}
\label{fig:Amr19a}}}
\centerline{\subfigure[]{\includegraphics[width=3.5in]{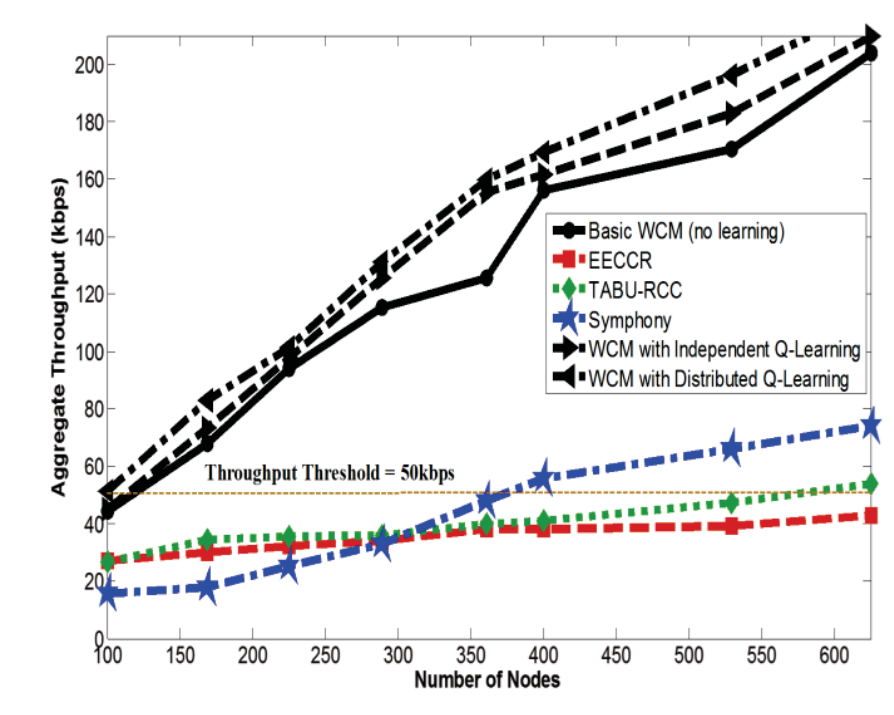}
\label{fig:Amr19b}}}
\caption{Simulation results of the learning protocol: (a) Network lifetime; (b) Network throughput.}
\label{fig:Amr19}
\end{figure}

In the simulation results shown in Fig. \ref{fig:Amr19}, the mobility of the nodes is fixed at 1.5 m/s. In addition, the learning protocol is compared to the energy-efficient coverage and connectivity routing (EECCR) protocol \cite{Saleet1}, as well as TABU-RCC, and Symphony. Moreover, an independent learning protocol is simulated, where CHs do not exchange values and do not use the DVF. This will give an idea about how the cooperation of CHs can enhance performance. This experiment considers an application with target network lifetime of at least 1500 cycles and target throughput of at least 50kbps, while in the second experiment assumes that the objective is to maximize network lifetime with no regard to throughput.

It can be seen that the learning protocol improves the performance of the WCM system. This is because of the ability of the learning protocol to accumulate knowledge that allows the WCM system to make choices for the reconfiguration actions to take place. Without the learning protocol, these choices are made blindly. In addition, distributed learning achieves better performance than independent learning, illustrating the importance of CH cooperation. Furthermore, the performance improvement is higher in larger networks. This is because larger networks have more options for the WCM to choose from.

\subsubsection{Complexity}
The WCM system imposes additional computational complexity and communication overhead on the regular WSN. However, it is important to notice that there are no loops in the execution of WCM algorithms. In addition, the WCM is an event-based system and the algorithms are executed upon detecting a significant change in one of the environment variables such as PLR or buffer capacity. Algorithms in other protocols that we have been using for comparison such as TABU-RCC or EECCR are executed periodically. This means that there is a chance they execute without there being a real need (e.g., when there are no changes in the network). With regards to communication overhead, the WCM system piggybacks control information on data packets. Special packets may be transmitted only if required, for example in the event of node failure, where immediate action is needed. This minimizes overhead and ensures robustness without incurring the risk of missing or delaying actions. Furthermore, the WCM system does not require centralized implementation. As we have seen, it can be implemented in CHs, with each CH controlling its local area. Note that TABU-RCC and EECCR protocols have to be implemented at the centralized sink node. Centralized solutions have to collect network-wide data and thus may incur high delays and communication overhead. Moreover, WCM is an entirely software-based network reconfiguration solution. Thus, implementing it does not require dedicated nodes or hardware modifications. To conclude, even though the WCM system imposes additional complexity, like any network management system, this extra complexity is minimum and justified by the significant improvement in performance.

\subsection{Context-Awareness (CA)}
CA is a classical research area in pervasive computing that has recently been adopted in mobile and wireless networking \cite{Makris1,El-Mougy2}. Technically speaking, CA can be a part of any cognitive networking or cross-layer reconfiguration framework, which is why we think it deserves a section on its own. In addition, the growth of heterogeneous and ubiquitous networking makes CA quite important. CA targets the minimization of uncertainty in decision making by providing necessary information \cite{Makris1}. This improves the efficiency, confidence, and user experience in the application.  CA is basically a collection of measured and inferred knowledge that arises from the general activity of a system and can exist independently of any system interactions \cite{Makris1}.

There are four main types of uncertain context information that can appear in wireless networks \cite{Makris1}: Imperfect, ambiguous, wrong, or unknown context information. Imperfect context information can be used directly or can be used to infer other information. It can arise from inaccurate, incomplete, or outdated context information. Unreliable communication links can be a cause of that. For example, in high mobility situations such as VANETs, links may break frequently causing context information to be missed. Eventually, information within one node may become obsolete. The second type of uncertain context information is ambiguous information. This can be the outcome of operating in heterogeneous environments where conflicting policies exist. For example, one node may decide to act selfishly to maximize its own throughput, thus causing a negative impact on the overall network performance. The third type of uncertain context information is wrong information, resulting from erroneous, inadequate, or inapplicable information. This can result from delivering information to the wrong entity or delivering incomplete information due to channel unreliability. Finally, unknown context information is generally expected information that has not been delivered. This may result from failures or node isolations.

CA systems try to bridge some of these information uncertainties. For example, all three cognitive standards discussed in the previous section include different forms of CA \cite{Tsagkaris1}. RRS includes specifications for acquisition of long-term and short-term context information. On the long-term the network collects information such as network status and spectrum availability, while on the short-term the information includes channel conditions, neighbor nodes, QoS parameters, and other information that may be necessary for JRRM. Similarly, IEEE 1900.4 collects context information about available spectrum bands, policies and regulations, and status of RANs. In addition, SON has an extensive CA framework that addresses different cases. For example, to minimize inter-cell interference, eNodeBs exchange information about interference indicators and transmit power used in the uplink and downlink directions. In addition, eNodeBs may exchange information about their current traffic load in order to perform load balancing.

The importance and potential of CA has been recognized by many researchers. Thus, in \cite{Hasswa1} the mobility management protocol known as Tramcar is proposed, which uses context information to make vertical handover decisions. This protocol considered context information about the cost of wireless services, power consumption, resource availability, network conditions, and security issues. At the application layer, user preferences are first translated into networking parameters. Then, a handoff necessity estimator uses context information about location, time, and places of events to avoid frequent handovers. In addition, the transport layer detect the presence of other networks and implement the decisions of the protocols at the application layer.

With regards to mobility management, the work in \cite{Fernandes1} has defined twelve necessary requirements for a context-aware system. These are: The availability of reliable information from all layers, standardized network discovery (particularly based on the IEEE 802.21 Media Independent Handover system \cite{IEEE5}), post-handover self-management, consideration of the application profile, use of a single radio interface, pre-authentication (before handovers are executed), the consideration of multiple available networks, use of vendor independent handoff trigger indicators such as the number of packet retransmissions, it must take into account quality-of-experience (QoE) factors (such as costs and user preferences) as well as QoS factors, it must work with several protocols, admission control, and finally making the actual handover decision after considering all the parameters. Thus, the work in \cite{Fernandes1} provides a framework for any researcher wishing to propose context-aware mobility management systems. Other works include the framework for context management for next generation networks \cite{Baladron1}, the framework for context mediation proposed in \cite{Roy1}, and the framework for collaborative CA proposed in \cite{Anagnostopoulos1}.

CA has been utilized in ad hoc networks as well. However, CA systems in ad hoc networks usually take a narrower perspective (such as focusing on one particular issue or a small subset of objectives) due to the lack of a centralized node. Location awareness in particular has found extensive use in different types of ad hoc networks such as in VANETs, where it has been used for issues such as routing \cite{Al-Rabayah1}, QoS support \cite{Saleet1}, and security \cite{Lu_Rongxing1}. Mobility awareness has also been utilized for routing in mobile ad hoc networks \cite{Manjappa1}. In addition, CA has been used for data fusion in WSNs \cite{Ren_Fengyuan1} to improve the efficiency of the process.

\section{Conclusions and Open Research Topics}
The demand for sophisticated and ubiquitous applications has shifted the paradigm of wireless networks to reconfigurable wireless networks. Nevertheless, the scattered research in this area hinders the gains that can be achieved by reconfiguration. This paper presented an overview of reconfigurable wireless networks with an in-depth discussion of reconfiguration at all layers of the protocol stack. Reconfigurable networks have been classified in the paper into three levels: Reconfiguration at the physical layer, reconfigurable networking, and network intelligence. The paper focused on reconfigurable systems in emerging areas such as cognitive radio networks, cross-layer reconfiguration and software-defined networks. Moreover, the paper covered context awareness and cognitive networking as key enablers of reconfiguration in highly complex and dynamic networks. Furthermore, three case studies have been detailed, one for each level of reconfigurable networks.  The first targeted management of CRNs where centralized and distributed network-based reconfigurations were used. The second was dedicated to cross-layer reconfiguration in M2M networks based on game theory. The third case study covered network intelligence in WSNs where a reasoning and learning framework was established to enable effective reconfiguration in WSNs. Several examples and experimental results have been presented in the paper illustrating the superiority of reconfigurable networks over non-reconfigurable ones, especially in the case of highly complex and dynamic networks. These results show that the reconfigurable wireless network paradigm will be a key enabler for next-generation wireless communications networks and services.

Although research in reconfigurable wireless networks has been quite extensive, there remain areas that need more attention. In this section, some of these areas are briefly discussed.

\subsubsection{Learning}
This is quite a powerful tool with great potential, which is why it has been recommended by standards such as RRS. Nevertheless, work in this area still faces challenges. One issue is the learning speed. The learning system has to observe the environment in order to accumulate knowledge. The time this takes depends on the parameters being considered. In dynamic environments, the parameters may be changing too frequently, rendering the learning system useless. To address this challenge, researchers typically confine the focus of the learning system to only a few parameters that can guarantee convergence. Another solution is to implement learning in a distributed way, across multiple agents. In this case, each agent collects information only from a subset of its neighbors. This may lead to suboptimal solutions and requires collecting agents that can aggregate the learned information. It also may require asynchronous operation, where each agent executes its learning algorithm according to the parameters in its neighborhood. In addition, the performance of learning systems in noisy environments, interference, changing topologies, or node failures needs investigation. In these situations, the learning system may become unstable and may have a negative effect on network performance. Finally, combining learning tools with CA approaches can have great potentials. Learning algorithms can be adapted to choose the most important parameters, given the right context, while CA systems can get more information about the history of the data being considered.

\subsubsection{Coexistence and Interoperability}
The techniques discussed in this paper provide solutions for a wide variety of challenges. However, there is limited work on the performance of reconfigurable systems under the coexistence of multiple solutions. For example, a learning system that targets the minimization of energy consumption may guide nodes to reduce their transmit powers. This may have an impact on packet loss, interference, and routing. Thus, an independent routing solution that is not considered in the learning process may suffer negatively. Interoperability is also another challenge. The world of wireless networks now includes a variety of standards, protocols, vendors, operators, etc. These are not always compatible. SDN is a promising solution to address this challenge, as explained above. However, SDN requires hardware modifications that may be expensive. The alternative direction that is currently pursued is to build increasingly complex reconfigurable devices capable of dealing with the incompatibilities of the different network components.

\subsubsection{Intelligence in Reconfigurable CRNs}
Most of the research in CRNs still focuses on the PHY layer. As indicated in this paper, tangible performance gains can be achieved if network-based reconfiguration is applied. Such reconfiguration can be achieved globally via an FC (in centralized CRNs), or locally via learning and consensus-based schemes (in decentralized CRNs). For multiband sensing, the existing literature has been limited to homogenous multiband detectors (which consist of single-type sensors), and thus reconfiguration has been restricted to adapting the sensors' parameters. Nevertheless, as shown in this paper, further gains can be accomplished if more than one sensor type are available at the SU device. In particular, it has been shown that the RMD outperforms the well-known MJD in terms of the aggregate throughput of the CRN. This must motivate the research community to further investigate low-cost and reconfigurable heterogenous multiband detectors.

\subsubsection{Context-Awareness}
There are still open issues in wireless network reconfiguration based on CA. One important challenge is the tradeoff in considering higher level versus lower level context information. On one hand, lower level context information enables fine-grained adaptation of network parameters. However, the parameters available in this level are numerous and the effect of considering only a subset of parameters must be considered. On the other hand, higher level information has a better chance to directly address the QoE of the user, but may not lead to fine adaptations. A promising direction that needs further investigation is context sharing. In a cooperative network, exchanging context information may lead to the improvement of all cooperating nodes. However, this consumes bandwidth. Thus, deciding which information is worth sharing with which node is important. There is also a need for researching various reasoning and learning techniques, and how they may coexist especially in complex dynamic networks.

\subsubsection{Security and Privacy}
Research in this area is usually performed independently of research in other branches in networking. In today’s ubiquitous networking world, security has to be considered at the design stage, not as an afterthought. For example, CA requires gathering information about particular parameters. This may potentially violate certain privacy issues such as location, which is quite important in many networks such as VANETs. In addition, reconfiguration in heterogeneous environments has potential security risks. Not all networks have the same security guarantees. Thus, choosing a particular wireless service may introduce vulnerabilities. If the violated node includes critical information about another network (such as an SDN controller managing multiple networks), then the vulnerability will stretch across multiple networks. Thus, trust management and security have to be an integral part of the reconfiguration process.

\bibliographystyle{IEEEtran}
\bibliography{IEEEabrv,References}

\end{document}